%% file: webb2022_df44_sfhs.tex
\documentclass[fleqn,usenatbib]{mnras}

\usepackage{newtxtext,newtxmath}
\usepackage{placeins}
\usepackage{stfloats}

\usepackage[T1]{fontenc}

\DeclareRobustCommand{\VAN}[3]{#2}
\let\VANthebibliography\thebibliography
\def\thebibliography{\DeclareRobustCommand{\VAN}[3]{##3}\VANthebibliography}

\usepackage{graphicx}	
\usepackage{amsmath}	

\newcommand{\prospector}{{\sc Prospector}}
\newcommand{\alf}{{\texttt alf}}
\newcommand{\thh}{$^\mathrm{th}$}
\newcommand{\logzsol}{$\log(Z_\ast/\mathrm{Z}_\odot)$}
\newcommand{\dust}{$\hat{\tau}_\mathrm{dust,~diffuse}$}
\newcommand{\aD}{$\alpha_\mathrm{D}$}
\newcommand{\aDo}{$\alpha_\mathrm{D}=1$}
\newcommand{\aDt}{$\alpha_\mathrm{D}=0.2$}

\title[DF44's SFH]{Still at Odds with Conventional Galaxy Evolution: The Star Formation History of Ultra-Diffuse Galaxy Dragonfly 44}

\input{authors}

\date{Accepted XXX. Received YYY; in original form ZZZ}

\pubyear{2022}

\graphicspath{{./}{figures/}}

\begin{document}
\label{firstpage}
\pagerange{\pageref{firstpage}--\pageref{lastpage}}
\maketitle

\begin{abstract}
We study the star formation history (SFH) of the ultra-diffuse galaxy (UDG) Dragonfly 44 (DF44) based on the simultaneous fit to near-ultraviolet to near-infrared photometry and high signal-to-noise optical spectroscopy. 
In fitting the observations we adopt an advanced physical model with a flexible SFH, and we discuss the results in the context of the degeneracies between stellar population parameters.
Through reconstructing the mass-assembly history with a prior for extended star formation (akin to methods in the literature) we find that DF44 formed 90~per~cent of its stellar mass by $z\sim 0.9$ ($\sim 7.2$~Gyr ago).
In comparison, using a prior that prefers concentrated star formation (as informed by previous studies of DF44's stellar populations) suggests that DF44 formed as early as $z\sim 8$ ($\sim 12.9$~Gyr ago).
Regardless of whether DF44 is old or very old, the SFHs imply early star formation and rapid quenching. 
This result, together with DF44's large size and evidence that it is on its first infall into the Coma cluster, challenges UDG formation scenarios from simulations that treat all UDGs as contiguous with the canonical dwarf population. 
While our results cannot confirm any particular formation scenario, we can conclude from this that DF44 experienced a rare quenching event.
\end{abstract}




\section{Introduction}

Matching predictions to observations of how, and when, galaxies assemble serves as an important test for our greater understanding of cosmology and baryonic physics.
Modern theories that suggest galaxy evolution is determined by the growth of their dark matter haloes, as well as the regulation of their gas processes (i.e., infall and star formation histories; e.g., \citealt{white1991, schaye2010, dave2012, wechsler2018}), have successfully replicated some observed relations between galaxy properties -- 
for example, the tight connection between stellar mass and halo mass (i.e., the SMHM relation; \citealt{moster2010}).
A number of outstanding issues remain, however.
A particularly challenging problem is explaining the increasing number of galaxies that cease forming stars (i.e., `quench') over time \citep{renzini2006, faber2007}. 
While simulations correctly predict scaling relations for massive galaxies (e.g., the mass--metallicity relation; MZR, and star formation main sequence), there are still fundamental discrepancies at lower stellar masses.

In the low mass regime, observations have shown that quenched galaxies associated with massive host haloes are rare \citep{geha2012}, such that quenching at $z<1$ is thought to predominantly be a result of environmental effects \citep[e.g.,][]{boselli2006, fillingham2018, mao2021}. 
Rather than remain quenched, recent studies instead suggest that isolated quiescent dwarfs may in fact oscillate between `star forming' and `quenched' states \citep[e.g.,][]{polzin2021}.
Yet cosmological simulations typically over-predict the abundance of quiescent field dwarfs \citep[e.g.,][]{dickey2021}.

\vspace{0.2cm}
The recently discovered ultra-diffuse galaxies (UDGs) potentially exemplify our limited understanding of the true diversity of galaxy evolution and quenching.
UDGs were initially noted for their surprisingly large sizes given their low surface brightnesses ($R_\mathrm{eff}\geq1.5$~kpc and $\mu_\mathrm{0}(g)\geq24$~mag~arcsec$^{-2}$; \citealt{vandokkum2015_udg}) which, along with their red colours,
distinguished them from classical low surface brightness (LSB) galaxies \citep[e.g.,][]{dalcanton1997}. 

Several current cosmological model predictions suggest that conventional processes can explain the UDG population, thus maintaining standard dark matter halo occupancy relations \citep[e.g.,][]{tremmel2020}. 
Such models typically focus on the mechanisms which increase the size of otherwise canonical dwarf galaxies to make them `ultra-diffuse' (for a summary of UDG origins see \citealt{jiang2019a}).
Simulations have shown that unusual star formation or galaxy evolution processes can `puff up' canonical dwarfs (e.g., high-spin scenarios, \citealt{amorisco2016, rong2017}; energetic star formation feedback, \citealt{dicintio2017, chan2018, jackson2021b}) or dynamically redistribute their stellar populations (e.g., tidal heating and/or stripping; \citealt{jiang2019a, liao2019, carleton2019, sales2020}). 
Alternatively, UDGs may represent the tail of galaxy evolution processes, such that only minor differences in their evolution (e.g., when they infall or have major mergers) distinguish their final properties from normal dwarfs \citep[e.g.,][]{tremmel2020, wright2021}.

Despite these differences, nearly all models rely on environmental processes to explain the lack of star formation in the subset of UDGs that are quiescent \citep[e.g., via ram pressure stripping;][]{yozin2015a, rong2017, chan2018, tremmel2020}.
Accordingly, all of the scenarios follow a dichotomy related to when UDGs infall into a cluster environment: whether the proto-UDGs surpassed the size-threshold prior-to or post infall, is tied to whether they infall `late' or `early'. 
While UDGs are found both in the field and clusters, those that are quiescent are usually located in clusters \citep[the few exceptions may be on backsplash orbits; e.g.,][]{papastergis2017, benavides2021}.
Explaining the origin of UDGs and the diversity of their properties in the context of their environments remains a key question in understanding galaxy formation and evolution. 

Testing the predicted UDG properties (e.g., kinematics, \citealt{amorisco2016}; stellar populations, \citealt{rong2017, ferremateu2018}; globular cluster (GC) properties, \citealt{carleton2021}; infall versus quenching times, \citealt{gannon2022}) from these scenarios against the observed properties, however, has revealed a number of discrepancies. 
And while some UDGs are found with very large sizes ($R_\mathrm{eff} > 4.5$~kpc), these exotic objects are beyond the predictions of most models \citep[][]{dicintio2017, carleton2019}. 
Along the same lines, models which accurately predict the distribution of UDG sizes fail to reproduce the distribution of sizes among normal dwarfs \citep[e.g.,][]{rong2017, jiang2019a, tremmel2020}.

On the other hand, \citet{vandokkum2015_udg} proposed that some UDGs originate similar to today's massive galaxies (and have sizes reflecting their massive haloes), but lost their gas early in their histories. 
As a result of their early quenching, these `failed' galaxies did not build up the stellar mass expected for their haloes. 
This scenario deviates from the expected galaxy--halo connection, in that either these failed galaxies do not follow the SMHM relation or at least have a larger scatter than the standard relation. 

\vspace{0.2cm}
A particularly interesting UDG is Dragonfly~44 (DF44) which is the largest galaxy in the original \citet{vandokkum2015_udg} sample, with {$R_\mathrm{eff}=4.7\pm0.2$~kpc} \citep{vandokkum2017}.
High S/N spectroscopy has revealed an extremely old and metal-poor stellar population \citep[$\sim 2.3\sigma$ below the canonical dwarf MZR;][]{villaume2022}, implying that DF44 quenched very early and over a short time-scale. 
Moreover, while DF44 appears to have very low rotation \citep{vandokkum2019} characteristic of dwarf spheroidal galaxies, the stellar population gradients are `inverted' compared to the gradients typical of dwarf spheroidals \citep{villaume2022}.
Regardless of whether DF44 has an over-massive halo or not \citep[][]{vandokkum2017, wasserman2019, bogdan2020, lee2020, saifollahi2021}, this UDG is inconsistent with the majority of UDG formation models. 

Late-quenching (after infall into a dense environment) scenarios can be ruled out for DF44 given its old age \citep[e.g.,][]{rong2017, chan2018, liao2019, jiang2019a, jackson2021b}. 
Moreover, DF44's low rotation conflicts with high-spin scenarios (e.g., \citealt{rong2017}; although the rotation could increase at larger radii, \citealt{grishin2021}).
Yet, given the uncertainty in establishing the cluster infall time for an individual galaxy, we cannot preclude early-infall scenarios \citep[e.g.,][]{yozin2015a, liao2019, carleton2019, carleton2021, tremmel2020}. 
While some evidence \citep[e.g.,][]{{alabi2018, vandokkum2019}} suggests that DF44 is on its first infall into Coma, this is difficult to prove. 

\vspace{0.2cm}
There is more to be learned, however, as UDG formation scenarios can be tested via their inferred star formation histories (SFHs). 
The time-scales of star formation reveal important epochs (e.g., mergers, infall, and/or quenching), which can be compared against observations.
A number of studies have investigated the ages and mass assembly histories of UDGs, relying either only on broadband colours, or low to moderate S/N spectroscopy ( e.g, \citealt{kadowaki2017, ferremateu2018, gu2018b, pandya2018, ruizlara2018, martinnavarro2019}; Buzzo et al. 2022, submitted).
While these studies provide important first steps, comparisons with predictions are not necessarily straightforward.
This is primarily because constraining the detailed shape of a galaxy's SFH is a complex problem.

Several galaxy properties can conspire to alter the spectral energy distribution (SED) in similar ways (e.g., stellar age, metallicity, and dust), which are particularly difficult to disentangle with low spectral resolution data \citep[e.g., with photometry alone;][]{bell2001}. 
Recovering the SFHs for old stellar populations is particularly difficult -- the integrated spectrum evolves non-linearly with age \citep{serra2007} such that old populations appear relatively similar \citep[for a complete discussion see the review by][]{conroy2013a}. 
Moreover, a late burst of star formation can `outshine' a (dominant) older population \citep[e.g.,][]{papovich2001, allanson2009}. 
While broad wavelength coverage is needed to precisely determine the dust absorption (and emission, with mid-infrared coverage), high resolution data of select spectral features are needed to precisely constrain the stellar metallicity and age.
Both observations are necessary to break the degeneracy between these parameters \citep[e.g.,][]{vazdekis1999, trager2000b}. 
Using spectra that span a relatively wide wavelength range, full-spectrum fitting has proven to be effective in this respect \citep[e.g.,][]{macarthur2009, sanchezblazquez2011}. 
However, this technique requires a well-calibrated spectral continuum. 
Simultaneously fitting photometry and spectra can bypass this issue, as the photometry provides a means to fit the continuum\footnote{In practice it is generally easier to calibrate photometry to standard filters than to calibrate a spectrum.} 
and increases the wavelength coverage. 

In fitting the data it is necessary to impose `prior knowledge',\footnote{`Prior' here is used in the Bayesian sense, where the probability of a model given the data (i.e.,the `posterior') is proportional to both the likelihood of the data (given the model) and the prior knowledge about the model.} 
such as the flexibility of the SFH. 
The choice of a prior for the shape of the SFH can significantly impact age estimates, particularly for older stellar populations, and for low resolution and/or low S/N data  \citep[as shown in, e.g.,][]{maraston2005, leja2017, leja2019a, han2018, carnall2019a}. 
In order to draw connections between the predicted and observed properties of UDGs it is necessary to give due attention to the choice of a prior. 
While it is advantageous to use flexible models together with physically motivated priors, a `good prior' is not necessarily known a priori. 
Therefore, results should be discussed in the context of the prior used (which may not be as `uninformative' as intended; e.g., \citealt{leja2019a}). 

\vspace{0.2cm}
In this work we simultaneously fit near-ultraviolet (NUV) to near-infrared (NIR) photometry (nine bands) with high S/N ($\sim 96$~\AA$^{-1}$) rest-frame optical spectroscopy (from KCWI, the Keck Cosmic Web Imager). The same data set was used in \citet{vandokkum2019} and \citet{villaume2022} to study the stellar kinematics and populations of DF44. 
We adopt flexible SFHs in our fiducial model which do not assume a certain
shape with time. 
Moreover, we compare the results between SFH priors of different degrees of `smoothness' in order to identify which results are fully constrained by the observations. 
We address the unique stellar population properties of this UDG, and its epoch of formation and quenching, in order to test models of UDG formation.

The data are described in Section~\ref{sec:data}, and Section~\ref{sec:fitting} details how we fit the data with an advanced physical model. 
In Section~\ref{sec:results} we discuss the results, and put the results in the context of the literature. 
What our results imply about the origins of DF44 in the context of theoretical models is discussed in Section~\ref{sec:discussion}. 
A summary of the key results is provided in Section~\ref{sec:conclusion}. 
The SFHs of DF44 determined by this work are listed in full in Appendix~\ref{app:fsfh_for_comparison}. 
We provide additional details on the above discussion in the Appendix, touching on systematic biases in measuring SFHs in Appendix~\ref{app:sfh_biases}, and degeneracies between dust extinction and the flux from old stellar populations in Appendix~\ref{app:old_v_dust}.

The magnitudes reported follow the AB magnitude system. We
use a Chabrier (2003) initial mass function (IMF), and adopt a flat $\Lambda$ cold dark matter ($\Lambda$CDM) cosmology with $\Omega_m=0.3$ and $ H_0 = 70~\mathrm{km}~\mathrm{s}^{-1}~\mathrm{Mpc}^{-1}$.

\section{Data} \label{sec:data}

Our data for DF44 include both rest-frame optical spectroscopy and NUV to NIR photometry, shown in Fig.~\ref{fig:spectrum_with_lines}, and described in more detail below.
We assume the spectroscopic redshift measured by \citet{vandokkum2017}: $z=0.02132\pm0.00002$.

\subsection{Spectroscopy} \label{sec:data_spectrum}

The spectroscopy is described in detail in \citet{vandokkum2019} and analysed further in \citet{villaume2022}; we summarise the relevant details here. 

Of particular note is the sky subtraction, as the sky is much brighter than the UDG. Sky exposures were obtained 1\arcmin.5 away from DF44 intermittently between DF44 observations. 
The wavelength-dependent time variation in the sky spectrum was obtained from the spatially collapsed individual sky spectra, as parameterised by principal component analysis (PCA).
The sky in each science cube was determined from a linear combination of templates, where the bestfit sky spectrum for the given exposure was subtracted from each spatial pixel. 
Additional details are provided in \citet{vandokkum2019}.

\begin{figure*}
  \begin{center}
    \includegraphics[width=\linewidth]{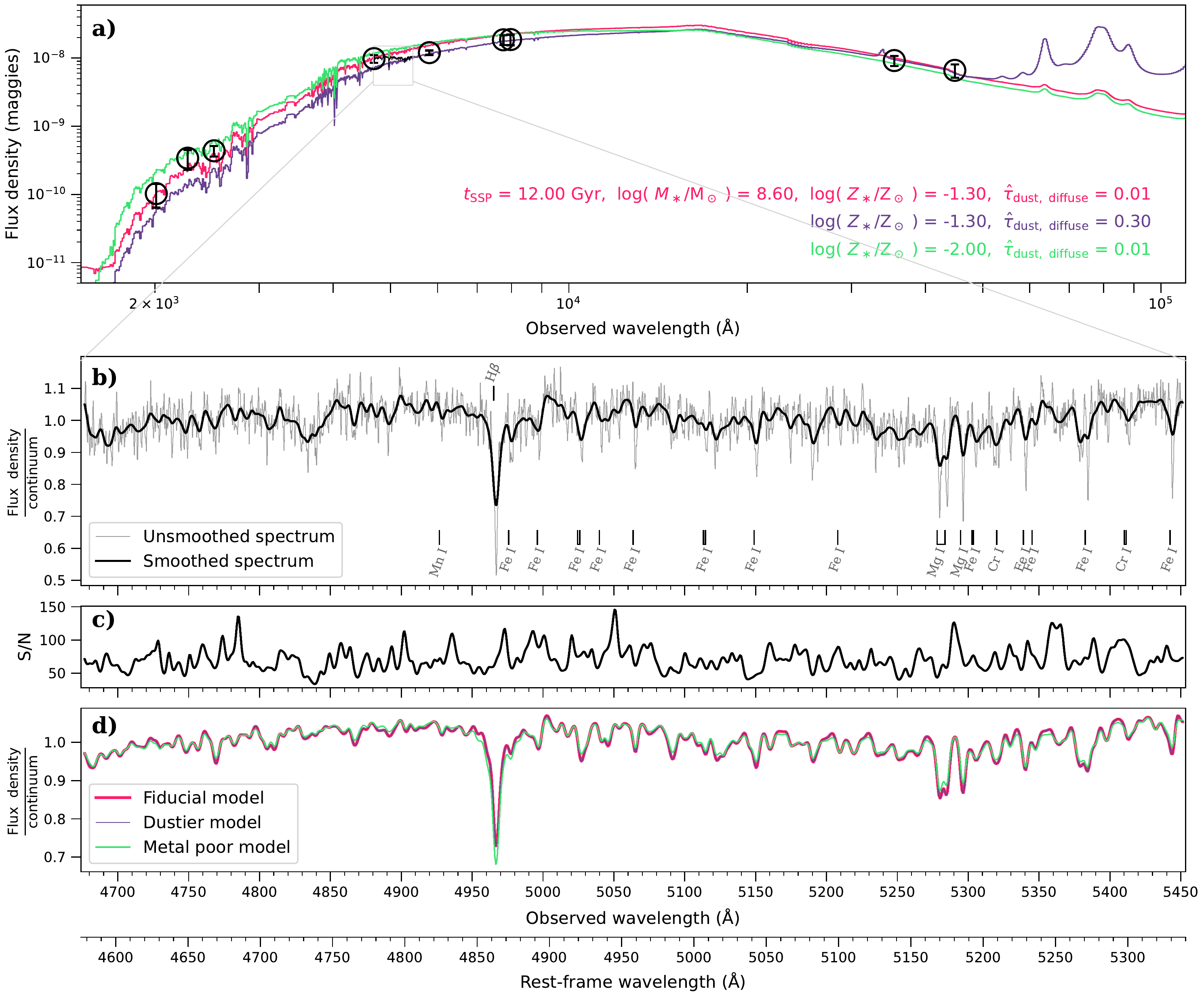}
  \end{center}
  \caption{Observations of DF44 and three simple stellar populations (SSPs) shown to briefly illustrate the effect of the dust attenuation and metallicity on the shape of the SED. 
  {\it (a)} The photometry with error bars (black markers) and model SEDs (coloured lines) with listed parameter values. 
  {\it (b)} The original (continuum removed) KCWI spectrum (thin grey line) and the smoothed spectrum (thick line) for fitting purposes. Significant spectral features are labelled for reference.
  {\it (c)} The S/N of the smoothed spectrum. 
  {\it (d)} The (continuum removed) model spectra, shown with the same resolution as the smoothed KCWI spectrum. 
  While changing the dust attenuation affects the shape of the overall continuum, changing the metallicity affects both the continuum shape and the absorption features.
  } \label{fig:spectrum_with_lines}
\end{figure*}

KCWI integral field spectroscopy was obtained for DF44 and spectra were extracted in nine elliptical apertures after masking the ten brightest point sources. 
The apertures were sized $9\arcsec \times 6\arcsec$, to match the UV photometry; see the following section.
The integrated spectrum was determined through bootstrapping the individual spectra, where we used the 50\thh\ percentile of the bootstrapped flux distribution and the average of the 16\thh\ and 84\thh\ percentile as the uncertainty.
With 17 hours of exposure on-target, the integrated spectrum reaches a $\mathrm{S/N}\sim 96$~\AA$^{-1}$ (see the third panel in Fig.~\ref{fig:spectrum_with_lines}). 

The KCWI Medium slicer with BM grating was used, yielding a spectral resolution of $\mathrm{R}\sim 4000$. 
After masking and interpolating over regions badly affected by sky transmission, the spectrum was smoothed to a resolution of 110~km~s$^{-1}$, for the purpose of later comparing with templates at this resolution. 
The final spectrum is shown in Fig.~\ref{fig:spectrum_with_lines} (the unsmoothed spectrum shown with grey lines), covering 4578--5337~\AA\ rest-frame, with notable absorption features labelled. 
Also shown is the S/N of the spectrum as a function of wavelength. 

Given the challenge of precisely flux-calibrating the spectrum (e.g., due to residuals from the spectral extraction), we instead rely on the calibration of the photometry to provide constraints on the SED continuum when fitting the galaxy properties and SFH (see Section~\ref{sec:fitting}). 
For this reason we do not flux calibrate the spectrum, and the continuum shape therefore reflects primarily the instrument response function and not the galaxy SED. 
We then effectively flatten the continuum by dividing through by a polynomial fit. In the fitting routine we therefore need to marginalise over the shape of the spectral continuum in comparing the models to the observations (see Section~\ref{sec:fitting_physical_model_speccal}). 

Lastly, we chose to mask the spectrum between 4700--4750~\AA\ rest-frame where there is a broad dip in the spectrum that does not appear in the models. We note that the blue end of the spectrum ($\lesssim$4800~\AA) was not fitted by either \citet{vandokkum2019} or \citet{villaume2022}. 
Our results are not impacted by masking this region of the spectrum, although the $\chi^2$ values are slightly higher without masking. 

\subsection{Photometry} \label{sec:data_phot}

Photometry in all the broadband images was performed by measuring fluxes within a $9\arcsec \times 6\arcsec$ elliptical aperture, with a position angle of 65 degrees, to be consistent with the UV photometry reported by \citet{lee2020}.  
As this is significantly larger than the image resolution in all filters, no point spread function homogenisation was applied, though appropriate aperture corrections are made to the {\it Spitzer} and {\it GALEX} images to account for light lost outside the aperture due to the point spread function. 
Details on the reduction and analysis of each image is described in more detail, below.
The photometric measurements in each broad-band filter were
corrected for foreground extinction in the Milky Way in the
direction of the Coma Cluster using the website \url{http://argonaut.skymaps.info/usage} and Table~6 of \citet{schlafly2011} with $R_\mathrm{V}=3.1$.

\subsubsection{Spitzer-IRAC Near-Infrared (NIR) Imaging}

{\it Spitzer}-IRAC \citep[][]{fazio2004, werner2004} observations of DF44 were taken on 2017 May 12 starting at 07:19 (UT). 
Both 3.6 and 4.5 $\mu$m (channels 1 and 2, respectively) observations were taken. 
50 medium-scale (median dither separation 53 pixels) cycling dither pattern 100 second frames were taken in each channel.
The total exposure time was $93.6\times50 = 4680$~s in channel 1 and $96.8\times50=4840$~s in channel 2.

We removed the `first-frame correction' (to address imperfect bias subtraction; see Section 5.1.10 of the IRAC Instrument Handbook). 
The rectification of each individual data frame for history effects in the IRAC arrays was performed in two steps that are explained in detail in \citet{pandya2018}.
In short, we first performed a per pixel correction that was based on IRAC idling time characteristics in the IRAC skydarks, matched to those that took place before our observations.
The typical magnitude of the per pixel correction was about $4$~kJy~sr$^{-1}$ in channel 1 and $1$~kJy~sr$^{-1}$ in channel 2. 
The typical corrections are much smaller than the read noise error and we do not add any systematic magnitude uncertainties due to these first-frame corrections.

In the second step, a mean background is calculated for each frame, and a function fitted to these means is subtracted.
The typical function consisted of a constant term plus terms that are declining exponentially with time.
The uncertainties in these first-frame effect corrections are negligible compared to other sources of systematic error.
We also formed a median image after doing a $3\sigma$ clipping from all the frames on the source in each channel and subtracted that median image separately in each frame.
Such a median image will subtract the residual images that have been formed on the detector from previous observations.
We determined that the uncertainty in the final magnitudes added by this step is less than $0.01$~mag.

The DF44 frames include a point source on top of the faint galaxy.
We used {\it Spitzer} Science Center provided software {\sc MOPEX}, specifically the APEX and APEX-QA modules, to subtract this point source using point response function (PRF) fitting.
The estimated uncertainty due to this step is about 0.5 micro-Jy in both channels.

We used the contributed {\it Spitzer}/IRAC software {\sc IMCLEAN} \citep{imclean} to remove leftover column pulldown artefacts from the CBCD frames.
We then used the {\it Spitzer} custom software package {\sc MOPEX} to create mosaics of the 50 frames in each channel, using the default parameters and the North up, East left orientation.
Before mosaicking we ran the overlap correction module to adjust for background offsets among the CBCD frames (one number per frame).
We used only the multiframe outlier rejection scheme in {\sc MOPEX} to reject outlier pixels in the input frames.

Next we manually created masks of other sources (including point-like sources on the galaxy) in both channels with the custom software {\sc GIPSY} \citep{gipsy}. 
We then measured the `sky background' in five empty areas of sky close to DF44 in channels 1 and 2, and from the results we estimated an average sky background (0.00408 and 0.00415 MJy sr$^{-1}$ in channels 1 and 2, respectively) to be subtracted at the position of DF44, applying the mask and using {\sc Astropy} Python library commands in a $9\arcsec \times 6\arcsec$  (P.A. +$65^{\circ}$) aperture centred on the coordinates given by \citet{vandokkum2015_udg}: $\mathrm{R.A.}=13^\mathrm{h}00^\mathrm{m}58^\mathrm{s}.0$, $\mathrm{Dec.}=26^{\circ}58\arcmin35\arcsec$. 
We corrected the results with the appropriate aperture corrections from the IRAC Instrument Handbook.

The uncertainty in aperture photometry was estimated by performing aperture photometry on several positions in empty sky and taking the rms scatter in these measurements.
This gave 0.05~and 0.10~mag in IRAC channels 1 and 2, respectively.
We estimated the uncertainty due to masking by replacing the pixel values under the masks by the average pixel values within the unmasked aperture, and performed the photometry again, and took the difference between this measurement and the measurement using the masks as the uncertainty.
The channel 1 masking uncertainty is thus $0.14$~mag, and $0.18$~mag for channel 2. 

The sky background subtraction uncertainty is 
estimated by taking the maximum difference in the sky background measurements in three areas of empty sky around DF44 in the images and adding this difference to all the pixels within the photometry aperture and summing them up.
This method gives $0.01$~mag and $0.11$~mag as the sky uncertainty in channels 1 and 2, respectively.

The calibration uncertainty was estimated to be $2$~per~cent in IRAC channels 1 and 2, amounting to $0.02$~mag in systematic uncertainty.
There is an additional uncertainty of $9$~per~cent in channel 1 and $2$~per~cent fractional flux in channel 2 due to the uncertainty in integrated aperture flux correction factor (limiting case is infinite aperture). 
These convert to $0.09$ and $0.02$~mag in channels 1 and 2. 
In addition there is the point source subtracting uncertainty of $0.01$~mag.

We list the final AB magnitudes for channels 1 and 2 and their respective uncertainties from the quadrature sum of the magnitude uncertainties in Table \ref{table:mags}.

\subsubsection{Gemini GMOS g- and i-Band Imaging}

DF44 was observed on 2017 May 12 with the Gemini Multi-Object Spectrometer (GMOS) for a total of 3000~s in both the $g$- and $i$-bands. 
The observations have been described by \citet{vandokkum2016}.
We flux-calibrated the images with SDSS, accounting for a $g-i$ colour term and using four SDSS catalogued stars in our images.
The data were obtained in photometric conditions, and we adopt an absolute calibration magnitude uncertainty to be $3$~per~cent, amounting to $0.03$~mag in the $g$- and $i$-bands, based on  \url{https://www.gemini.edu/instrumentation/gmos/calibrations}.
The sky background uncertainty was calculated as above for the IRAC channels, and amounted to $0.03$~mag in the $g$-band and $0.09$~mag in the $i$-band. 
Aperture photometry was performed using the coordinates from \citet{vandokkum2015_udg} and the {\sc Astropy} Python library commands. 

We list the final AB magnitudes for the $g$- and $i$-bands and the respective uncertainties in Table \ref{table:mags}.

\subsubsection{HST/WFC3/UVIS F606W and F814W imaging}

Additional visual images of DF44 were taken on 2017 April 23 with the {\it Hubble Space Telescope} using the WFC3 camera and its UVIS detector and broadband filters F606W and F814W.
\citet{vandokkum2017} reported $5\sigma$ AB depths of F606W$=28.4$ and F814W$=26.8$ for DF44. 
A total of 2430~s and 2420~s were spent on the source in F606W and F814W filters. 
In both filters we calculated the sky mode in five different `empty' regions of the sky and took an average and subtracted those values from the images.
We also manually masked out point sources in the images. 
We used the image headers to calculate the conversion from electrons/s to AB magnitudes and performed elliptical aperture photometry within the same apertures as mentioned above for IRAC.

The uncertainties were estimated in the following way: we estimate a photometric calibration offset uncertainty of $0.03$~mag, and the uncertainty due to background subtraction (estimated as above) is $0.05$~mag in F606W and $0.13$~mag in F814W.
The uncertainty due to masked point sources within the aperture is estimated to be $0.03$ and $0.01$~mag in F606W and F814W.
The uncertainty in performing aperture photometry was estimated as above and results in an additional $0.05$ and $0.14$~mag in F606W and F814W. 

We list the final AB magnitudes for F606W and F814W and the respective uncertainties in Table \ref{table:mags}.

\subsubsection{Ultraviolet}

The UV data reduction and analysis was presented in \citet{lee2020}.  
This consists of two filters observed with {\it Swift} UVOT (UV1 at 2600~\AA\ and UV2 at 1928~\AA), and {\it GALEX} NUV images.  
The UVOT data include a correction for red leakage and scattered light, where the correction (14~per~cent) was comparable to the flux uncertainty. 
Again we list the final results in Table~\ref{table:mags}.

\begin{table} \footnotesize
\centering
\caption{ DF44 Photometry. 
}
\label{table:mags}
\begin{tabular}{lrr}
\hline
     Filter &        m$_0$ (AB) & $\lambda_\mathrm{eff}$ (\AA) \\ 
\hline
UVOT UV1            &   $23.40\pm0.19$ 	& 2516.7 \\
UVOT UV2        	&   $24.97\pm0.41$ 	& 2010.4 \\
{\it GALEX} NUV     &   $23.67\pm0.35$ 	& 2271.1 \\
GMOS g\_G0301    	&  $20.02\pm0.14$  	& 4687.6 \\
GMOS i\_G0302   	&  $19.33\pm0.18$  	& 7751.6 \\
WFC3 F606W  		&  $19.80\pm0.08$ 	& 5813.0 \\
WFC3 F814W   	    &  $19.32\pm0.19$  	& 7972.9 \\
IRAC1               &  $20.09\pm0.18$  	& 35439.4 \\
IRAC2               &  $20.45\pm0.24$  	& 44840.9 \\
\hline
\end{tabular}
\end{table}

\section{Stellar population modelling and fitting} \label{sec:fitting}

\begin{table*} \footnotesize
\centering
\caption{SFH parameters and priors. Notes: 
1) Fraction of SFR in a given time bin, where the SFH is a piece-wise constant function with $N$ parameters ($N-1$ free parameters). The prior is a Dirichlet function, controlled by the parameter \aD, see Section~\ref{sec:fitting_physical_model_sfh}.
2) Redshift, with a tight prior about the measured spectroscopic redshift, $z_\mathrm{spec}$.
3) Total stellar mass is the integral of the SFH, which includes the mass lost to outflows. To convert to stellar mass \textit{remaining} at the time of observation we regenerate the spectral templates and subtract the mass lost as calculated by {\sc FSPS}. 
4) The total stellar metallicity where scaled-Solar $\alpha$-abundance is assumed.
5) Parameters for the two-component \citet{charlot2000} dust absorption model, with an adjustable attenuation curve slope from \citet{noll2009} with a UV bump based on \citet{kriek2013}. 
6) Parameters for the \citet{draine2007} dust emission model.
7) The uncertainty on the spectra can be increased by a given factor, with a likelihood penalty for factors giving reduced $\chi^2{<}1$.
8) An outlier pixel model can increase the errors for individual pixels by a factor of 50, to accommodate for poor matches between the data and spectral templates.
9) A fourth degree Chebyshev polynomial is fit (via optimisation) to the residual of the normalised ratio between the observed spectrum and the proposed model spectrum and multiplied out prior to each likelihood calculation. This effectively accounts for the lack of flux-calibration in the spectrum. 
}
\label{tab:params}
\begin{tabular}{p{0.03\linewidth} p{0.13\linewidth} p{0.36\linewidth} p{0.38\linewidth}}
\hline
Note & Parameter	& Description & Prior \\
\hline
& \multicolumn{2}{l}{SFH} \\
\hline
1 & $f_n$
					             	& sSFR fraction.  & Dirichlet(\aD) 		\\
2 & $z_\mathrm{obs}$				& Redshift					& Uniform(min$=z_\mathrm{spec}-0.01$, max$=z_\mathrm{spec}+0.01$)		\\

3 & $\log\left(M_\ast/\mathrm{M}_{\sun}\right)$	& Total stellar mass formed			& Uniform( min$ = 8$, max$ = 12$)		\\
4 & $\log\left(Z/\mathrm{Z}_{\sun}\right)$	& Stellar metallicity			& Uniform(min$ = -2$, max$ = 0.19$)		\\
\hline

& \multicolumn{2}{l}{Dust attenuation}  \\
\hline
5 & \dust		& Diffuse dust optical depth (eq.~\ref{eqn:dust_diffuse_2}) 	& Uniform(min$ = 0$, max$ = 1.5$)		\\
 & ${\hat{\tau}_\mathrm{young}}/{\hat{\tau}_\mathrm{dust,~diffuse}}$	    	& Ratio of diffuse to birth-cloud dust optical depth (eq.~\ref{eqn:dust_diffuse_1}) 	& Clipped Normal( $\mu=1$, $\sigma=0.3$, min$ = 0$, max$ = 1.5$)		\\
& $n_\mathrm{dust}$	    	& Diffuse dust attenuation index 	& Uniform(min$=-2$, max$=0.5$)		\\
 \hline
& \multicolumn{2}{l}{Dust emission}   \\
\hline
6 & $Q_\mathrm{PAH}$   		  & Percent mass fraction of PAHs in dust 	& Uniform(min$=0.5$, max$=7$)		\\
 & $U_\mathrm{min,dust}$   	  & Minimum starlight intensity to which the dust mass is exposed 	& Uniform(min$=0.1$, max$=25$)		\\
 & $\gamma_\mathrm{dust}$     & Mass fraction of dust in high radiation intensity 	& LogUniform(min$=0.001$, max$=0.15$)		\\
 \hline
& \multicolumn{2}{l}{Noise model}   \\
\hline
7 & spec\_jitter				    & Multiplicative spectral noise inflation term  	& Uniform(min$ = 1$, max$ = 3$)		\\
8 & $f_\mathrm{outlier,~spec}$	    & Fraction of spectral pixels considered outliers 	        & Uniform(min$ = 10^{-5}$, max$ = 0.5$)		\\
\hline 
& \multicolumn{2}{l}{Spectrophotometric calibration}   \\
\hline
9 & $c_n$               & Chebyshev polynomial coefficients, $n=4$ 	& 		\\
 \hline 
\end{tabular}
\end{table*}

In this section we describe how we fit the DF44 observations using the fully Bayesian inference code \prospector\footnote{https://github.com/bd-j/prospector} \citep[v1.0][]{leja2017, johnson2019, johnson2021}. 
The photometry and spectroscopy are fitted simultaneously, incorporating the information on the stellar properties and SFH from both data sets. 
In Section~\ref{sec:fitting_physical_model} we describe the advanced physical model, which includes a non-parametric SFH and a flexible dust attenuation law.  
We additionally include a white noise and spectral outlier model described in Section~\ref{sec:fitting_physical_model_noise_and_outliers}, and a spectrophotometric calibration model which marginalises out the shape of the spectral continuum, in Section~\ref{sec:fitting_physical_model_speccal}. 
A summary of the parameters and priors of our physical model is shown in Table~\ref{tab:params}. 
Section~\ref{sec:fitting_sampling} briefly describes the sampling method.

\subsection{The physical model}\label{sec:fitting_physical_model}

The physical models are based on the stellar population synthesis (SPS) models from the Flexible Stellar Population Synthesis library \citep[{\sc FSPS};][]{conroy2009, fsps} with {\sc MESA} Isochrones and  Stellar Tracks ({\sc MIST}; \citealt{choi2016, dotter2016}, based on the {\sc MESA} stellar evolution code; \citealt{paxton2011,paxton2013,paxton2015,paxton2018}), and {\sc MILES}\footnote{http://miles.iac.es/} spectral templates \citep{sanchez-blazquez2006}. 

The dust is modelled with the two-component dust attenuation model from \citet{charlot2000}, which separates the dust components between those associated with the birth-cloud, and a uniform dust screen. 
While we expect DF44 to have an old stellar population with very little dust content, we prefer to include a flexible dust model and marginalise over the parameters rather than assume a simplistic model. 
This avoids the assumption that dust attenuation in DF44 is the same as dust attenuation in the local Universe. 
The birth-cloud dust acts to only attenuate stellar emission for stars younger than 10~Myr,
\begin{equation}\label{eqn:dust_diffuse_1}
 \tau_\mathrm{dust,~birth}(\lambda) = \hat{\tau}_\mathrm{dust,~birth} \left( \frac{\lambda}{ \text{5500~\AA} }\right)^{-1} 
\end{equation}
\noindent while the diffuse-dust acts as a uniform screen with a variable attenuation curve \citep{noll2009},
\begin{equation}\label{eqn:dust_diffuse_2}
 \tau_\mathrm{dust,~diffuse}(\lambda) =\frac{\hat{\tau}_\mathrm{dust,~diffuse}}{4.05} \left( k^\prime(\lambda) + D(\lambda) \right) \left(\frac{ \lambda }{ \text{5500~\AA} }\right)^{n} 
\end{equation}
\noindent where $n$ is the diffuse dust attenuation index, $k^\prime(\lambda)$ is the attenuation curve from \citet{calzetti2000}, and $D(\lambda)$ describes the UV bump based on \citet{kriek2013}. 
The diffuse dust is given a uniform prior (min$=0$, max$=1.5$). 
We note that the diffuse dust optical depth is related to the dust extinction via $A_\lambda = 2.5~\log_{10}(e)~\tau_\lambda$, where $\tau_\lambda$ is the sum of the diffuse and birth dust components.

We use a joint prior for the ratio of diffuse to birth-cloud dust, rather than a direct prior on birth-cloud dust, to avoid degeneracies between the two parameters. 
The prior for ${\hat{\tau}_\mathrm{young}}/{\hat{\tau}_\mathrm{dust,~diffuse}}$ is a clipped normal with $\mu=1$, $\sigma=0.3$, min$=0$, and max$=1.5$, which broadly follows results from the literature for massive galaxies while allowing some variation. 
Lastly the prior on the attenuation index is uniform (min$=-2$, max$=0.5$).

Dust emission is calculated assuming energy conservation, i.e., all the energy attenuated by dust is re-emitted at infrared wavelengths \citep{dacunha2008}. 
As our photometry is limited to $<4.4~\mu$m (rest-frame) there is no significant information in the SED constraining dust emission. 
We chose to include the full dust model and marginalise over the unconstrained parameters, rather than a more simplistic model, in order to avoid biasing the result. 

The stellar metallicity is a free parameter; however we assume a constant metallicity for all the stars and for the entire history of the galaxy. 
This single metallicity has a uniform prior in \logzsol\ (min$=-2$, max$=0.19$), where $\mathrm{Z}_\odot = 0.0142$ \citep{asplund2009}. 
In addition, we assume scaled-Solar abundances, which is a current limitation of the {\sc FSPS} models.
Lastly, a \citet{chabrier2003} IMF is used. 

\subsubsection{Non-parametric SFH}\label{sec:fitting_physical_model_sfh}

\begin{figure}
  \begin{center}
    \includegraphics[width=\linewidth]{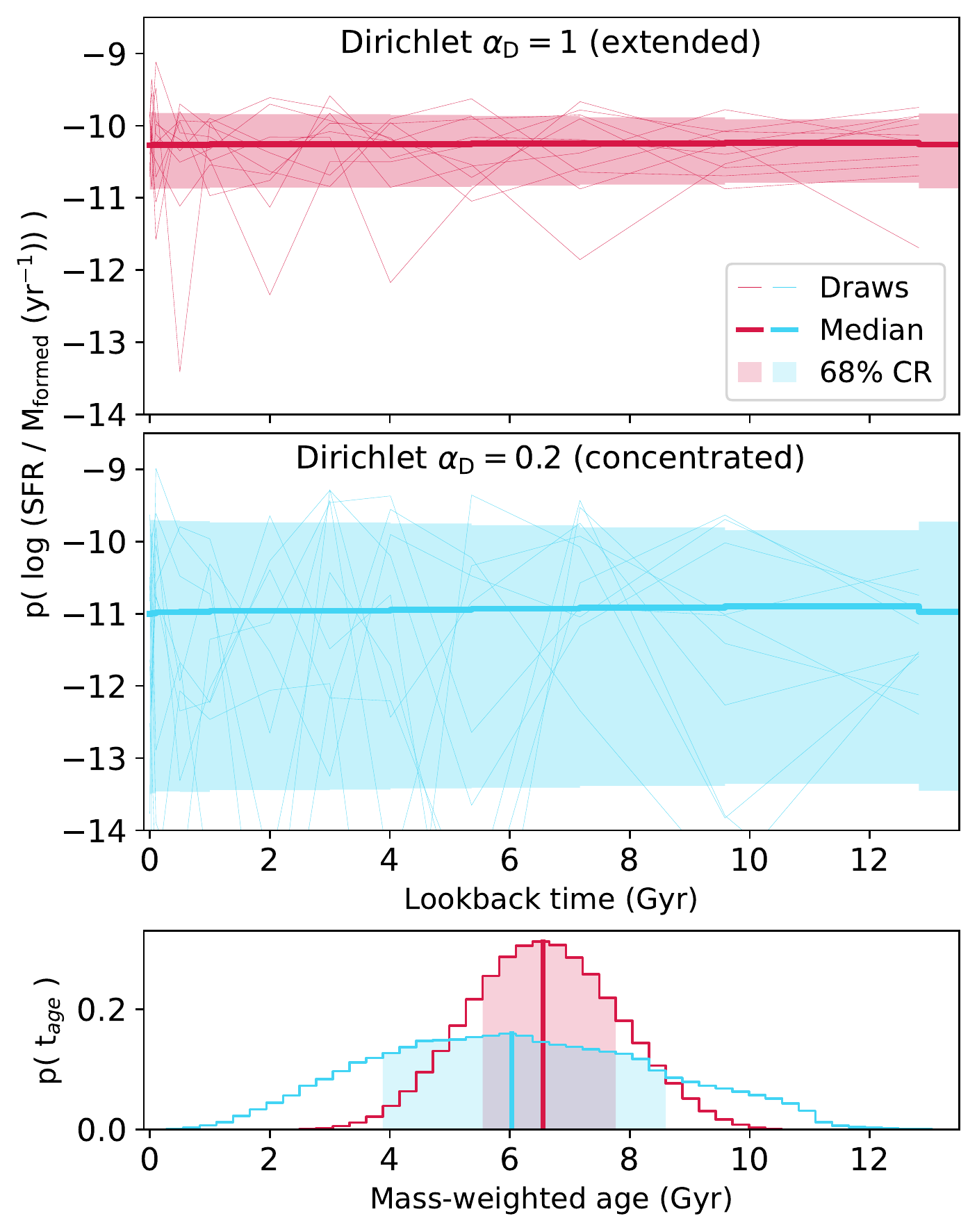} 
  \end{center}
  \caption{ Comparison of SFH priors between specific SFR (SFR per unit mass; first and second panels) and corresponding mass-weighted age (third panel). 
    The \aDo\ SFH prior prefers solutions where the star formation is equally weighted between the time bins, hence we label it `extended'.
    In comparison, the \aDt\ SFH prior prefers solutions where the star formation is unequally distributed between time bins, which we label as `concentrated'.
    The sSFR is shown as a function of lookback time for ten random draws (thin lines) from the Dirichlet SFH priors with \aDo\ (extended; red) and \aDt\ (concentrated; blue). The medians and 68~per~cent CRs of the prior are indicated with thick lines and shaded regions, respectively. The implicit mass-weighted age priors are shown in the lower panel, with vertical lines indicating the median, and shaded regions indicating the 68~per~cent CRs.
  } \label{fig:sfh_priors}
\end{figure}

To characterise the SFH we use a non-parametric\footnote{\textit{Non-parametric} here means that the SFH has no specified functional form.} 
model of the form of a piece-wise constant function with $N=12$ time bins. 
The benefits of such a flexible SFH (relative to parametric functions, e.g., declining exponential or log-normal) have been well characterised by \citet{leja2019a} and \citet{lower2020}, among others. The time bins are defined in lookback time, spaced so that the first seven bins correspond to 0--30~Myr, 30--100~Myr, 100--500~Myr, 500~Myr--1~Gyr, 1.0--2.0~Gyr, 2.0--3.0~Gyr, and 3.0--4.0~Gyr. There are four bins spaced logarithmically between 4~Gyr to 0.95$\times~t_\mathrm{univ}$ (4.0--5.4~Gyr, 5.4--7.2~Gyr, 7.2--9.6~Gyr, and 9.6--12.6~Gyr), and the last bin covers 0.95$\times t_\mathrm{univ}$--$t_\mathrm{univ}$, where $t_\mathrm{univ}$ is the age of the Universe at the time of observation. 
Defining the time bins this way reflects the non-linear evolution in the SEDs: the narrower time bins at recent lookback times allow a sufficient precision in capturing recent star formation, while the wider bins at later lookback times reflect the modest evolution of older stellar populations. 
The last time bin is included to permit a maximally old population. 

Fitting SEDs to recover SFHs is an ill-defined problem, and prone to overfitting \citep[e.g.,][]{moultaka2000, moultaka2004, ocvirk2006a}.
In order to recover a physically plausible SFH it is common to invoke `regularisation.'
There a number of ways that this can be done, which differ in technical detail. 
One approach is to impose Gaussian-like priors on the SFH and/or the age-metallicity relation \citep[e.g., as in the commonly used code {\sc steckmap};][]{ocvirk2006a, ocvirk2006b}, and another is to penalise sharp transitions in the SFH \citep[e.g., the continuity prior;][]{leja2019a}.
In this work we use a third method, which is to control the degree of concentration of fractional specific SFR (sSFR) between the time bins of the nonparametric function. 
While these approaches differ in detail, they all attempt to avoid nonphysical solutions by imposing constraints on the variability of the SFH over time.

We adopt a Dirichlet prior which includes a concentration parameter, \aD, that controls the preference to distribute the fractional sSFR in one bin ($\alpha_\mathrm{D}<1$) or evenly between all bins ($\alpha_\mathrm{D}\geq1$), respectively.
A detailed description of this prior is provided in \citet{leja2017}.
Without direct physical motivation to inform a choice of \aD, we consider both \aDo\ and \aDt\ as valid options, labelling them as `extended' and `concentrated' versions of the SFH prior.
In comparing the results produced from these two choices of SFH prior, we explore the dependence of the results on the degree of regularisation.

Fig.~\ref{fig:sfh_priors} shows random draws (thin lines) for priors with \aDo\ (extended) and \aDt\ (concentrated), with the time bins as defined above.
The median and 68~per~cent credible regions (CRs) of the priors are shown with a thick line and shaded regions, respectively.
The corresponding implicit prior on the mass-weighted age is shown in the bottom panel for reference.
The mass-weighted stellar age ($t_\mathrm{ age}$, sometimes referred to as the mean stellar age, broadly describes the average formation time of stars in a given galaxy in units of lookback time) is calculated from the SFH:
 \begin{equation}\label{eqn:mwa}
  t_\mathrm{age} = \frac{\int_{t_\mathrm{obs}}^{0} t~~\mathrm{SFR}(t)~\mathrm{d}t}{\int_{t_\mathrm{obs}}^{0} ~\mathrm{SFR}(t)~\mathrm{d}t}
 \end{equation}
 \noindent where $t_\mathrm{obs}$ is the age of the Universe at the time of observation. 
The implicit age prior for an extended SFH is centred at half the age of the Universe with a 99.9~per~cent CR between 3.08--9.98~Gyr, 
and thus is a strong prior against both very old and very young ages. 
In comparison, the concentrated SFH also peaks around half the age of the Universe (although offset given the varying widths of the time bins), but the prior is not as tight (99.9~per~cent CR between 0.83--12.17~Gyr) such that old ages are less strongly disfavoured.

\subsection{Noise and outlier models}\label{sec:fitting_physical_model_noise_and_outliers}

A noise model is used to account for possible under- or over-estimates of the spectral uncertainties, where the noise is uniformly inflated (or deflated). 
This effectively modifies the spectral uncertainty by a multiplicative factor, but is counterbalanced by a penalty in the likelihood calculation for larger uncertainties. 
This down-weights spectra where the uncertainties are otherwise low, but there is a mismatch between the spectrum and the models. 
The minimum uncertainty in the photometry is 7\%, and as we expect this to be large enough to account for deviations with the template SEDs, we do not include a noise model for the photometry in our model.

A mixture model is used to identify and mask pixels in the spectra which have large deviations from the model. 
The purpose here is to avoid being overly sensitive to outlier pixels in the spectrum. 
This is again relevant where the S/N is large and significant residuals can result from poor matches to the models where the model itself is inaccurate (due to differences in, for example, $\alpha$-enhancement). 
\prospector\ uses the mixture model approach described in \citet{hogg2010b}.

The spectral outlier model finds that less than 1~per~cent of the pixels are inconsistent with the model templates beyond the specified uncertainty. 
Note that the spectral white noise model prefers to inflate the uncertainties by $\sim 1$--3~per~cent, which is not unexpected given that the S/N of the spectrum is high, 40--140 (median 96) and that the models are not flexible enough to precisely match the metallicity- and $\alpha$-abundance sensitive spectral features (i.e, the Mg triplet). 

\subsection{Spectrophotometric calibration}\label{sec:fitting_physical_model_speccal}

We rely on the calibration of the photometry to constrain the shape of the SED continuum. 
The DF44 spectrum is not flux-calibrated such that neither the normalisation nor the shape of the spectral continuum provides information about the stellar properties. 
In fact, the spectrum was flattened prior to fitting (see Section~\ref{sec:data_spectrum}). 
For this reason we ignore the shape of the spectrum when computing the likelihood of the SED model (relative to the spectrum). 
We do this by following the routine provided through \prospector\ which fits (via optimisation) a polynomial to the residual between the spectrum and the model, which is then multiplied to the model. 
We use an $n=(\lambda_\mathrm{max}-\lambda_\mathrm{min})/100~$\AA\ $\sim 8$ order Chebyshev polynomial, which is flexible enough to remove the broad continuum shape without over-fitting absorption features \citep[e.g.,][]{conroy2018}. 
We test our results using several different orders of the polynomial, and find that we are generally insensitive to the choice of $n$ as long as $n>4$ (otherwise the dust attenuation pdf is skewed). 

\subsection{ Sampling } \label{sec:fitting_sampling}

The complete model includes 19 free parameters (11 of which describe the shape of the SFH), which are summarised in Table~\ref{tab:params}. 
We follow the sampling procedure outlined in \citet{johnson2021} \citep[see also ][]{tacchella2021}, using the dynamic nested sampling \citep{skilling2004, higson2019} algorithm {\sc dynesty}\footnote{https://dynesty.readthedocs.io/en/latest/} \citep{speagle2020} to efficiently sample the high-dimensional parameter space of the model and build posterior pdfs. 
This approach provides full posterior distributions of the model parameters together with their degeneracies. 
A useful primer on Bayesian methods can be found in \citet{vandeschoot2021}.

Throughout this work we report the uncertainties as 68~per~cent CRs (which corresponds to the 16\thh\ to 84\thh\ percent range) of the posterior pdfs as the majority of the distributions are non-symmetric. 

\subsection{Simultaneously fitting the photometry and spectroscopy}\label{sec:fitting_physical_model_both}

In fitting both the photometry and spectroscopy we consider the log-likelihood of the model, conditioned on the observation, to be the sum of the two individual likelihood functions: 
\begin{equation}
    \ln \mathcal{L}( d_s, d_p | \theta, \phi, \alpha ) = 
    \ln \mathcal{L}( d_s | \theta, \phi, \alpha ) + 
    \ln \mathcal{L}( d_p | \theta ) 
\end{equation}
\noindent where $d_s$ is the spectroscopic data, $d_p$ is the photometric data, the parameters $\theta$ describe the physical model used in \prospector, the parameters $\alpha$ describe the spectroscopic noise model (Section~\ref{sec:fitting_physical_model_noise_and_outliers}), and the parameters $\phi$ include the spectro-photometric calibration (Section~\ref{sec:fitting_physical_model_speccal}). 
The parameters of the physical model are summarised in Table~\ref{tab:params}.
We apply no relative weighting between fitting the spectroscopy and photometry in assessing the match between the observations and SEDs.

The basic likelihood calculation is effectively a $\chi^2$ calculation for both the spectral and the photometric data.
We alter the likelihood calculation for the spectroscopy to include the noise model and outlier model described in Section~\ref{sec:fitting_physical_model_noise_and_outliers}, following the procedure outlined in Appendix~D of \citet{johnson2021}.

\section{Results}\label{sec:results}

\begin{table} \footnotesize
\centering
\caption{Summary of results. 
Time-scales $t_x$ correspond to the time at which $x$ percent of stellar mass had formed, in units of Gyr since the Big Bang. Given that our SFH is a step function, we interpolate to estimate $t_{x}$. We provide the 16\thh, 50\thh, and 84\thh\ percentiles of the posterior (i.e., the 68~per~cent CR) as an estimate of the uncertainty, although this is likely an underestimate given the width of the time bins in our SFH. Given the observed redshift of DF44 and the adopted cosmology, the age of the universe is 13.47~Gyr. Time-scales in units of lookback time are therefore $t_\mathrm{Lookback}=13.47~\mathrm{Gyr}-t_x$. The fractional SFH within the time bins of the non-parametric model (i.e., not interpolated) are listed in  Table~\ref{tab:results_fsfh} in Appendix~\ref{app:fsfh_for_comparison}. 
}
\label{tab:results}
\begin{tabular}{ccccccc}
\hline
Time-scale   & \multicolumn{3}{c}{Extended}  & \multicolumn{3}{c}{ Concentrated} \\
(Gyr)  & \multicolumn{3}{c}{SFH prior}  & \multicolumn{3}{c}{ SFH prior } \\
\hline
 & 16 & 50 & 84  & 16 & 50 & 84 \\
\hline 
$t_{10}$ & 0.60 & 0.87 & 1.12 & 0.068 & 0.068 & 0.069 \\
$t_{20}$ & 1.18 & 1.39 & 1.64 & 0.135 & 0.136 & 0.137 \\
$t_{30}$ & 1.63 & 1.79 & 2.33 & 0.203 & 0.204 & 0.206 \\
$t_{40}$ & 2.07 & 2.33 & 2.92 & 0.271 & 0.272 & 0.274 \\
$t_{50}$ & 2.55 & 2.85 & 3.53 & 0.339 & 0.340 & 0.343 \\
$t_{60}$ & 2.93 & 3.41 & 4.17 & 0.406 & 0.408 & 0.411 \\
$t_{70}$ & 3.37 & 4.07 & 4.92 & 0.474 & 0.475 & 0.480 \\
$t_{80}$ & 3.86 & 5.39 & 5.70 & 0.542 & 0.543 & 0.549 \\
$t_{90}$ & 5.73 & 6.32 & 7.01 & 0.610 & 0.611 & 0.617 \\
$t_{95}$ & 6.25 & 7.57 & 7.93 & 0.643 & 0.645 & 0.652 \\
 \hline 
\end{tabular}
\end{table}

\begin{figure*}
  \begin{center}
    \includegraphics[width=\linewidth]{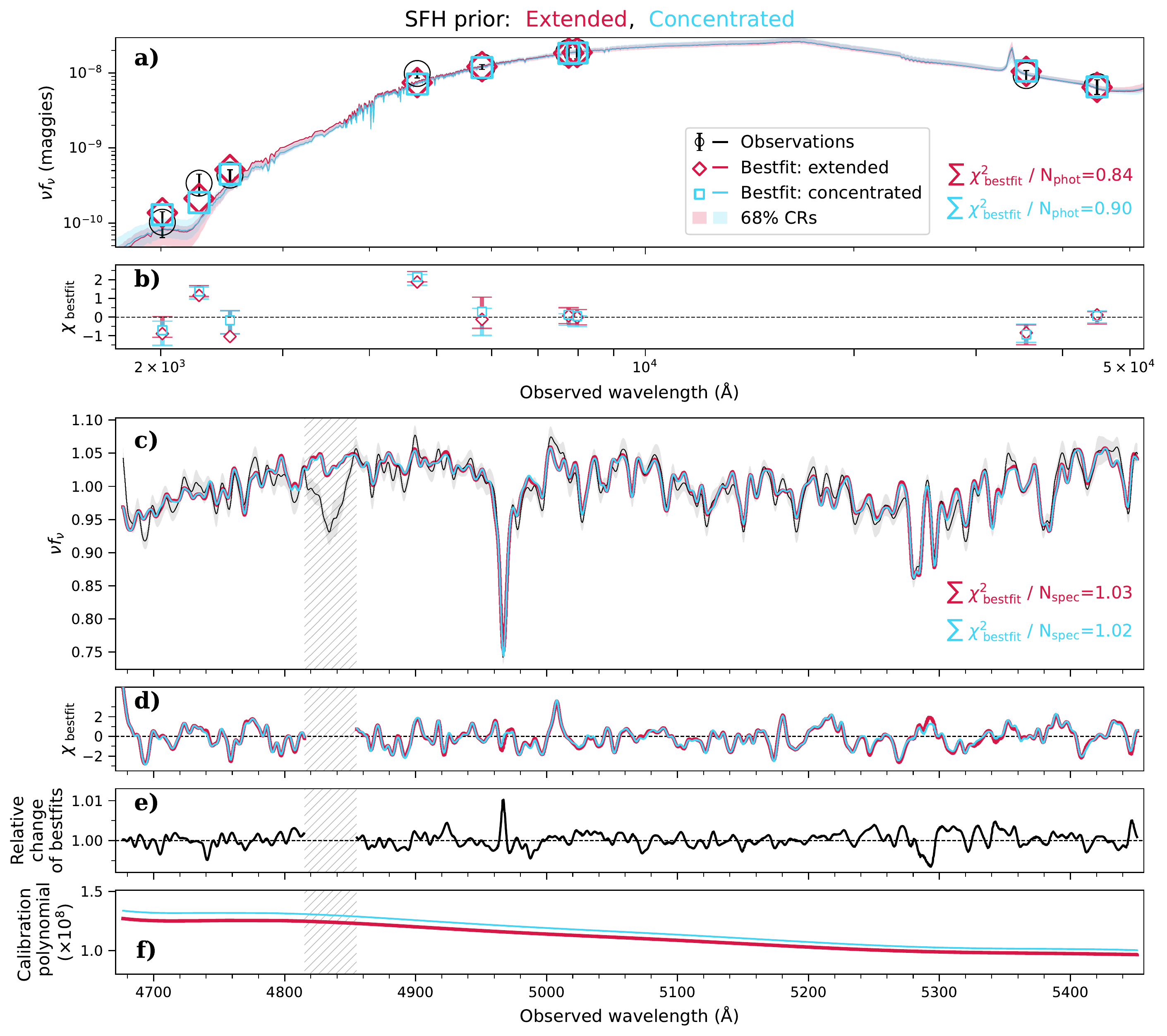} 
  \end{center}
  \caption{ Summary of the fitting results for DF44. 
  The observed data (black) is compared to the bestfit models (red, extended SFH prior; blue, concentrated SFH prior) and the 68~per~cent CR of 500 randomly drawn models from the posteriors (red/blue shaded). The corresponding posteriors are shown in Figs.~\ref{fig:summary_fit_corner} and \ref{fig:summary_fit_sfhs}. 
  {\it (a)} The observed (circles) and bestfit (diamonds and squares) photometric points, where the $\chi^2/N_\mathrm{data}$ of the bestfit SED is listed and 
  {\it (b)} shows $\chi$ ($[\mathrm{data}-\mathrm{model}]/\sigma$) of the bestfit points. 
  {\it (c)} The observed (uncertainties shown in grey) spectrum and bestfit spectra (multiplied by the spectrophotometric calibration polynomial). The light grey region indicates the spectral region masked throughout the fitting process. 
  {\it (d)} The $\chi$ of the bestfit spectra as a function of wavelength. 
  {\it (e)} The relative change of the bestfit models, i.e., the ratio of the two bestfit spectra. 
  {\it (f)} The spectrophotometric calibration polynomials. 
  } \label{fig:summary_fit}
\end{figure*}

Given the sensitivity of modelling ages of old stellar populations, and their dependence on both the flexibility of the assumed SFH and the choice of SFH prior \citep[e.g.,][]{leja2017, leja2019a}, we present the results for two `extremes' of the SFH prior: i) an `extended' SFH, preferring equal distribution of fractional sSFR between the time bins (\aDo), and ii) a so-called `concentrated' SFH, preferring an unequal distribution of fractional sSFR between time bins (\aDt). 
The difference between these priors is discussed in Section~\ref{sec:fitting_physical_model_sfh}.

In assuming the SFH is extended, there is a preference for ages of half the age of the Universe and against old ages ($\gtrsim 10$~Gyr; see Fig.~\ref{fig:sfh_priors}).
However, the results of \citet{villaume2022} suggest that DF44 formed its stellar population early, and shortly thereafter rapidly quenched, as determined from its inverted stellar population gradients and low iron metallicity for its mass.
The `concentrated' SFH prior has a higher likelihood for such an SFH. 
Moreover, the concentrated SFH prior has an overall broader implicit prior on mass-weighted age as there is no preference for where the fraction sSFR is concentrated between the time bins.

\begin{figure*}
  \begin{center}
    \includegraphics[width=\linewidth]{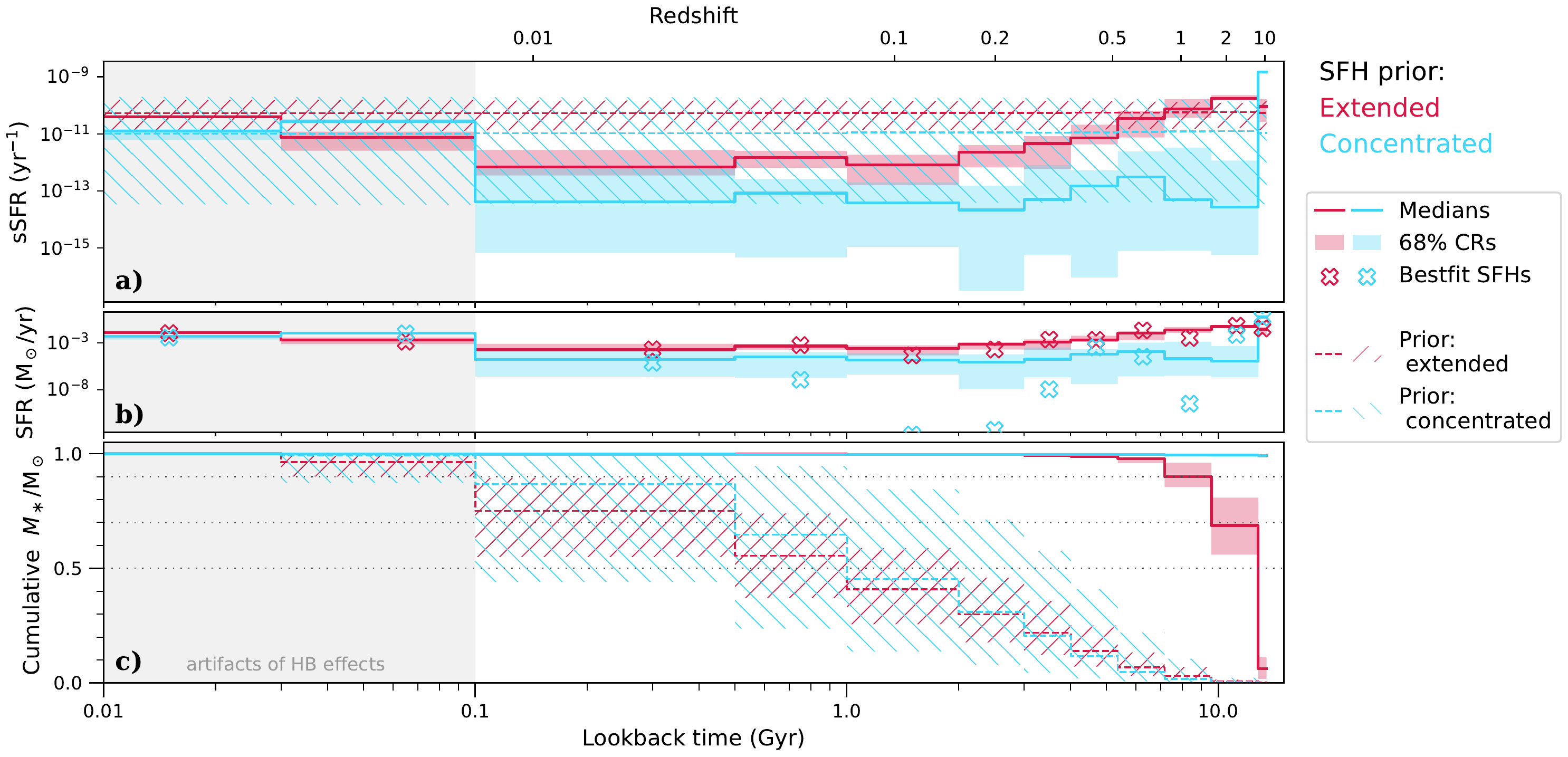} 
  \end{center}
  \caption{ Posteriors for the ({\it a}) sSFR, ({\it b}) derived SFR, and ({\it c}) derived cumulative stellar mass for the fits shown in Fig.~\ref{fig:summary_fit}. The results and priors shown in red (blue) colours correspond to the fit with an extended (concentrated) SFH prior. Solid lines indicate the median, and the shaded regions the 68~per~cent CR. These values are listed in Table~\ref{tab:results_fsfh} in Appendix~\ref{app:fsfh_for_comparison}. The SFHs of the bestfit models are indicated with open crosses. The median (dashed line) and 68~per~cent CRs (hatched) of the priors are shown for reference. Note that the cumulative mass and mass-weighted age priors are implicit, as they are derived from the sSFR prior. Dotted lines are drawn at 50, 70, and 90~per~cent of the  cumulative mass for reference. 
  The last 100~Myr are shaded grey to indicate that the SFH is affected by artefacts such as HB stars (see text).
  } \label{fig:summary_fit_sfhs}
\end{figure*}

\begin{figure*}
  \begin{center}
    \includegraphics[width=0.8\linewidth]{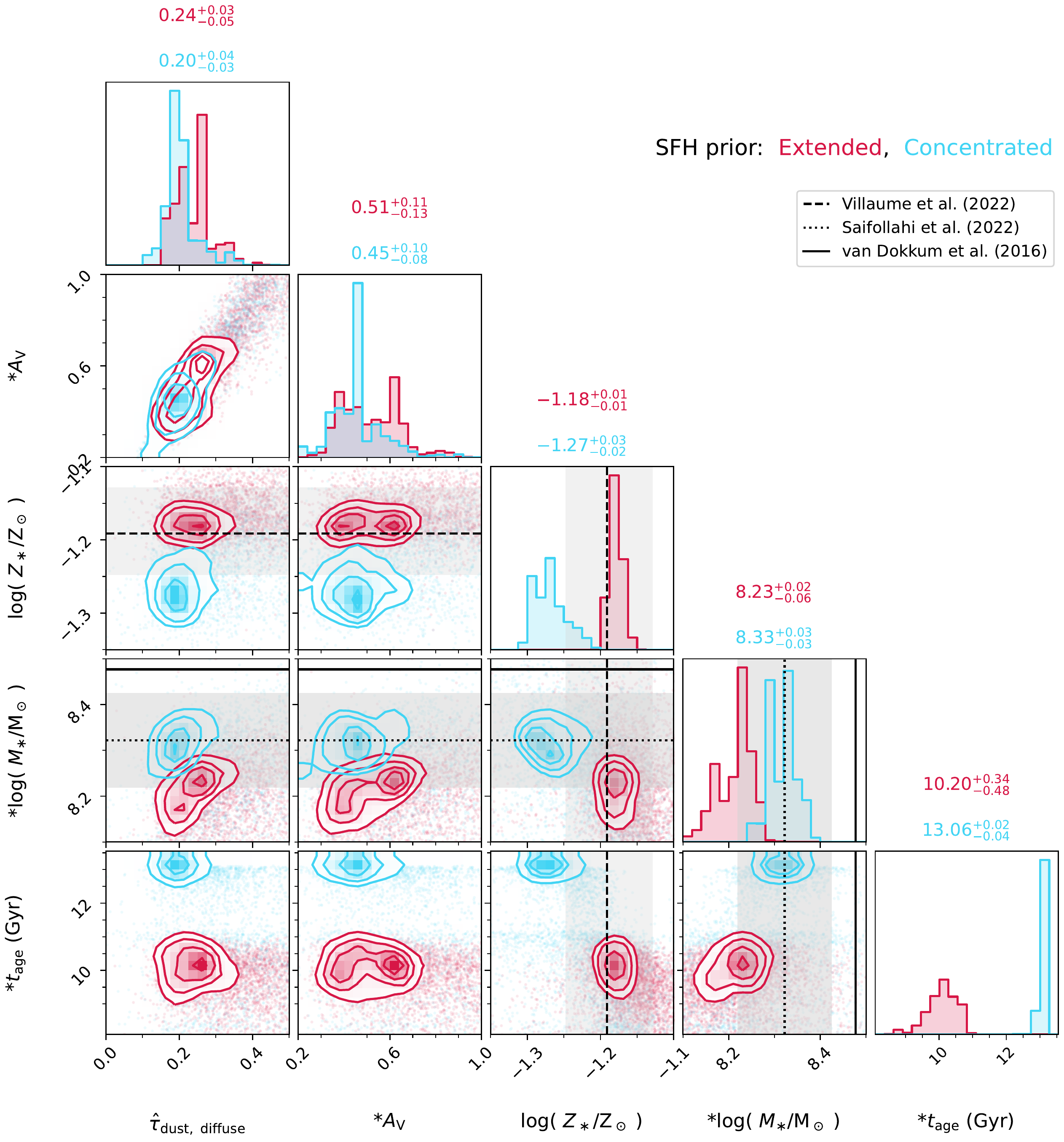}
  \end{center}
  \caption{ Posteriors of selected fitted and derived parameters (indicated with an asterisk) for the fits shown in Fig.~\ref{fig:summary_fit}. Contours are shown smoothed with a $n=1$ Gaussian kernel, where red (blue) contours show the fits with an extended (concentrated) SFH prior. Black lines denote the expected results given the analysis by \citet{villaume2022} for \logzsol, and stellar mass from \citet{vandokkum2016} and \citet{saifollahi2022}. 
  Grey shaded regions indicate the uncertainties on these values.
  The median and uncertainties from the 68~per~cent CR for our results are listed along the top of the one-dimensional histograms. 
  } \label{fig:summary_fit_corner}
\end{figure*}

\vspace{0.2cm}
The results from the full-spectral modelling of DF44 are shown in Figures~\ref{fig:summary_fit}--\ref{fig:summary_fit_corner}, where the fits for the extended SFH prior are shown in red, and for the concentrated SFH prior in blue. 
In Fig.~\ref{fig:summary_fit} the observations are shown with the `bestfit' models (the maximum a-posteriori model; i.e., with the highest probability of the set of samplings) and the 68~per~cent CR of 500 random draws from the posteriors.

Overall the fits to the photometry are similar between the two SFH priors; the extended SFH model has marginally smaller residuals at NUV wavelengths. 
Similarly, the bestfit model spectra (multiplied by the spectrophotometric calibration polynomial) compared to the spectroscopy are nearly identical, with differences only at the $<1$~per~cent level. 
Given the degeneracy between age, dust, and metallicity, the subtle differences in these features lead to the differences in the predicted stellar population parameters.

\subsection{Star formation history and stellar population parameters at z=0} \label{sec:results_sfh}

Fig.~\ref{fig:summary_fit_sfhs} shows the median (solid line) and 68~per~cent CR (shaded) of the posterior pdfs for the sSFR, and corresponding SFR and mass-assembly history. 
Similarly, the median (dashed line) and 68~per~cent CRs (hatched) for the explicit and implicit priors are shown (see also Fig.~\ref{fig:sfh_priors}). 
The SFHs of the bestfit models (shown in Fig.~\ref{fig:summary_fit}) are indicated with open crosses. 
Dotted lines are drawn at the 50\thh, 70\thh, and 90\thh\ percent levels of the cumulative stellar mass for reference. In Table~\ref{tab:results} we summarise the two SFH results by listing the times at which different percentiles of the final stellar mass were in place. 
The SFH within the time bins of the non-parametric model are provided in Appendix~\ref{app:fsfh_for_comparison}.

The SFHs determined from both priors suggest that DF44 formed early, having 90~per~cent of its stellar mass in place at least $\sim 7.2$~Gyr ago ($z\sim 0.9$). 
Using the extended SFH prior, we find that it took $\sim3.5$~Gyr for DF44 to assemble between 50~per~cent to 90~per~cent of its mass, suggesting a relatively fast transition between star forming and quiescent states.
The SFH determined with the concentrated prior is extreme in that more than 90~per~cent of the mass formed within the first time bin, i.e., $\sim12.8$~Gyr ago ($z\sim 8$).
During the last 5~Gyr the two results are otherwise similar, with low levels of star formation until the last 100~Myr.\footnote{Additional testing of the prior-sensitivity of the SFH showed that using $\alpha_\mathrm{D} = 0.5$ (mildly concentrated) produced parameter values between the results from \aDt\ and \aDo, as expected. The mass-weighted age was found to be 11.9~Gyr, which indicates that the very old age is not overly sensitive to the choice of the \aD\ value.}

{\vspace{0.2cm}} 
A curious feature of both SFHs is the rise in SFR within the last 100~Myr (corresponding to the first two time bins and shaded grey in Fig.~\ref{fig:summary_fit_sfhs}; by 1.8--2.4~dex).
Although residual star formation appears to be common for massive early type galaxies, where $\sim$0.5 per cent of their mass formed within the last 2~Gyr, the fraction decreases at lower stellar masses, consistent with galaxy `downsizing' \citep[e.g.,][]{salvador-rusinol2020}. 
The recent rise in DF44's SFH accounts for $\lesssim 1$~per~cent of the total stellar mass, assuming either SFH prior.
While DF44 shows no indication of recent star formation from the photometry, and similarly lacks emission lines in the spectrum, it is possible that $\mathrm{H}\alpha$ emission (perhaps related to star formation ignited by a late infall in to the Coma cluster) recently stopped.
This is perhaps unlikely, however, given the lack of blue regions within the galaxy.
\citet{lee2020} concluded, based on the difference in NUV and UVW2 bands, that the light traces older stars (on $\sim$~Gyr time-scales, as opposed to young stars which evolve on the order of $\sim$~Myr time-scales).
The `recent burst' is not a consequence of an artefact in the KCWI spectrum; the same feature is apparent when fitting the MaNGA data from \citet{gu2018b}.
Rather, we expect this recent star formation to be an artefact of the stellar models not being flexible to the contribution of blue horizontal branch (HB) stars (discussed in Appendix~\ref{app:sfh_biases_bhb}) or non-solar Mg-abundances.

We none the less test the sensitivity of the models to the presence of a very young stellar population by re-defining the time bins of our SFH, only allowing for star formation older than 1~Gyr. 
This places a strong prior against recent star formation (SF) to counteract the inability of the SPS models to correctly model the influence of the blue HB stars.
In excluding star formation younger than 1~Gyr, the models are better able to recover the shape of the SED, particularly in the NUV, but are marginally worse in matching the spectrum.  
With this revised model we recover SFHs equivalent to that of our primary results (at times $>1$~Gyr), with statistically consistent but less dust and metallicity, and slightly higher stellar mass. Interestingly, with the extended SFH prior, the revised age estimate is $\sim$2.4~Gyr older. 
With the concentrated SFH prior, the revised age estimate is unchanged from that of our main result.
As such, we conclude that the presence of the `recent burst' of SF does not affect our conclusion that DF44 formed and quenched very early in the history of the Universe.

\vspace{0.2cm}
Fig.~\ref{fig:summary_fit_corner} shows the posteriors for the normalisation of the diffuse dust attenuation curve, stellar metallicity, stellar mass, and mass-weighted age.
The parameters marked with an asterisk are not directly fit in our physical model, but derived from the posterior distributions.
We calculate the dust extinction following equations~(\ref{eqn:dust_diffuse_1}) and (\ref{eqn:dust_diffuse_2}) in the $V$-band, where we use $\lambda=5500~$\AA.
We note that `total stellar mass formed' is a free parameter in our model, which we convert to `stellar mass' by subtracting the mass lost throughout the SFH, as calculated by {\sc FSPS}.
The median and uncertainties of the marginalised posteriors for extended (concentrated) SFH priors are:
\begin{enumerate}\centering
    \item[] $\hat{\tau}_\mathrm{dust,~diffuse}={0.24}_{{-0.05}}^{{+0.03}}$ ~$\left({0.20}_{{-0.03}}^{{+0.04}}\right)$,
    \item[] *$A_\mathrm{V}={0.51}_{{-0.13}}^{{+0.11}}$ ~$\left({0.45}_{{-0.08}}^{{+0.10}}\right)$,
    \item[] $\log(Z_\ast/\mathrm{Z}_\odot)={-1.18}_{^{-0.01}}^{_{+0.01}}$ ~$\left({-1.27}_{{-0.02}}^{_{0.03}}\right)$, 
    \item[] *$\log(M_\ast/\mathrm{M}_\odot)={8.23}_{{-0.06}}^{{+0.02}}$ ~$\left({8.33}_{{-0.03}}^{{+0.03}}\right)$, 
    \item[] *$t_\mathrm{age}~/~ \mathrm{Gyr}={10.20}_{{-0.48}}^{{+0.34}}$ ~$\left({13.06}_{{-0.04}}^{{+0.02}}\right)$,
\end{enumerate}
as labelled above the one-dimensional histograms.
In both cases DF44 has a very old, modestly dusty, and metal-poor stellar population. 

\vspace{0.2cm}
Contrary to our expectation that old \citep[e.g.,][]{peroux2020} and metal-poor \citep[e.g.,][]{galliano2018} populations are devoid of dust \citep[see also][]{barbosa2020}, DF44 appears to have a non-negligible amount: the normalisation of the diffuse-dust attenuation curve is $\hat{\tau}_\mathrm{dust,\ diffuse}\gtrsim 0.2$ and $A_\mathrm{V}\gtrsim0.5$. 
The origin of such dust is not clear, however, Buzzo et al. 2022 (submitted) recently measured similar extinction values from optical to mid-infrared photometry for a sample of quiescent UDGs.
The overall shape of the SED constrains the dust content, however there are degeneracies with both metallicity and age. 
If we instead fix $\hat{\tau}_\mathrm{dust,\ diffuse}=0$ and refit DF44 (with an extended SFH prior), the posterior pdfs are statistically consistent with that of our main result, although we note that the age increases (as expected) by $\sim0.23$~Gyr.
In Appendix~\ref{app:sfh_biases_need_both} we discuss the fit to just the photometry, which prefers an even dustier solution ($\hat{\tau}_\mathrm{dust,\ diffuse}\sim 0.36$ and $A_\mathrm{V}\sim0.8$, although the photometry provides no direct constraint for the metallicity, and little constraint for the age).
While the spectroscopy breaks the degeneracy between dust and metallicity, the degeneracy with age remains; adding either more dust or a stellar population older than $\sim$3~Gyr lowers the flux at wavelengths $<5000$~\AA\ (see Appendix~\ref{app:old_v_dust}). 
Additional observations in the mid-infrared would provide better constraints on the dust content, as age and dust affect the flux in opposite directions at this wavelength range. 

Other than \dust, the posteriors of the dust model parameters largely reflect their priors -- which is to be expected given the lack of constraining data.
None the less, to check that our results do not depend on the particular dust model, we also fit the data with the dust model of \citet{gordon2003} based on the SMC Bar (thought to have similar dust properties to dwarf elliptical galaxies, i.e., without a UV bump in the extinction curve), and find no change to our result.
A degeneracy between the dust normalisation and stellar mass can be seen in the joint posterior in Fig.~\ref{fig:summary_fit_corner}, where an increase in dust suggests a higher stellar mass. 
As a point of comparison, a solid black line indicates the estimated stellar mass from \citet[][]{vandokkum2016}, and a dotted black line indicates that measured by \citet[][]{saifollahi2022} with uncertainties reflecting the systematics of the model fitting. 
Both of our fits produce stellar masses lower than (and statistically inconsistent within their 68~per~cent CRs) with the \citet{vandokkum2016} value, but consistent with \citet{saifollahi2022}.
Given that the photometry included in our fits is measured within an aperture, and thus does not include all of the light of the galaxy, it is not unexpected that the stellar mass we recover underestimates that from the literature.

\vspace{0.2cm}
There is a $\sim0.1$~dex difference in \logzsol\ between the fits with an extended or concentrated SFH prior, where the sense of the metallicity difference is consistent with that of the age difference ($\sim2.9$~Gyr) with respect to the age--metallicity degeneracy.
This indicates that we are not able to fully break the age--metallicity degeneracy with the data at hand.
While in Fig.~\ref{fig:summary_fit_corner} we show the stellar `isochrone' metallicity measured by \citet{villaume2022} as a black dashed line for comparison, there are several caveats to their comparison which are discussed in the following section.

At this point the dichotomy of DF44 being `old' or `very-old' is subject to the choice of SFH prior. 
We remind the reader that the extended SFH prior behaves analogously to regularisation methods used throughout the literature.
While the concentrated SFH prior provides more flexibility to better recover the short and early star formation expected for DF44, it is not necessarily a `good' prior; we provide no physical information for the shape of the SFH. 
We simply tune the prior such that it prefers to distribute the SF within fewer time bins (see Section~\ref{sec:fitting_physical_model_sfh}).

\begin{figure}
  \begin{center}
        \includegraphics[width=\linewidth]{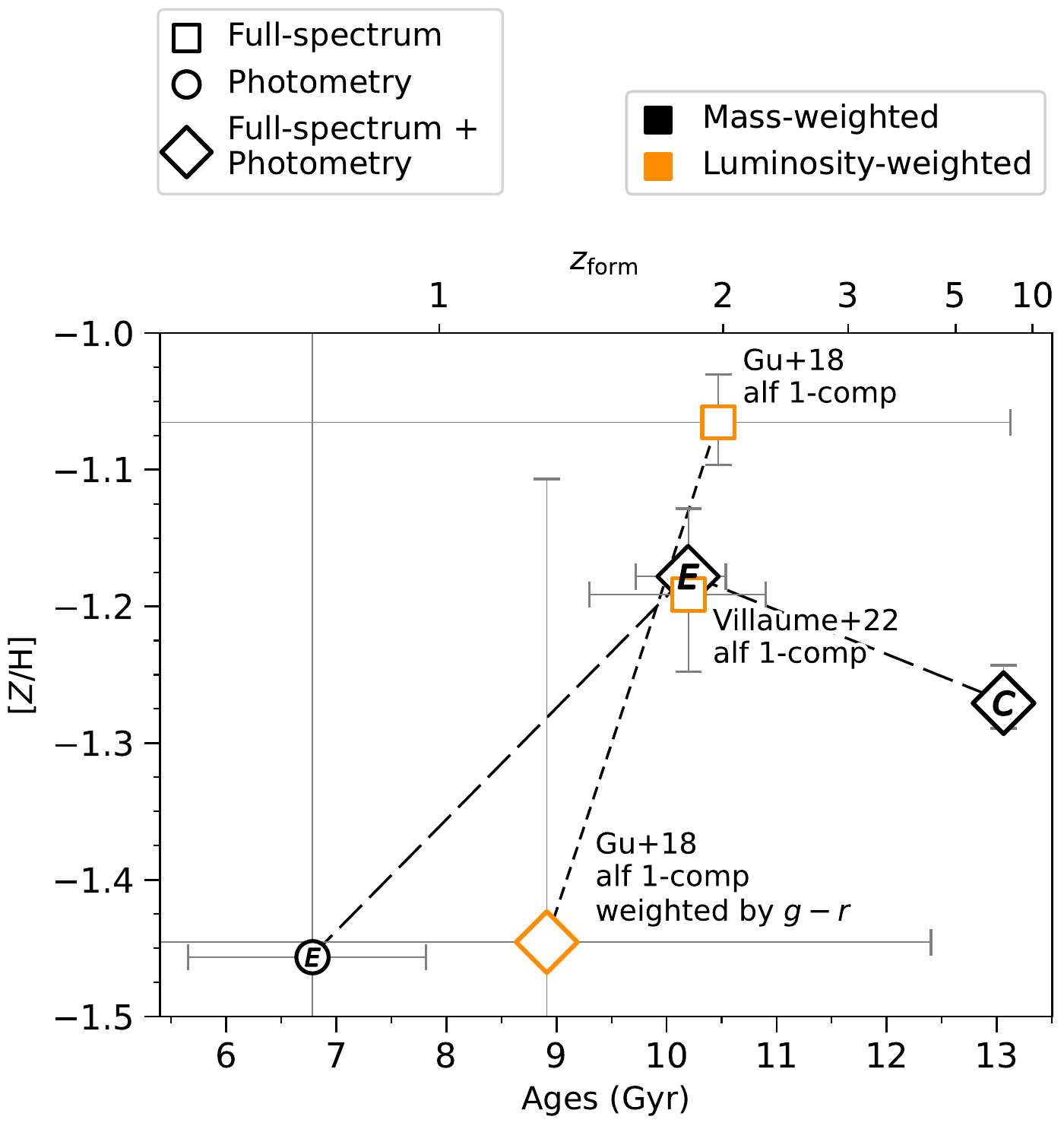}
  \end{center}
  \caption{ 
  Comparison of stellar metallicity and age for DF44 from this work, \citet{gu2018b}, and \citet{villaume2022} (the latter using the same spectroscopic data set as ours).
  Both results from the literature derived values using \alf, and thus are not directly comparable to our results using \prospector\ (see text).
  Black coloured points show mass-weighted ages, while orange points show luminosity-weighted ages.
  Marker shapes indicate the data used in fitting the stellar properties. 
  Dashed lines connect results obtained from the same study. 
  We mark the results from this work derived with an extended SFH with an `E', or with the concentrated SFH with a `C'.
     } \label{fig:lit_compare_df44}
\end{figure}

\vspace{0.2cm}
This prior-dependency problem is exacerbated with less complete or lower S/N data sets. 
As a brief example, 
in Fig.~\ref{fig:lit_compare_df44} we compare the stellar metallicities and ages determined through fitting both the spectrum and photometry (diamond), with that fitted to only the photometry (circle) for the extended SFH prior (points marked with an `E'). 
While the NUV--NIR photometry provides information on the dust in DF44 (see Appendix~\ref{app:sfh_biases}), the age estimate is more heavily weighted by the SFH prior than are the full spectrum fitting results.
Accordingly, the photometry-only fit gives a median age $\sim 3.4$~Gyr younger than the fit to the spectrum and photometry together.\footnote{
If instead of the non-parametric model, we assume the SFH follows a delayed exponential form (a common parametric model adopted within the literature) we find similar results. 
With a logarithmically uniform prior on the $e$-folding time, $\tau$, and linearly uniform prior for the delay time, $t_\mathrm{age}$, the implicit age prior has a complex form with 16\thh, 50\thh, and 84\thh percentiles of 1~Gyr, 3.8~Gyr, and 8.4~Gyr respectively -- preferring younger ages than the extended SFH prior results. 
The implicit age skews even younger if instead $\tau$ is linearly sampled. 
Fitting the photometry of DF44 suggests the age is $\sim8.2$~Gyr, and slightly less dusty than using the extended SFH model. 
Fitting both the photometry and spectroscopy suggests the age is $\sim13.6$~Gyr, and slightly less dusty and more metal poor than our main result. 
We note that the photometry-only results with the delayed parametric model appear particularly sensitive to the S/N -- if we inflate the photometric uncertainties by a factor of two, the age posterior decreases by $\sim2$~Gyr.
The same is not true when using the non-parametric models.
}

\subsection{Which SFH prior is preferred?} \label{ sec:results_compare_lit}

There is little statistical evidence to decide whether the results from either SFH prior better reflects the `true' properties (or SFH) of DF44.\footnote{Comparing the Bayesian evidence of the two fits (as derived from the nested sampling described in Section~\ref{sec:fitting_sampling}) we find a strong preference (according to the Jefferys scale, see for example \citealt{kass1995}) for the concentrated SFH prior ($\ln Z_\mathrm{concentrated} = 62590$ is much larger than $\ln Z_\mathrm{extended}=62542$), where here $Z$ is the Bayesian evidence. However, this likely reflects the fact that the old age of DF44 is more disfavoured by the extended SFH prior (see Section~\ref{sec:fitting_physical_model_sfh}) more than a preference of the data itself. }
The distributions of SED models shown in Fig.~\ref{fig:summary_fit} are similar between the fits with each prior, and the models have similar residuals.

There are subtle differences, however, particularly around the H$\beta$ and Mg~{\sc II} features where the concentrated SFH gives a (statistically) lower $\chi^2$.
The H$\beta$ line is sensitive to recent star formation (and to HB stars, as discussed in Section~\ref{sec:results_sfh}), while Mg~{\sc II} is to sensitive the $\alpha$-abundance of the stellar population. 
The FSPS models that we use are currently limited to fixed solar $\alpha$-abundance.
However, \citet{villaume2022} found that DF44 has [Mg/Fe]$=0.11^{+0.06}_{-0.04}$ through fitting the same spectrum of DF44 as this work with the full-spectrum fitting code \alf\ \citep[][]{conroy2018}, which includes response functions to measure the non-solar chemical abundance variations. 
Given the relationship between both features and the age of the stellar population, this points to the need to include more complex stellar populations variables, e.g., $\alpha$-abundance, in models
in order to break this degeneracy.\footnote{While {\sc FSPS} does include an option to set the fraction of blue HB stars, for technical reasons we cannot include it as a free parameter in our models.}

Fig.~\ref{fig:lit_compare_df44} compares the stellar metallicities and ages measured for DF44 by this work, \citet{villaume2022}, and \citet{gu2018b}.\footnote{The stellar `isochrone' metallicity (distinct from that which includes the response function for individual elements) [$Z$/H] values from \citet{villaume2022} and \citet{gu2018b} were provided via private communication.\label{footnote:alf_values}}
Both previous studies fitted rest-frame optical spectra of DF44 with the full-spectrum fitting code \alf. 
We caution that there are fundamental differences between \alf\ and \prospector\ which make their results only broadly comparable: e.g., the inclusion of non-solar abundance patterns (as mentioned above), and \alf\ fits a single-age stellar component (with a uniform prior with minimum age of 1~Gyr) rather than an SFH.
That said, the luminosity- and mass-weighted ages should be comparable given that DF44 is old. 

\citet{villaume2022} fitted the same KCWI spectrum as this work, while \citet{gu2018b} fitted a MaNGA spectrum which covers a broader wavelength range (including several additional age diagnostics: H$\delta$, H$\gamma$, Ca~II H and K, and $G$-band). 
The MaNGA spectrum has $\mathrm{S/N}\sim 8$~\AA$^{-1}$, however, which is only $\sim 12$~per~cent the S/N of the KCWI spectrum. 
Despite differences in data, the two studies both found the age of DF44 to be $\sim$10.5~Gyr, although the stellar metallicities are formally discrepant.\footnote{\citet{villaume2022} considered the presence of a second young population (aged 1--3~Gyr), which lowers their age estimate by 0.6~Gyr but is consistent with their fiducial fit.} 
Notably, \citet{gu2018b} also considered the $g-r$ colour of DF44 from Dragonfly imaging, and re-weighted their posteriors, 
which considerably lowers their metallicity value (and is then consistent with \citealt{villaume2022} owing to its large uncertainty).

Considering that we fit DF44 in a completely independent way compared to these studies, it is at least encouraging that the results are fairly similar. 
Significant variations among age and metallicity measurements for the same object, measured between different studies, is not unique to DF44. 
In Appendix~\ref{app:sfh_biases_compare_lit} we outline two additional examples and discuss the reasons behind their differences.

The comparison shown in Fig.~\ref{fig:lit_compare_df44} demonstrates the difficulty in measuring the stellar properties of old stellar populations, related both to limitations of data and modelling. 
As discussed in the previous section, a solution is within reach as the inclusion of a variable $\alpha$-abundance or the addition of mid-IR photometry would help to break degeneracies between the stellar population properties.

We conclude that DF44 has an age of $\sim$10--13~Gyr.
Without clear statistical evidence to favour one SFH model over the other, throughout the remainder of this work we present both sets of results. 
In the next section, we discuss the implications of such a large sized galaxy having formed the bulk of its stellar mass very early.

\section{Discussion}\label{sec:discussion}

In this work, we sought to measure the detailed SFH of DF44 as a means to distinguish between UDG formation scenarios, which predict a variety of quenching times (i.e., SFHs).
The consistent narrative among theoretical simulations is that UDGs are contiguous with the canonical dwarf population.
However, \citet{villaume2022} established that DF44 is dissimilar to canonical dwarf galaxies with respect to both the stellar population gradients, stellar metallicity, and kinematics. 
In measuring the SFH of DF44 we can further test this scenario.

Previous analyses of DF44 found that its stellar population is old, having an age of $\sim 10$~Gyr \citep[see Fig.~\ref{fig:lit_compare_df44};][]{gu2018b, villaume2022}. 
In this work we have shown that DF44 formed the majority of its mass early, where we consider the galaxy `quenched' after it forms $\sim90$~per~cent of its mass. 
In using an extended SFH prior we obtain a lower limit of the quenching epoch of $z \sim 0.9$ ($\sim 6.3$~Gyr after the Big Bang).
Alternatively, in using a concentrated SFH prior (motivated by the results of \citealt{villaume2022}), we recover an extremely early quenching epoch of $z\sim8$ ($\sim0.6$~Gyr after the Big Bang).
In either case we find that DF44 is old, the distinction being that a concentrated SFH prior suggests that it is {\it very} old.
Without clear statistical evidence to favour one prior over the other (see Section~\ref{ sec:results_compare_lit}) we instead focus on providing a qualitative comparison of the implications of the two results. 

\begin{figure}
  \begin{center}
        \includegraphics[width=\linewidth]{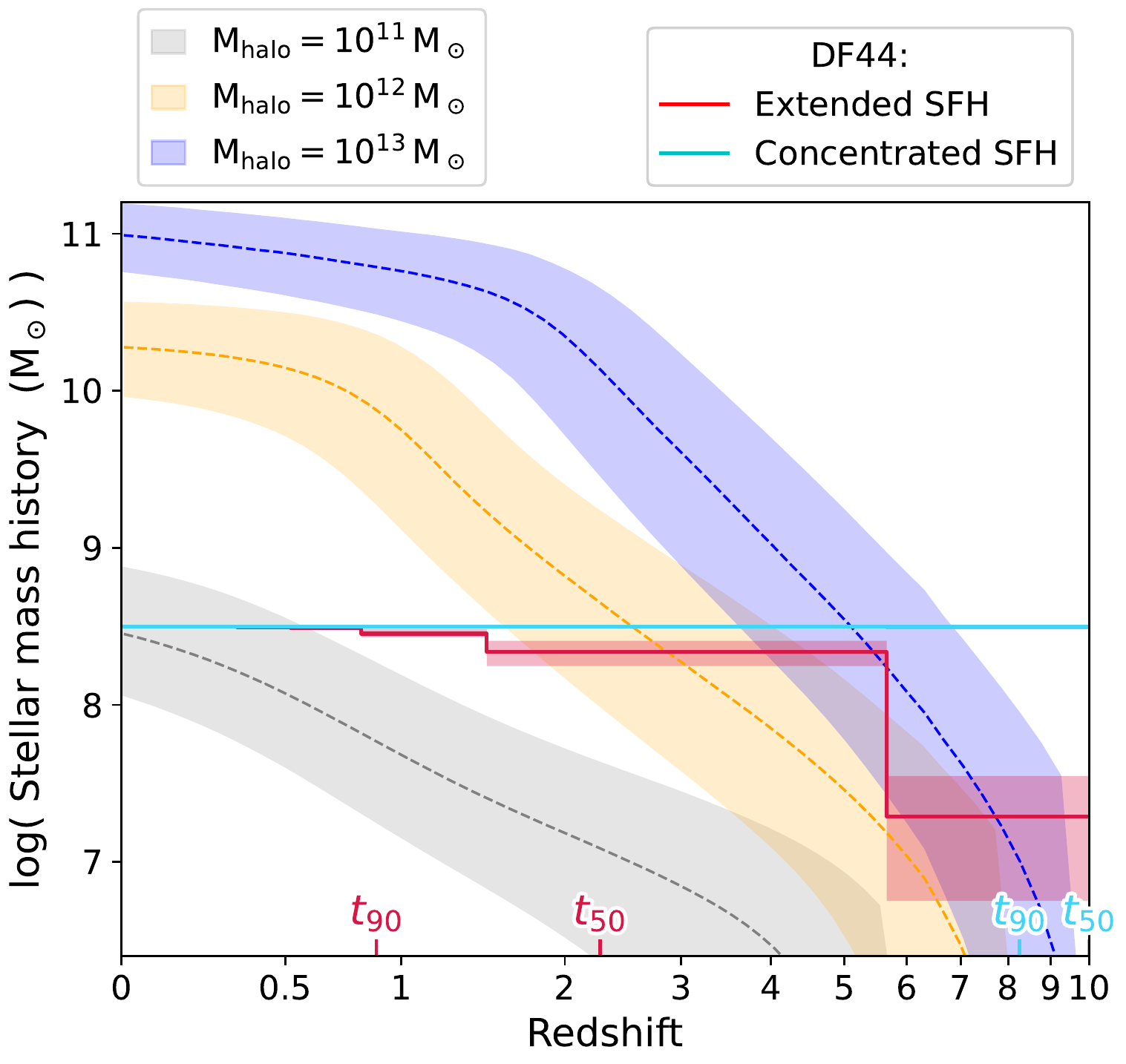}
  \end{center}
  \caption{ DF44's mass assembly history compared to total stellar mass histories of the main progenitor for haloes of several masses from the empirical model of \citet{behroozi2019b}. Lines indicate the median values, and shaded regions correspond to the 68~per~cent CRs. The halo mass bins have width of $\pm0.25$~dex.
  Time-scales of DF44's SFH are labelled on the plot: when 50~per~cent and 90~per~cent of the final mass had been formed.
  At the time that DF44 quenched ($z>0.9$) it was already more massive than expected for most normal dwarf galaxies.
     } \label{fig:mass_assembly}
\end{figure}

\vspace{0.2cm}
For either of our two results, the bulk formation of DF44 occurs during an epoch where the evolution of galaxies in dwarf-scale dark matter haloes ($\lesssim 10^{11}~\mathrm{M}_\odot$) significantly differs from that of galaxies in more massive haloes. 
The mass assembly histories expected for average galaxies with dark matter halo masses between $10^{11}$--$10^{13}~\mathrm{M}_\odot$ are shown in Fig.~\ref{fig:mass_assembly}, from the empirical model of \citet{behroozi2019b}.\footnote{The population averages likely overestimate the star formation time-scales of galaxies in dense environments \citep[e.g.,][]{thomas2010}. Indeed, dEs and UDGs in clusters are found to be old \citep[e.g.,][]{weisz2011a,ferremateu2018,ruizlara2018}, and satellite dwarfs in simulations appear to form much earlier than central dwarfs \citep{digby2019, garrisonkimmel2019, joshi2021}. However, DF44 is likely only on its first infall into the Coma cluster (discussed further below).} 
The mass assembly history of DF44 (as shown in Fig.~\ref{fig:summary_fit_sfhs}) is shown for comparison. 

While the current stellar mass of DF44 falls within the range expected for the $z= 0$ canonical central dwarf population, and its halo mass is in the neighbourhood of $\sim10^{11}~\mathrm{M}_\odot$ (e.g., \citealt{vandokkum2016, vandokkum2019, wasserman2019}; see also \citealt{bogdan2020}), its mass assembly history is not necessarily compatible with this population.
In fact, the mass of DF44 at $z\sim 8$ was typical for galaxies destined to become brightest cluster galaxies (BCGs) -- however, the mass growth was halted. 
This provides our first evidence that DF44 may not originate among the canonical field dwarf population. 

We now further investigate our results in the context of the predictions of UDG formation scenarios from theoretical work.

\begin{figure*}
  \begin{center}
    \includegraphics[width=\linewidth]{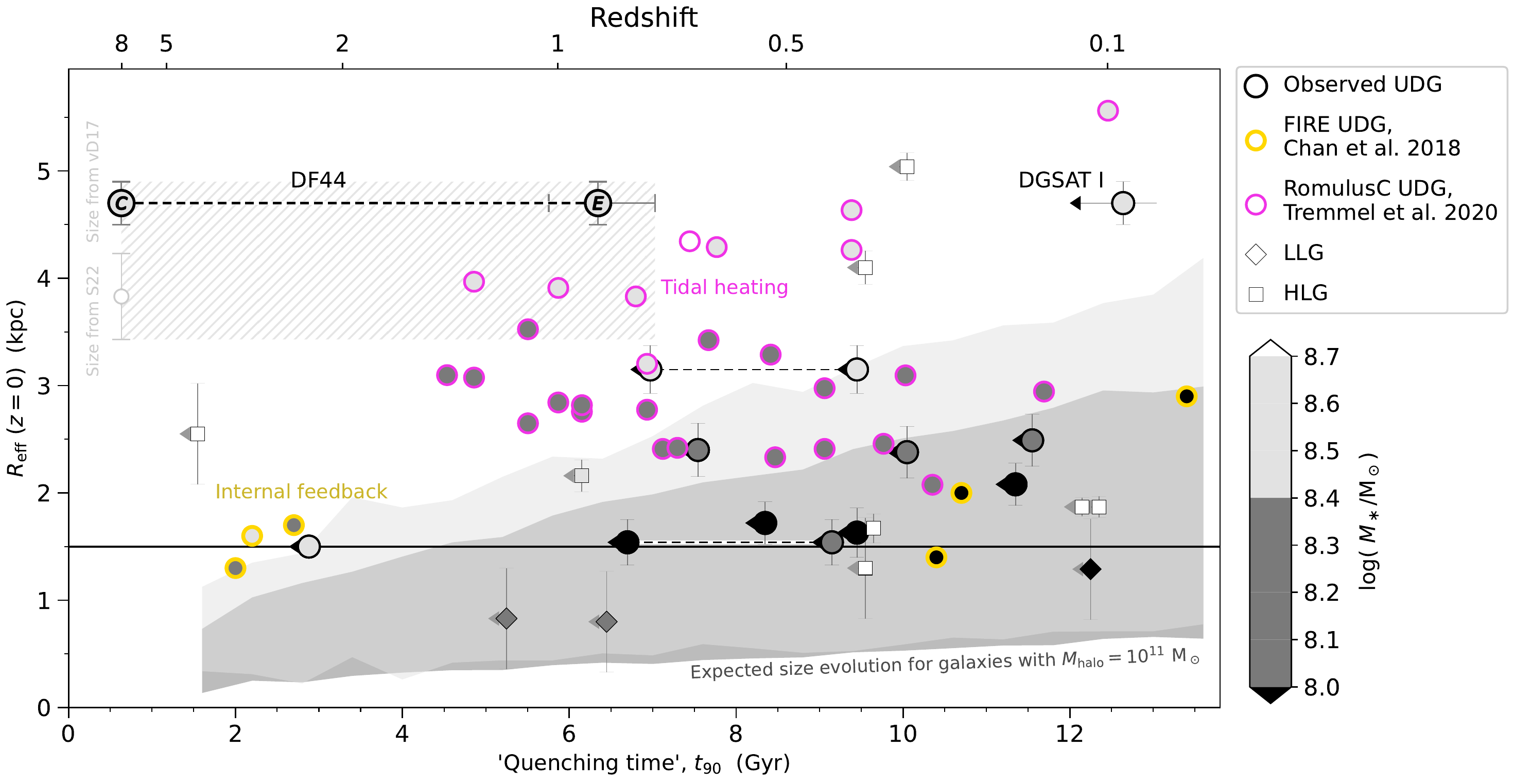}
  \end{center}
  \caption{ Comparison of effective radius and quenching time of UDGs (circles) between observations (black outlines) and predictions from two UDG formation scenarios (coloured outlines). 
  Quenching times are shown in units of age of the Universe, such that $t_{90}=0$ is the Big Bang.
  We show the quenching time results for DF44 as points with a `C' (concentrated SFH prior) and with an `E' (extended SFH prior). 
  A hatched region fills the parameter space covered by the range in DF44's size and quenching time measurements.
  The distribution of dwarf galaxies in two mass ranges is shown with grey shaded regions.
  A horizontal line shows the size threshold for UDGs, $R_\mathrm{eff}>1.5$~kpc.
  We show UDGs from two simulations: FIRE (yellow outlines; \citealt{chan2018}) and
  {\sc RomulusC} (magenta outlines; \citealt{tremmel2020}), which are described in the main text of Section~\ref{sec:discussion}.
  { \it The large size and early quenching of DF44 are inconsistent with the internal feedback model of FIRE, but is not inconsistent with the tidal-heating model of {\sc RomulusC}} (although see text).
  Observations from the literature are shown with arrows indicating that they may in fact be upper limits, given the potential bias in their measured SFH time-scales, see Section~\ref{sec:discussion_context}.
  Most of these UDGs fall within the expected size range of normal or tidally-enlarged dwarfs. 
  We also include low luminosity galaxies (LLGs; diamonds), and high luminosity galaxies (HLGs; squares) from the literature. 
  Dashed lines connect values obtained for the same object. 
  Points are coloured according to their stellar mass. 
     } \label{fig:lit_compare_size}
\end{figure*}

\subsection{DF44 in tension with UDG formation scenarios}

There have been many scenarios that have come out of cosmological simulations and SAMs which satisfy the size and surface brightness constraints of the observed UDG population.
As these scenarios all work under the same constraints of conventional galaxy evolution physics, they follow the same dichotomy: whether the significant size growth necessary to transform a canonical dwarf galaxy into a UDG occurs pre- or post-infall (which is to say, whether the cluster environment is necessary to the process) is related to whether they are late or early infallers into that environment.  
Nearly\footnote{The exception is \citet{wright2021} in which a small fraction of isolated UDGs which experience major mergers are quenched at $z=0$. This scenario is discussed further below.}
all of these models require the infall into the dense, hot environment of a cluster to quench the UDGs such that the quenching time is directly linked to the infall time.   

This dichotomy is demonstrated in Fig.~\ref{fig:lit_compare_size} where we show the relation between the effective radius and the quenching time for three different scenarios: 
i) what is typically expected under `normal' conditions (grey shaded regions)\footnote{The expected size growth was determined from the stellar mass assembly histories of \citet{behroozi2019b} and the size--mass relation of \citet{sales2020}.}, 
ii) the predictions for the `internal feedback' scenario from the FIRE simulations \citep[gold outlined circles;][]{chan2018}, and 
iii) the predictions from the RomulusC simulations \citep[magenta outlined circles;][]{tremmel2020}\footnote{These values were taken from Figures 2 and 11 of \citet{tremmel2020}.}.
The symbols are additionally colour-coded by stellar mass. 

We show our quenching time results for DF44 in this figure with the size measured by \citet[][]{vandokkum2017} based on the deepest imaging available ($R_\mathrm{eff}=4.7\pm0.2$~kpc; with black outlines). However, \citet{saifollahi2022} obtained a smaller size ($R_\mathrm{eff}=3.83\pm0.4$~kpc; grey point) based on the same data. 
A hatched region covers the parameter space of the various quenching times from our results, and the two size measurements.

The FIRE UDGs largely follow the expected size growth trend, where the distinction is that their internal feedback causes bursts of SF (which puff up the sizes of the galaxies, prior to quenching) place them on the top end of the size distribution. 
While \citet{chan2018} predicted that there are objects with quenching times as early as measured for DF44, these objects barely reach the nominal size of UDGs ($R_\mathrm{eff}<2$~kpc).
Indeed, they could not reach the size of DF44 without significantly more time to form stars, which would then violate the stellar mass/surface brightness constraint. 

A formation scenario that can explain both large and early-quenched UDGs is tidal heating, where the expected size--quenching time trend is the exact opposite of the FIRE scenario, i.e., the earliest infallers/quenchers will be the largest because they have spent the longest time expanding due to the cluster environment. 
We see this effect demonstrated in the RomulusC points.\footnote{We note that other simulations and SAMs which invoke tidal heating \citep[e.g.,][]{carleton2019, jiang2019a, liao2019, sales2020} with slightly different prescriptions (i.e., cuspy vs. cored dark matter haloes, how early satellite infall begins) could change the exact predictions of the sizes of UDGs. 
We note that RomulusC appears to over-predict the sizes of cluster dwarfs by a factor of $\sim 2$ (compared to observations from \citealt{eigenthaler2018}).
}
While \citet{tremmel2020} show that some objects reach the nominal sizes of UDGs prior to infall into a cluster, 
in order to reach the large end of UDG sizes requires the additional effect of tidal heating from the cluster environment. 
With the extended SFH prior, the quenching time of DF44 is reasonably consistent with the tidally heated RomulusC UDGs. 

Certainly tidal heating is happening on some level to some galaxies in clusters. 
Evidence of such has been observed among proto-UDGs in clusters \citep[e.g.,][]{grishin2021}.
\citet{carleton2019} interpreted the radial alignment of UDGs in Coma (\citealt{yagi2016}, which includes DF44) as evidence that these galaxies have been tidally influenced.
While we cannot discount the tidal heating scenario in explaining the size and quenching times of DF44, this scenario conflicts with other properties of DF44. 

Measurements of the kinematics and dynamics of DF44 indicate that it has not been in the cluster environment long enough to be impacted by tidal effects.
Its position in phase-space points to a late infall into the Coma cluster \citep[$< 2$ Gyr ago;][]{alabi2018}. 
Moreover, DF44 appears to be part of a dynamically cold group that would have surely been disrupted if tidal heating had taken place \citep{vandokkum2019}, and there is no distortion in its ellipticity that would be a marker of tidal heating \citep{mowla2017}.  

Together with the above points, the SFH provides evidence that DF44 certainly quenched prior to cluster infall. 
This would suggest that its progenitor was larger than a dwarf galaxy or that a process unrelated to environment caused an expansion. 
This interpretation is consistent with the conclusion of \citet{saifollahi2022}, who find that the elevated GC populations at a given stellar mass ($N_\mathrm{GC}/M_\ast$) of large UDGs (including DF44) are inconsistent with scenarios which explain the sizes of UDGs via redistributing the stars to larger radii (i.e., tidal interactions, stellar feedback, or high-spin).
\citet{villaume2022} similarly ruled out such scenarios given DF44's `inside-out' stellar population gradients.
Therefore, {\it how} DF44 quenched is the crucial question to answer to understand its origins.

From simulations, only \citet[][based on {\sc Romulus25}; \citealt{tremmel2020}]{wright2021} have proposed a scenario, `early major mergers'\footnote{We note that \citet{saifollahi2022} refer to this scenario as `lack of late mergers'.}, 
in which UDGs can form and quench\footnote{Less than 5~per~cent of the simulated UDGs with masses $M_\ast>10^8~\mathrm{M}_\odot$ are quenched, in the sense that they are gas poor. This population is dominated by galaxies that have had an interaction with a more massive halo and/or AGN activity.}
without relying on environmental quenching mechanisms. 
The UDGs in {\sc Romulus25} had their star forming gas and star formation moved outwards from the central cores of the galaxies to larger radii by major mergers $\sim$8--11~Gyr ago.
For most of the simulated UDGs, star formation continued in the galaxy outskirts, while the central core passively dimmed, leading to negative radial age gradients.

Considering that DF44 quenched $\gtrsim7$~Gyr ago, this may suggest that a major merger is responsible for (or at least concurrent with) its quenching -- and that there would be a flat age gradient.
The central ($<0.5$~kpc) SFH predicted for {\sc Romulus25} UDGs is broadly consistent with DF44's SFH when assuming an extended SFH prior, although not when assuming a concentrated prior (which quenches much earlier).
\citet{villaume2022} measured a flat-to-negative [Mg/Fe] gradient out to $\sim$2.5~kpc, which taken as a proxy for an age gradient is not strictly inconsistent with this scenario.\footnote{While \citet{villaume2022} measured a flat age gradient, they note that given the limitations of modelling granular differences in old stellar populations, the [Mg/Fe] gradient is more sensitive to age variations.}

Further work is needed in order to establish whether DF44 is the product of an early major merger. 
For instance, the mechanism that quenches $\lesssim5$~per~cent of the {\sc Romulus25} UDGs is not fully described, providing no point of comparison with DF44's SFH or stellar population gradients.
Moreover, when this quenching occurs, or whether the galaxies remain quenched, is unclear.
While \citet{wright2021} and \citet{vannest2022} explored the predictions of `early major mergers' in differentiating average UDGs and non-UDGs, the fact that DF44 is a rare case warrants more detailed comparisons.

\vspace{0.2cm}
The results of this work show that DF44 has been shaped by some rare galaxy evolution process, no matter whether the `true' SFH resembles our result with an extended or concentrated SFH prior, or falls somewhere in between. 
As was shown in Fig.~\ref{fig:mass_assembly}, the early SFR of DF44 is more typical of normal (MW-like) star forming galaxies at $z>3$ \citep{rinaldi2021}. 
The implication is that it is not the early, extreme SFH that makes DF44 unusual among $z=0$ galaxies, but rather its sudden quenching. 
Given the lack of galaxies like DF44 in cosmological simulations, this would imply that galaxy evolution models are not capturing the true diversity of quenching mechanisms. 

In fact, cosmological simulations already struggle to reconcile the opposing stellar mass--effective radius constraints for objects like DF44 in the context of the broader galaxy population. 
A common problem among cosmological simulations is that they do not accurately reproduce the population of normal sized dwarfs (e.g., \citealt{chan2018, el-badry2016, lupi2017, tremmel2020, benavides2021}; see also \citealt{jiang2019a}).
Since this points to issues in the implementation of star formation and related feedback, the evidence from this work and \citet{villaume2022} that there are objects like DF44 that require even more intense star formation feedback exacerbates this problem.

Analytic and semi-analytic models can avoid such issues to some degree.
With respect to size, several UDG formation scenarios apply empirical distributions \citep[e.g.,][]{carleton2019, sales2020} but they are then subject to the likely bias of `getting out what they put in' \citep[see][]{jiang2019b}.
With respect to star formation and feedback,
\citet{danieli2021} analysed the large number of GC candidates hosted by NGC~5846\_UDG1 \citep{forbes2021} with a model that connects the evolution of a galaxy with its dark matter halo and GC populations \citep{trujillo-gomez2019} 
to show that it is plausible that clustered supernova feedback could significantly increase the mass loading factor of gas outflows.
However, these models miss an important component of galaxy evolution -- the impact of the different environments a galaxy moves through over its lifetime. 
DF44's very early quenching and relatively late infall into the Coma cluster invokes the question of what has it been doing for the last $\sim 10$ billion years? 
Given the potential `pre-processing' by group environments or filaments that can affect everything from the size of a galaxy's dark matter halo, to its SFH and present-day GC population, makes it vital to understand this aspect of galaxy evolution in general.

\subsection{DF44 in context}\label{sec:discussion_context}

The prior-dependence of the SFH for old stellar populations, even with high-S/N data, means that further work is needed to understand what `good' SFH priors are for these systems. 
The problem is amplified at lower S/N, where the prior will have a stronger influence on the posterior pdfs (see Appendix~\ref{app:sfh_biases_priors} for an example).
Consequently, it is not straightforward to compare results between studies in the literature. 
With this caveat in mind, we also show in Fig.~\ref{fig:lit_compare_size} the quenching times and sizes of UDGs from three studies \citep[][]{ferremateu2018, ruizlara2018, martinnavarro2019}, and for comparison high- and low-luminosity dwarfs in Coma \citep[squares and diamonds, respectively;][]{ferremateu2018}.
Arrows attached to these points indicate that they are perhaps upper limits, given potential biases from the use of regularised SFHs (akin to the extended SFH prior used in this work; see the discussion in Appendix~\ref{app:sfh_biases_compare_lit}).
We note that the UDGs from the literature are shown with effective radii from the catalogue of \citet[][]{alabi2020} when possible, where DF44 was found to have a size of $3.74\pm0.23$~kpc in
the Subaru/Suprime-Cam $R$-band.

Regardless of potential biases in the SFHs, there are still interesting conclusions to draw from this data set.
DF44 stands out as an outlier among the largest observed UDGs with an early quenching time, for any of the discussed quenching times or sizes. 
On the other hand, the UDG DGSAT~I stands out
with both the largest size and latest quenching time among the literature values shown in Fig.~\ref{fig:lit_compare_size}, and it is also the only non-cluster member. 
Unlike the rest of the UDGs, DGSAT~I is similar to a subset of the {\sc RomulusC} UDGs which follow a trend in size--quenching time in distinct disagreement with the standard expectations of tidal heating.
Its size is also well outside of what is plausible for the concentrated SFH scenario, or normal expectations of size growth given its late quenching time. 

While it is outside the scope of this work to examine DGSAT~I in detail, it is relevant to this discussion in that it further provides evidence that multiple observed objects, all of which are `UDGs,' in fact have distinct formation pathways. 

That DF44 attained a similar stellar mass and size as the other large galaxies, but much earlier, supports the idea that it is either the product of unconventional galaxy evolution processes, or it was interrupted from becoming a much more massive galaxy by some catastrophic quenching event. 
Speculation of the latter has also been drawn on the basis of the wide range of GC counts among UDGs, and the range of implied dark matter halo masses (with some having little to no dark matter).
This is the first time this diversity has been shown in the SFHs of the galaxies' field star populations.

\section{Summary}\label{sec:conclusion}

In this work we simultaneously fit NUV to NIR photometry and high S/N rest-frame optical spectroscopy of the UDG DF44 with an advanced physical model. Our model includes non-parametric SFHs, a flexible dust attenuation law, a white noise model, and an outlier model, which we fit to the observations in a fully Bayesian framework with \prospector. 

We find that DF44 formed the majority of its stellar mass ($>90$~per~cent) early, although how early is sensitive to the choice of the SFH prior and degeneracies between stellar population parameters.
Using an extended SFH prior akin to similar studies in the literature (which strongly favours ages of half the age of the Universe, and therefore disfavours very old ages) we find that DF44 formed by $z\gtrsim 0.9$.
If we instead adopt prior knowledge from DF44's stellar population gradients that the DF44 formed early and rapidly quenched \citep{villaume2022}, such that its SFH is concentrated within a short timescale, we find that DF44 assembled as early as $z\sim 8$.
Neither of these priors encode physical information of the shape of the SFH based on a priori knowledge, and thus neither are necessarily `good' priors. Further work is needed to understand what `good' SFH priors are for such old galaxies from a theoretical standpoint. 
Even with the high-S/N spectral data used in this work ($\sim 96$~\AA$^{-1}$) the data showed no statistical preference for either result.
Improved age constraints are possible with the inclusion of observations in the mid-infrared in that this would pin down the dust attenuation, which in the NUV is degenerate with the contribution of old stellar populations. 
Improvements in the models (e.g., including variable $\alpha$-abundance) to replicate old and complex stellar populations are also needed.  

DF44's early and short SFH determined from this work, together with previous results that DF44 is very metal poor for its mass, and that stellar population gradients indicate `inside-out' formation \citep[unlike kinematically- and morphologically-similar dwarfs;][]{villaume2022}, points towards an unusual origin, likely distinct from the canonical dwarf population. 
UDG formation scenarios outlined in simulations only predict the SFH and size of DF44 through invoking prolonged environmental effects, yet we conclude that DF44 quenched prior to accretion into the Coma cluster. 
While analysis of the {\sc Romulus25} simulation by \citet{wright2021} proposes early major mergers as a means to produce UDGs in the field, it is not yet clear if the properties of DF44 are fully consistent with this scenario. 
Instead, DF44 may be a `failed galaxy' with its initial size, or whatever processes that expanded it, being unrelated to its environment. 
In Summary, early quenching an late infall taken together rules out most UDG formation scenarios except for the failed-galaxy and early-major-mergers (with the caveats above).
Additional work is needed to explain the old quiescent UDGs from a theoretical standpoint, while reproducing the observed stellar properties beyond general size--mass trends.

\section*{Acknowledgements}

We thank Chris Lee for helpful discussions regarding the UV data of DF44. We would like to thank Meng Gu for providing the MaNGA spectrum of DF44, Josh Speagle for help with technical details in using {\sc dynesty}, and Joel Leja for help with technical details related to the SFH priors and \prospector.
We thank the anonymous referee’s helpful report that improved the quality of this paper.
This research is supported by the following grants: National Sciences and Engineering Research Council of Canada (NSERC) PGS award (KW), Discovery grants (MLB), Waterloo Centre Astrophysics Postdoctoral Fellowship (AV).
DAF thanks the ARC for financial assistance via DP220101863.
AJR was supported as a Research Corporation for Science Advancement Cottrell Scholar.
This work was partially supported by a NASA Keck PI Data Award, administered by the NASA Exoplanet Science Institute. The data presented herein were obtained at the W.~M.~Keck Observatory, which is
operated as a scientific partnership among the California Institute of Technology, the University of California, and the National Aeronautics and Space Administration. The Observatory
was made possible by the generous financial support of the W.~M.~Keck Foundation.
We recognise and acknowledge the significant cultural role and reverence that the summit of Mauna Kea has always had within the indigenous Hawaiian community. 
We are most fortunate to have the opportunity to conduct observations from this mountain.

This work made use of the following software: 
{\sc Astropy} \citep{astropy1, astropy2},
{\sc dynesty} \citep{speagle2020}
{\sc FSPS} \citep{conroy2009, fsps},
{\sc IPython} \citep{ipython},
{\sc matplotlib} \citep{matplotlib},
{\sc NumPy} \citep{numpy},
{\sc python-fsps} \citep{pythonfsps},
\prospector\ \citep{leja2017, johnson2019, johnson2021},
and
{\sc SciPy} \citep{scipy},

\section*{Data Availability}

The data underlying this article will be shared on reasonable request to the corresponding author.



\bibliographystyle{mnras}
\bibliography{webb2022_df44_sfhs} 




\appendix

\section{The SFH of DF44} \label{app:fsfh_for_comparison}

For comparison with future works, in Table~\ref{tab:results_fsfh} we provide the fraction of SF, and cumulative fraction of stellar mass formed, within the time bins of the non-parametric models. We list the 16\thh, 50\thh, and 84\thh\ percentiles of the distributions, where we note that the 50\thh\ percentiles of the fractional SFHs do not necessarily sum to unity. 

\begin{table*} \footnotesize
\centering
\caption{Summary of SFH results. The fraction of SF and the cumulative fraction of stellar mass formed are listed for each time bin of the non-parametric SFH model. The 16\thh, 50\thh, and 84\thh\ percentiles of the posterior (i.e., the 68~per~cent CR) are listed. We note that the 50\thh\ percentiles of the fractional SFH do not necessarily sum to unity. The SF time-scales listed in Table~\ref{tab:results} are interpolated from these step functions. 
}
\label{tab:results_fsfh}
\begin{tabular}{c|ccc|ccc|ccc|ccc}
\hline
Time bin   & \multicolumn{6}{c|}{Extended SFH prior}  & \multicolumn{6}{c}{ concentrated SFH prior } \\
(Gyr)  & \multicolumn{3}{c}{ SF Fraction }  & \multicolumn{3}{c|}{ Cumulative fraction of $M_\ast$ }  & \multicolumn{3}{c}{ SF Fraction }  & \multicolumn{3}{c}{ Cumulative fraction of $M_\ast$ } \\
\hline
 & 16 & 50 & 84  & 16 & 50 & 84 & 16 & 50 & 84  & 16 & 50 & 84 \\
\hline 
$10^{-9}$ -- 0.03	& 0.0703	& 0.0795	& 0.0967		& 1.0000	& 1.0000	& 1.0000		& 0.0042	& 0.0082	& 0.0099		& 1.0000	& 1.0000	& 1.0000		\\
0.03 -- 0.10		& 0.0050	& 0.0152	& 0.0240		& 0.9986	& 0.9988	& 0.9989		& 0.0160	& 0.0181	& 0.0221		& 0.9995	& 0.9996	& 0.9998		\\
0.10 -- 0.50		& 0.0007	& 0.0014	& 0.0061		& 0.9980	& 0.9983	& 0.9986		& 0.0000	& 0.0000	& 0.0004		& 0.9974	& 0.9977	& 0.9980		\\
0.50 -- 1.00		& 0.0014	& 0.0031	& 0.0049		& 0.9972	& 0.9979	& 0.9983		& 0.0000	& 0.0001	& 0.0002		& 0.9974	& 0.9975	& 0.9979		\\
1.00 -- 2.00		& 0.0003	& 0.0015	& 0.0039		& 0.9964	& 0.9970	& 0.9975		& 0.0000	& 0.0000	& 0.0001		& 0.9973	& 0.9974	& 0.9978		\\
2.00 -- 3.00		& 0.0014	& 0.0046	& 0.0086		& 0.9949	& 0.9962	& 0.9970		& 0.0000	& 0.0000	& 0.0001		& 0.9971	& 0.9974	& 0.9978		\\
3.00 -- 4.01		& 0.0011	& 0.0089	& 0.0158		& 0.9917	& 0.9932	& 0.9955		& 0.0000	& 0.0000	& 0.0005		& 0.9969	& 0.9973	& 0.9977		\\
4.01 -- 5.36		& 0.0091	& 0.0145	& 0.0435		& 0.9833	& 0.9886	& 0.9925		& 0.0000	& 0.0001	& 0.0004		& 0.9964	& 0.9971	& 0.9973		\\
5.36 -- 7.16		& 0.0166	& 0.0702	& 0.1351		& 0.9591	& 0.9790	& 0.9847		& 0.0000	& 0.0002	& 0.0016		& 0.9943	& 0.9969	& 0.9971		\\
7.16 -- 9.57		& 0.0722	& 0.1548	& 0.3310		& 0.8543	& 0.8993	& 0.9606		& 0.0000	& 0.0000	& 0.0022		& 0.9917	& 0.9950	& 0.9969		\\
9.57 -- 12.80		& 0.3036	& 0.3726	& 0.5208		& 0.5596	& 0.6872	& 0.8085		& 0.0000	& 0.0000	& 0.0008		& 0.9851	& 0.9938	& 0.9958		\\
12.80 -- 13.47		& 0.0583	& 0.1918	& 0.3111		& 0.0179	& 0.0618	& 0.1116		& 0.9678	& 0.9714	& 0.9737		& 0.9822	& 0.9916	& 0.9947		\\
 \hline 
\end{tabular}
\end{table*}

 \begin{figure*}
   \begin{center}
     \includegraphics[width=0.87\linewidth]{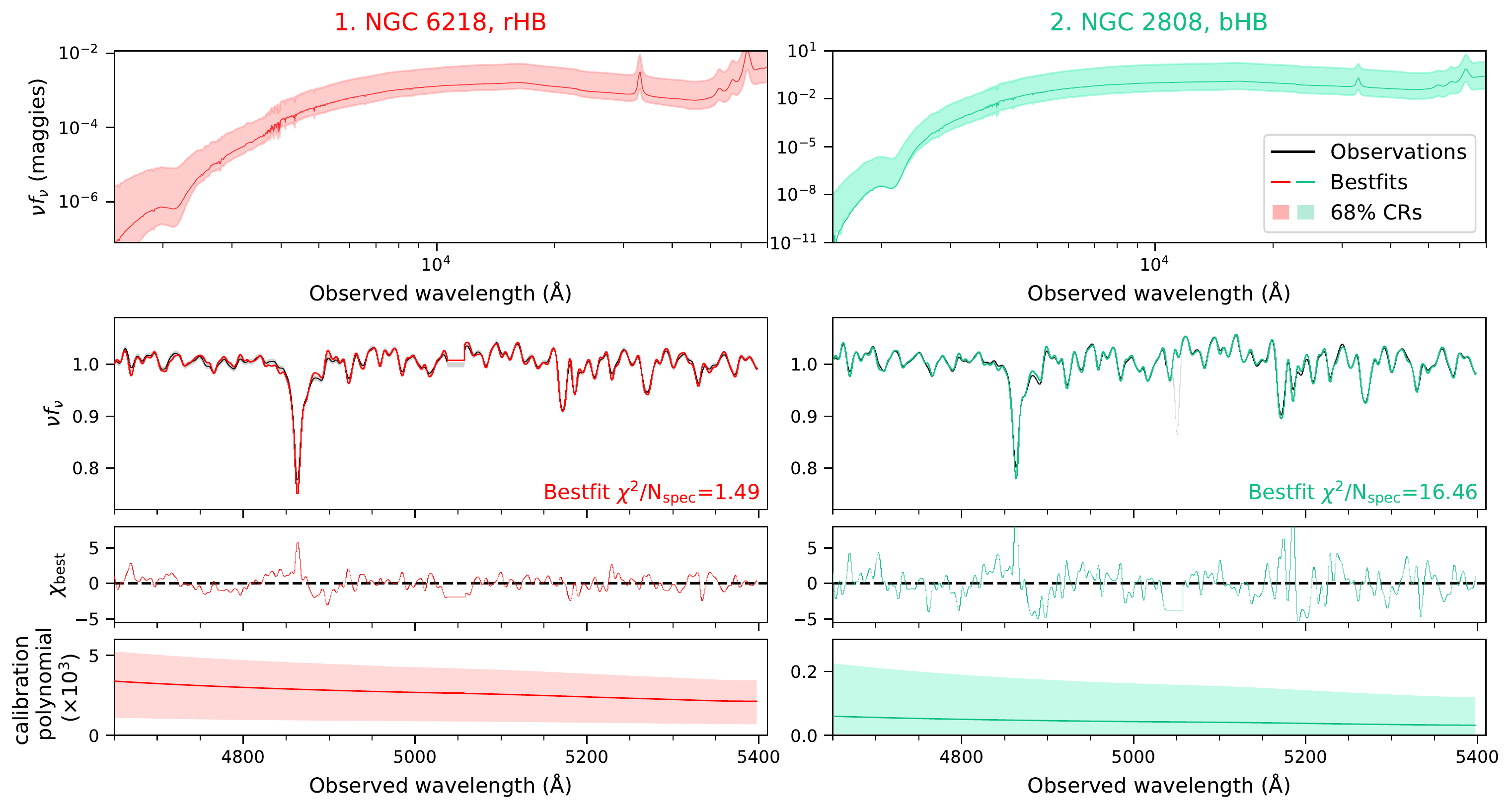} \\
     \includegraphics[width=0.87\linewidth]{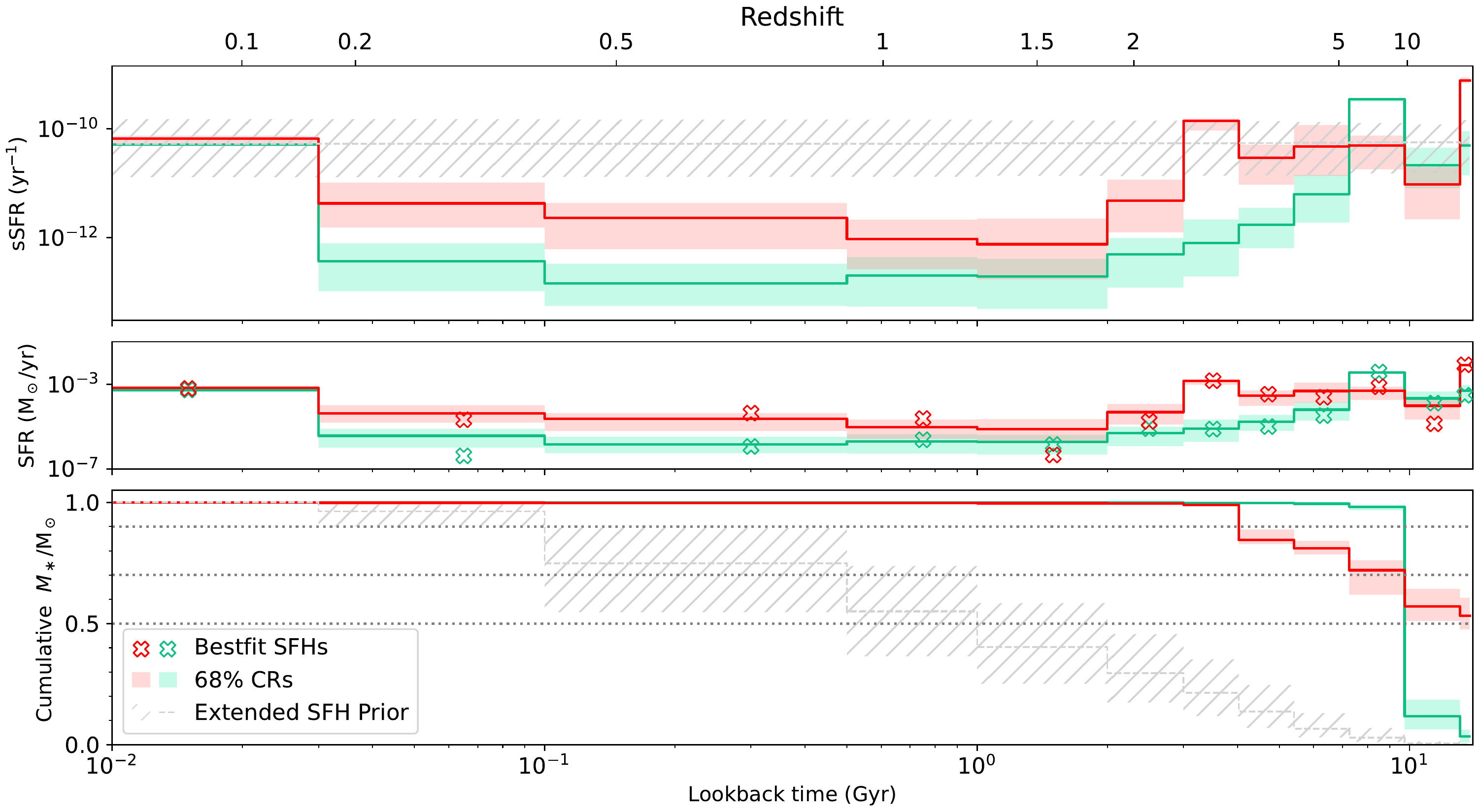}
   \end{center}
   \caption{ Summary of the fitting results for two Milky Way globular clusters selected to have a similar metallicity as DF44: 
   NGC~6362 (redder horizontal branch) and
   NGC~2808 (bluer horizontal branch). 
   \textit{Top:} Fits to observations, similar to Fig.~\ref{fig:summary_fit} for DF44. 
   \textit{Bottom:}  Posterior distributions for star formation and mass growth, similar to Fig.~\ref{fig:summary_fit_sfhs} for DF44.  
   Both clusters are fitted with a fixed mass, $\log(M_\ast/\mathrm{M}_\odot)=8$. 
   } \label{fig:compare_gcs}
 \end{figure*}
 
\section{Systematic biases in measuring SFHs} \label{app:sfh_biases}

\subsection{SFH biased by blue horizontal branch stars} \label{app:sfh_biases_bhb}

The use of integrated light to reconstruct stellar populations has the caveat that multiple types of stars can share spectral signatures. 
This is the case for young, massive main-sequence stars and old, metal poor stars on the blue side of the horizontal branch (HB); both act to amplify the equivalent width of the Balmer lines. 
A population of blue HB stars produces a flux shortward of 3000~\AA\ which increases with decreasing metallicity due to a hotter main-sequence turnoff.
Neglecting to include a blue HB population in models can lead to predictions of unrealistically young ages \citep[e.g.,][]{worthey1994, schiavon2004, thomas2005, schiavon2007}. 
The difficulty of distinguishing between these two stellar populations has been noted in GCs and dwarf galaxies \citep[e.g.,][]{monaco2003, schiavon2004, conroy2018, cabrera-ziri2022}, as well as in elliptical galaxies \citep{maraston2000}.

In the SFH fit to DF44 (for which the primary age indicator is the H$\beta$ absorption line) we see a rise in the SFR in the two most recent time bins corresponding to the last 100~Myrs (by ${1.8}_{^{-0.4}}^{_{+0.2}}$~dex for the extended SFH, ${2.5}_{^{-1.1}}^{_{+1.5}}$~dex for the concentrated SFH) -- yet there are no corresponding emission lines to suggest the presence of a young stellar population. 
Given the low metallicity of this UDG ($\log(Z_\ast/\mathrm{Z}_\odot)\sim -1.2$), the presence of a blue HB population would not be unexpected. 

While the bias between the age and the blue HB stars is well known when fitting simple stellar populations \citep[SSPs;][]{conroy2018}, it is not yet well studied for non-parametric SFHs.
\citet{ocvirk2010} provided a first look at the impact of blue HB stars on linear combinations of SSP models, finding that the presence of blue HB stars can be inferred as a recent burst of star formation at $\sim 100$~Myr, contributing less than around 10~per~cent of the total stellar mass. 
While this provides a promising explanation for the apparent star formation bursts we observe in the SFH of DF44, we follow a similar test using the non-parametric SFH described in Section~\ref{sec:fitting_physical_model}.

In order to investigate if our SFHs are affected by the presence of blue HB stars which mimic a burst of SF within the last 100~Myrs, we fit the SFHs of two Galactic GCs, one with a known blue HB and the other without. 
We select the GCs from \citet{schiavon2005}, with metallicities similar to DF44: 
NGC~2808 has [Fe/H] $= -1.29$ and $(B-R)/(B+V+R)=-0.49$ (bluer HB), and 
NGC~6218 (M12) has [Fe/H] $= -1.32$ and $(B-R)/(B+V+R)=0.97$ (redder HB).
Given that GCs are reasonable approximations of SSPs, we expect an early single burst of star formation only. 
Fits to the spectra\footnote{Downloaded from \url{http://www.noao.edu/ggclib}.} of NGC~2808 and NGC~6218, over the same wavelength range as DF44, following the procedure described in Section~\ref{sec:fitting}, are shown in Fig.~\ref{fig:compare_gcs}. 
The top panels summarise the comparison between the observations (black) lines, and models (coloured lines). 
Similar to Fig.~\ref{fig:summary_fit_sfhs}, the bottom panels show the sSFR, SFR, and mass assembly histories.
An extended SFH was assumed, and the total stellar mass was fixed to $10^{8}~\mathrm{M}_\odot$.

In both cases we find an increase in the SFR within the last 100~Myr, although to a larger extent for the GC with the blue HB stars (by ${1.5}_{^{-0.3}}^{_{+0.5}}$~dex for NGC~6218, and by ${2.6}_{^{-0.4}}^{_{+0.4}}$~dex for NGC~2808).  
In addition, we see that both SFHs are early and short-lived, although there are modest levels of star formation at $>2$~Gyr, which likely results from the models being unable to precisely match the high S/N spectra ($\sim 180$ and $\sim 480$, respectively). 
We conclude from this comparison that some component of the recent SF burst we measure for DF44 could plausibly be related to a population of blue HB stars.

\begin{figure*}
  \begin{center}
    \includegraphics[width=\linewidth]{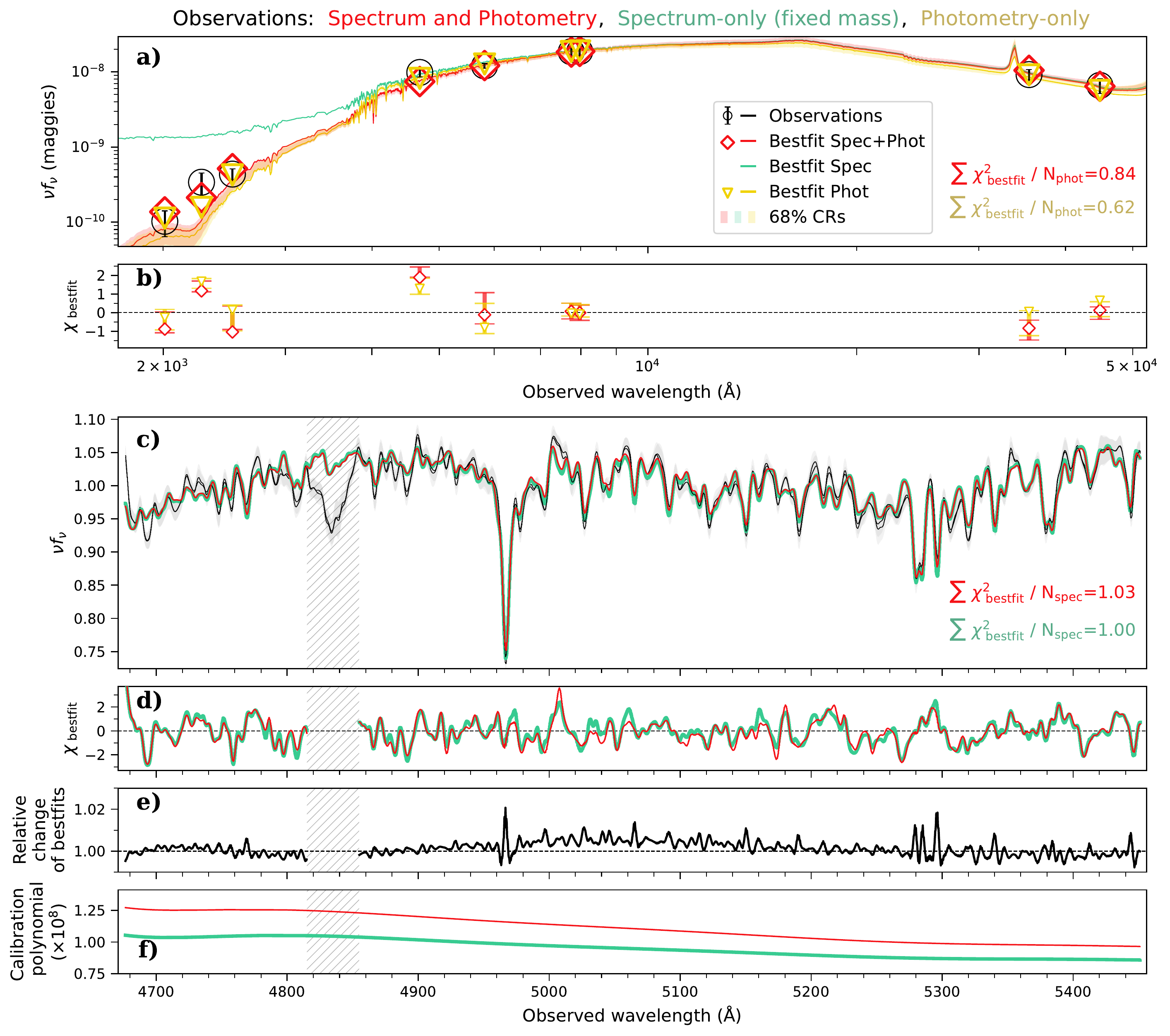} 
  \end{center}
  \caption{ Comparison of fits with the spectrum and photometry (red), spectrum only \citep[green, with mass fixed to the value from][]{vandokkum2016}, and photometry only (yellow), assuming an extended SFH prior. 
  The observed data (black) is compared to the bestfit models (coloured lines) and the 68~per~cent CR of 500 randomly drawn models from the posteriors (shaded coloured regions). 
  The corresponding posteriors are shown in Fig.~\ref{fig:compare_data}.
  {\it (a)} The observed (circles) and bestfit (diamonds and triangles) photometric points, where the reduced $\chi^2/N_\mathrm{data}$ of the bestfit SED are listed. 
  {\it (b)} The $\chi$  ($[\mathrm{data}-\mathrm{model}]/\sigma$) of the bestfit photometric points. 
  {\it (c)} The observed spectrum (uncertainties shown in grey) and bestfit spectra (multiplied by the spectrophotometric calibration polynomial). The hatched grey region indicates the spectral region masked throughout the fitting process.
  {\it (d)} The $\chi$ of the bestfit spectra as a function of wavelength.
  {\it (e)} The relative change of the bestfit models, i.e., the ratio of the two bestfit spectra. 
  {\it (f)} The spectrophotometric calibration models, with 68~per~cent CRs shown as shaded regions.
    } \label{fig:compare_data_fits}
\end{figure*}

\begin{figure*}
  \begin{center}
    \includegraphics[width=0.8\linewidth]{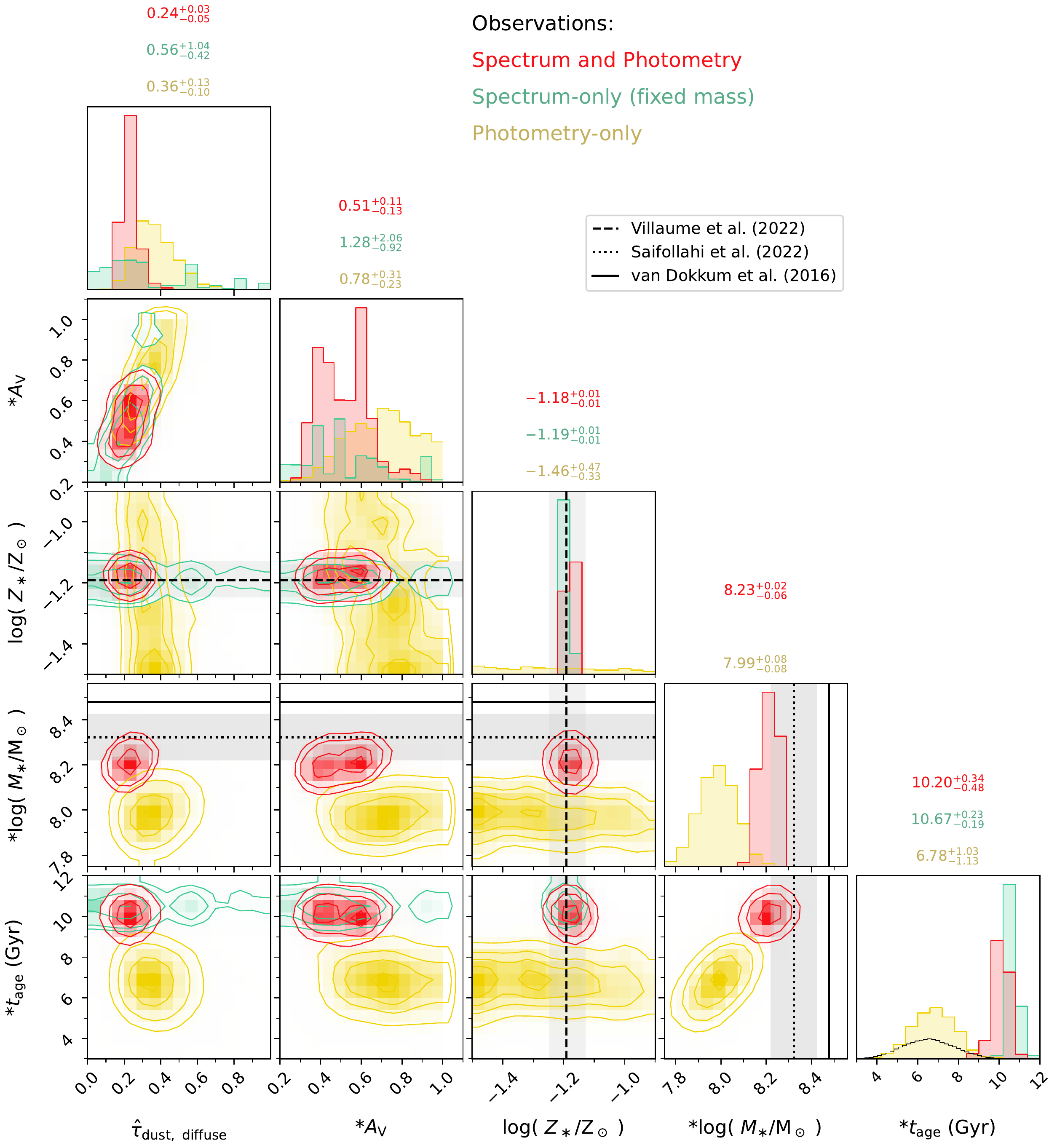}
  \end{center}
  \caption{ 
  Comparison of posterior pdfs derived with the photometry-only (yellow), spectrum-only \citep[green, with mass fixed to the value from][]{vandokkum2016}, and both spectrum and photometry (red), assuming an extended SFH prior. 
  Fits are shown in Fig.~\ref{fig:compare_data_fits}.
  Posteriors of selected fitted and derived (marked with asterisks) parameters are shown. 
  Contours are shown smoothed with a $n=1$ Gaussian kernel. 
  Black lines denote the expected results from the literature: stellar metallicity from \citet{villaume2022}, and stellar mass from \citet{vandokkum2016} and \citet{saifollahi2022}. 
  The median and uncertainties from the 68~per~cent CR are listed along the top of the one-dimensional histograms. 
  The implicit age prior is shown as a black histogram for reference. 
      } \label{fig:compare_data}
\end{figure*}

\subsection{Fitting the spectroscopy and photometry together vs separately} \label{app:sfh_biases_need_both}

Figs.~\ref{fig:compare_data_fits} and \ref{fig:compare_data} show the results of fitting the models to observations of DF44, where we include the following input: 
i) using only the photometry (yellow), 
ii) using only the spectroscopy (green), and 
iii) using both the photometry and spectroscopy (red),
and assuming an extended SFH prior.
We note that the stellar mass is fixed \citep[to the value reported by][]{vandokkum2016} for the spectrum-only fit as the continuum was subtracted from the spectrum.

Similar to Fig.~\ref{fig:summary_fit} discussed in Section~\ref{sec:results}, in Fig.~\ref{fig:compare_data_fits} the observations (black lines and markers) are shown relative to the bestfit models (coloured lines and markers, where the colours denote which observations were fit). 
Shaded coloured regions indicate the 68~per~cent CRs from sampling the posterior pdfs, where the grey shaded region indicates the uncertainties in the spectrum. 

Both bestfit SED models match the photometry with reasonable $\chi_\mathrm{bestfit}$. 
In comparison, the UV flux is significantly overestimated when fitting only the spectroscopy.
Since the UV provides information about recent star formation, and the UV to optical colours constrain the dust attenuation, we do not expect to constrain these properties from the spectrum alone. 

A comparison of the observed spectrum with the bestfit models is also shown in Fig.~\ref{fig:compare_data_fits}, with the $\chi_\mathrm{bestfit}$ as a function of wavelength, and the spectrophotometric calibration polynomial (see Section~\ref{sec:fitting_physical_model_speccal}). 
The ratio of the two bestfit models, shown flattened by dividing through by a polynomial, shows that the fits are similar at the 2~per~cent level. 
The only notable differences between the two bestfit models are around the H$\beta$ line and Mg~{\sc II} features at $\sim5285$~\AA~--~5305~\AA\ (observed-frame). 
The positive ratio of the H$\beta$ line between the spectrum-and-photometry fit over the spectrum-only fit is consistent with the UV flux being constrained for the former, such that the absorption line is preferentially shallower. 
The difference in the Mg~{\sc II} lines reflects the difference in metallicities predicted for each fit, as well as the inability of the (fixed scaled-solar abundance) models to be flexible to such features.

\vspace{0.2cm}
Fig.~\ref{fig:compare_data} compares the basic stellar properties (normalisation of the diffuse dust attenuation curve, $V$-band extinction, stellar metallicity, stellar mass, and mass-weighted age) for the fits to the three sets of observations.
This figure is akin to Fig.~\ref{fig:summary_fit_corner}, discussed in Section~\ref{sec:results}.
For comparison, black lines indicate values measured in the literature: dashed lines indicate the stellar isochrone metallicity measured by \citet[][]{villaume2022}, while dotted and solid lines indicate the stellar mass measured by \citet{vandokkum2016} and \citet{saifollahi2021}, respectively. 
For reference, the prior on the age (which is implicit, as age is determined by the time bin widths and SFH) is shown as a black histogram.

The broadband NUV to NIR photometry (yellow) and continuum normalised spectroscopy (green) carry different information about the galaxy properties.
The broad (yet coarse) photometry provides a tighter constraint on the dust attenuation, while the spectroscopy constrains the metallicity. 
The dust attenuation cannot be determined from the spectroscopy alone because of the lack of continuum information; the spectrophotometric calibration marginalises over the continuum shape, and is degenerate with both the stellar mass and dust attenuation. 
On the other hand, the metallicity is tightly constrained by the spectroscopy as there is detailed information among the numerous absorption lines. 

Despite the formal consistency of the dust and metallicity parameters between these two fits (given the large uncertainties), the age posteriors are significantly different. 
The age posterior from the photometry largely traces the (implicit) prior. 
A tighter pdf for the stellar metallicity provides a more precise estimate of the age, as expected given the degeneracy between these two parameters. 

Simultaneously fitting the photometry and spectroscopy (shown in red) constrains the full set of parameters.
In the particular case of DF44, the results are largely informed by the spectroscopy, which covers a broad range in metallicity and age features -- the inclusion of the photometry only modestly affects the posteriors.
The stellar mass derived from the combined data sets is consistent with that of \citet{saifollahi2022}, while the photometry-derived posterior is skewed lower by $\sim0.23$~dex, which is likely also related to the lower estimate for the stellar metallicity. 
The combined result shows DF44 to be very old, metal poor, and perhaps with some small amount of dust.

\begin{figure*}
  \begin{center}
    \includegraphics[width=0.85\linewidth]{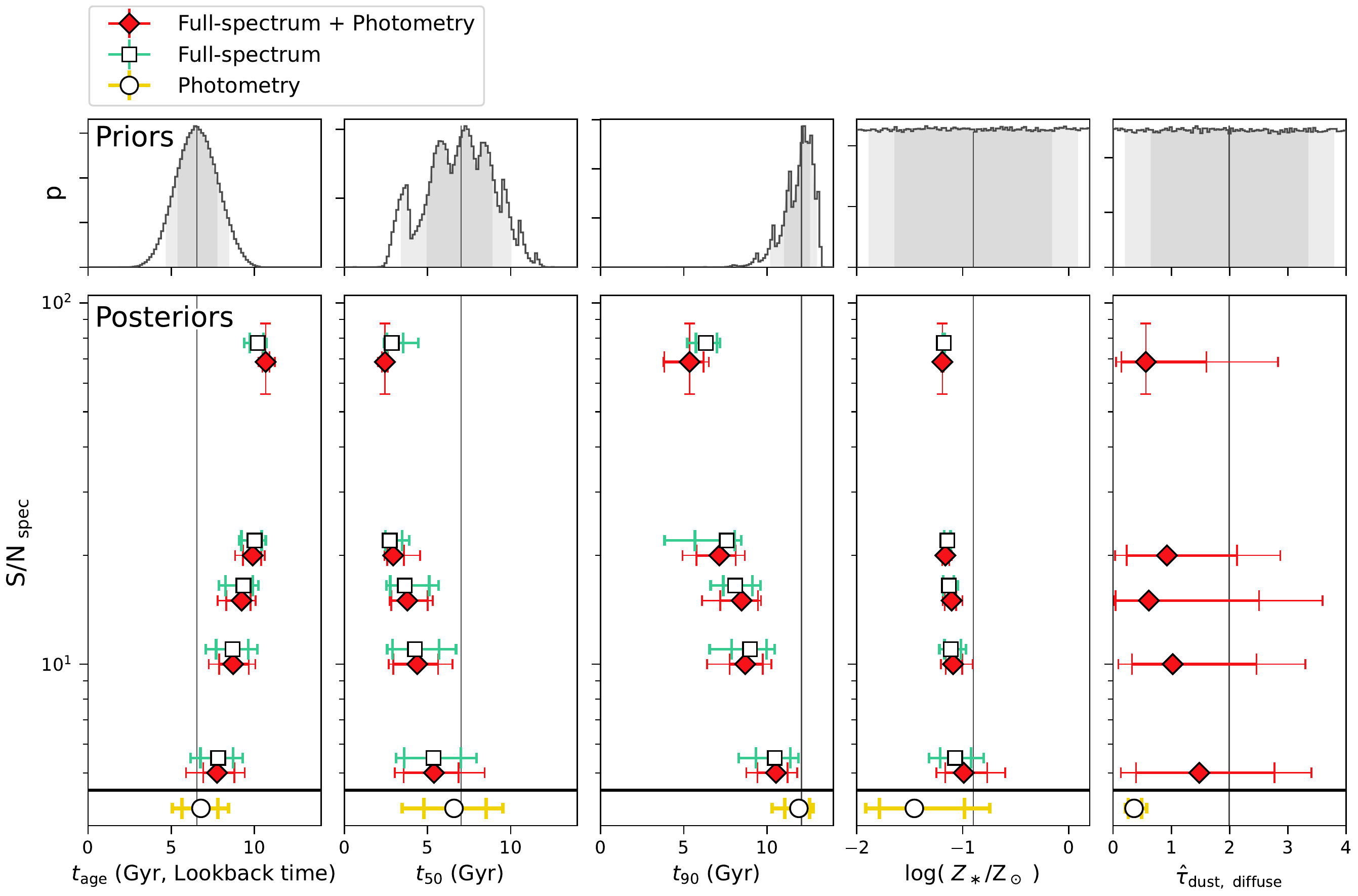} 
  \end{center}
  \caption{ Using the extended SFH prior, we compare the priors (top panels) and posterior parameter estimates (bottom panels) as a function of S/N. Values determined from full-spectrum fits with (diamonds) and without (squares; vertically offset for clarity) including the photometry are shown, as well as from photometry alone (circles). 
  Points are shown with error bars corresponding to the 68~per~cent (thick and wide) and 95~per~cent (thin and narrow) percentiles of the posteriors. Similarly, for the prior we show the median with a grey line, and 68~per~cent and 95~per~cent CRs with shaded regions. 
  While mass-weighted ages are shown in units of lookback time, the time-scales $t_{50}$ and $t_{90}$ have units of time since the Big Bang ($t=0$ is the Big Bang).
  } \label{fig:parameter_vs_snr}
\end{figure*}

\subsection{SFH biased by choice of prior} \label{app:sfh_biases_priors}

Fig.~\ref{fig:parameter_vs_snr} demonstrates the S/N dependence of the bias imposed by the choice of SFH prior, which in this case is an extended SFH in describing a very old stellar population. 
We refit the KCWI spectrum of DF44 with the extended SFH prior (\aDo), successively increasing the uncertainties of the spectrum such that the $\mathrm{S/N}_\mathrm{spec}=5$, 10, 15, and 20.
The medians of the recovered pdfs are shown for the mass-weighted age (in lookback time), $t_{50}$ and $t_{90}$ (in time since the Big Bang), \logzsol, and diffuse dust, with error bars corresponding to the 68~per~cent (thick and wide) and 95~per~cent (thin and narrow) CRs. 
Points mark the results from fitting the spectrum and photometry simultaneously (diamonds), the spectrum alone (squares, offset vertically for clarity), and the photometry alone (circles). 
The prior distributions are shown in the top panels. 
Note that because the implicit priors for the SFH time-scales depend on the widths of the SFH time bins (a step function), the distributions are not necessarily smooth. 

The SFH time-scales are more heavily weighted by the SFH prior at low S/N.
This is particularly true for $t_{90}$, which we use as a proxy of the quenching time. 
In contrast, neither the stellar metallicity nor the dust is significantly biased, or at least the offsets are well within the (large) uncertainties. 
While having a complete set of observations informs many of the galaxy properties, the choice of a `good' SFH prior is important.

\subsection{Comparing results between studies -- prior and data dependence} \label{app:sfh_biases_compare_lit}

\begin{figure}
  \begin{center}
        \includegraphics[width=\linewidth]{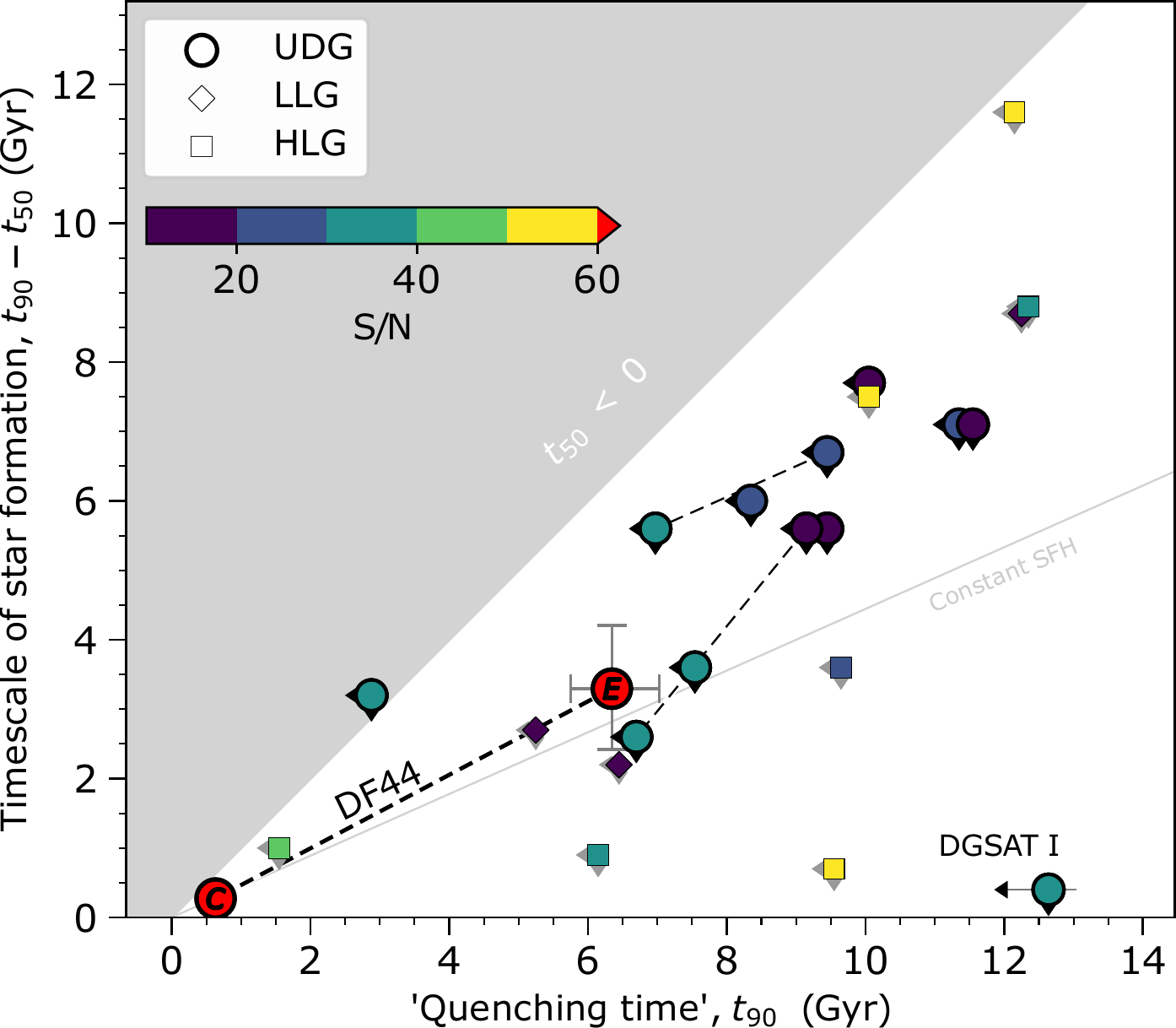}
  \end{center}
  \caption{ Star formation time-scales of UDGs (circles) and dwarfs (low luminosity and high luminosity galaxies; squares and diamonds, respectively) for observations from the literature.
  We approximate the quenching time as when 90~per~cent of the stellar mass is in place ($t_{90}$), while the time-scale $t_{90}-t_{50}$ gives a sense of the duration of star formation, i.e., how concentrated/extended the SFH is. 
  Other than DF44, we show the points from observations with arrows indicating that they are upper limits (see text).
  Points are coloured according to their S/N, where DF44 has a mean S/N of 96~\AA$^{-1}$.
  Dashed lines connect points measured for the same object, but from different studies. 
  Sources: \citet{ferremateu2018}, \citet{ruizlara2018}, and \citet{martinnavarro2019}.
  The points from \citet{ruizlara2018} are shown with S/N$=32$~\AA$^{-1}$, the median of the reported range in values.
     } \label{fig:lit_compare_times}
\end{figure}

Fig.~\ref{fig:lit_compare_times} shows a comparison of the star formation time-scales of UDGs (circles) and dwarfs (squares and diamonds) for observations from the literature
(for Coma galaxies in almost all cases).
We compare the time at which we consider the galaxy quenched, $t_{90}$, with how extended the SFH is, $t_{50}-t_{90}$.
The grey shaded region denotes the parameter space where ages ($t_{50}$) are older than the Universe (e.g., OGS1 from \citealt{ruizlara2018}).
We show the results from the literature as upper limits given the possible biases in SFH time-scales discussed above related to the S/N, and choice of SFH priors.

Except for DF44, all the literature values were measured using the full-spectrum fitting code {\sc steckmap}.
Notably {\sc steckmap} smooths the SFHs via (tuneable) regularisation
akin to Gaussian priors on the SFH and age--metallicity relations (see the discussion in Section~\ref{sec:fitting_physical_model_sfh}). 
The details of the regularisation differ between all studies, where for example, \citet{ruiz-lara2015} present the outcome of averaging several results with various smoothing parameters.
\citet{martinnavarro2019} show in their appendix~A the difference in their regularised and un-regularised results to be $\sim 1$~Gyr in $t_{50}$ and $\lesssim 0.4$~Gyr in $t_{90}$.

\citet{ferremateu2018} compared their SFH time-scales derived from {\sc steckmap} with those from an alternative fitting code, {\sc starlight}, which does not impose regularisation but does require relative-flux calibrated spectra. 
Between the two fitting approaches, \citet{ferremateu2018} found consistent results in that the SFHs are extended and had similar quenching times.
That said, {\sc starlight} preferred starting star formation $\sim 2$~Gyr later, such that the ages were younger and star forming time-scales were shorter. 
In contrast, the `burstier' prior used in this work produced earlier star formation and quenching. 

Because of the difficulties in determining the ages of old stellar populations, even subtle differences in data or analysis can impact results beyond the expected uncertainties. 
As an example, we can compare measurements for two UDGs, DF26/Yagi93 and Yagi418, both studied by \citet{ferremateu2018} and \citet{ruizlara2018}; the values are connected with dashed lines in Fig.~\ref{fig:lit_compare_times}. 
Each author used rest-frame optical spectroscopy (where \citealt{ruizlara2018} reported higher S/N and had a wider wavelength coverage) and they used the same code ({\sc steckmap}). 
However, the median mass-weighted ages differ by $\sim 1$~Gyr (uncertainties were not reported, but the luminosity weighted ages are formally consistent). 
In both cases the higher S/N data provided a solution shifted in the expected direction (i.e., towards older and less-extended SFHs).

While DF44 appears to have (one of) the shortest SFHs and earliest quenching times, we caution that a detailed comparison should consider priors and the S/N. 
A poorly chosen SFH prior will have a stronger bias at a low S/N. 
For example, in using an extended SFH prior with the DF44 KCWI spectrum degraded to $\mathrm{S/N}=20$, we recover $t_{50}\sim 2.9\pm0.5$~Gyr and $t_{90}\sim 7.1\pm1.2$~Gyr (see Fig.~\ref{fig:parameter_vs_snr} in Appendix~\ref{app:sfh_biases_priors}), which overlaps with the lower end of UDGs in Fig.~\ref{fig:lit_compare_times}. 
This suggests that some of these objects could be older, and have less-extended SFHs.

Along the same lines, we do not include photometry-derived results in Fig.~\ref{fig:lit_compare_times} as the comparison can be misleading given the different choices (and relative contributions) of SFH priors. 
In the preceding sections we have shown that the photometry-derived ages are younger than the spectroscopy- or combined-derived ages.
There is a similar difference between the results of 
\citet[][with optical to NIR photometry; not shown in Fig.~\ref{fig:lit_compare_size}]{pandya2018} and \citet[][with rest-frame optical spectroscopy, S/N$\sim 10$~\AA$^{-1}$]{martinnavarro2019}.
Both studied the UDG DGSAT~I, although using different fitting methods and assuming different SFHs. 
\citet{pandya2018} fitted their photometry (via MCMC) to a delayed-exponential model, while \citet{martinnavarro2019} fitted their spectroscopy with {\sc steckmap}. 
We note that in this example the priors are considerably different.
For a delayed exponential model with linearly uniform priors with $\tau=0.1$--10~Gyr and $t_\mathrm{age}=0.1$--14~Gyr, the implicit prior on the mass-weighted age has a median of 3.2~Gyr. In comparison, a constant SFH has a median age of half the age of the Universe, $\sim 6.8$~Gyr (see also the discussion in \citealt{johnson2021}).
While the luminosity-weighted ages are similar ($\sim 3$~Gyr), their mass-weighted ages are discrepant by $>1$~Gyr 
($t_\mathrm{age}$ in the delayed-exponential model is the onset of star formation, where for a $\tau>3$ this corresponds to ages considerably younger than $t_\mathrm{age}$). 
The metallicities are also discrepant by $>1$~dex, although \citet{martinnavarro2019} found that DGSAT~I is unusually $\alpha$-enhanced.
Several other studies have studied UDGs from photometry alone \citep[e.g.,][]{greco2018b, barbosa2020}, and have similarly noted younger ages than spectroscopy-derived results.

We additionally note that \citet{martinnavarro2019} uses a set of SSP models different than used in both this work and the other UDGs studies discussed here.
Neither the choice of SSP models or appilcation of regularisation would explain the significant offset between the SFHs of DGSAT~I and the other UDGs, however.

 \newpage
 \FloatBarrier
\section{Degeneracy between dust attenuation and flux from old stellar populations in the NUV} \label{app:old_v_dust}

\begin{figure*}
  \begin{center}
        \includegraphics[width=\linewidth]{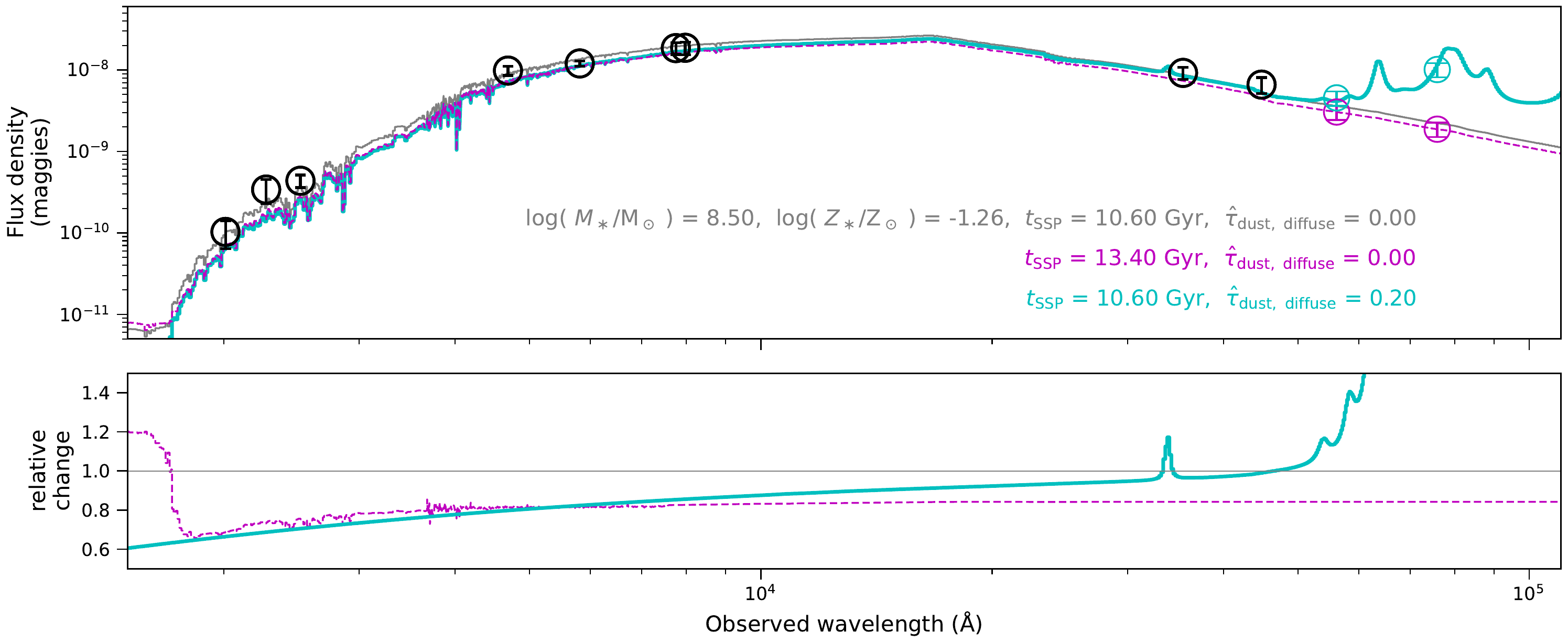}
  \end{center}
  \caption{ A brief demonstration of the degeneracy between dust attenuation and age on the shape of SEDs. 
  {\it Top:} Photometry of DF44 (black markers) and models (coloured lines) for three SSP populations. Photometric points corresponding to the `old' (purple dashed) and `dusty' (solid cyan) models are shown as measured by the {\it JWST} F560W and F770W filters (coloured markers), with $\mathrm{S/N}\sim5$. 
  {\it Bottom:} The relative change between the fiducial (grey) and older or dustier models. While the effect of either increasing the age or dust acts similarly at wavelengths $<1~\mu$m, the effect acts in the opposite sense in the mid-infrared.   
     } \label{fig:obs_models_age_v_dust}
\end{figure*}

The normalisation of the dust attenuation curve (\dust) and the fraction of old stars, both parameters of our physical model, are degenerate at optical and UV wavelengths. 
As a brief example of this degeneracy, Fig.~\ref{fig:obs_models_age_v_dust} shows the photometry for DF44 (black points) relative to three model SEDs with simple stellar populations (i.e., not the results of fitting the physical model described in Section~\ref{sec:fitting}).
Taking the grey model as the `fiducial' model, slight variations in age and dust are shown by the purple and cyan models, respectively.
While the 2.8~Gyr age increase or 0.2~dex increase in diffuse dust produces an equivalent effect in the NUV, they have opposing effects at wavelengths $>1~\mu\mathrm{m}$.
Coloured markers show the expected photometry in two {\it JWST} filters in the mid-infrared, with $\mathrm{S/N}\sim5$ to reflect the average uncertainty of the IRAC data. 
In this example, the `old' and `dusty' models are slightly distinguishable in F560W ($\Delta m_\mathrm{AB} \sim 0.6 \sigma_m$) but very different in F770W ($\Delta m_\mathrm{AB} \sim 3 \sigma_m$).
The inclusion of mid-infrared data to our data set would allow us to assess whether DF44 is as dusty as our results suggest or a product of the complex degeneracies between physical parameters (see Section~\ref{sec:results_sfh}).

\bsp	
\label{lastpage}
\end{document}

%% file: authors.tex
\author[K. Webb et al.]{
\newauthor
Kristi A. Webb,$^{1,2}$\thanks{E-mail: kristi.webb@uwaterloo.ca} 
Alexa Villaume,$^{1,2}$ 
Seppo Laine,$^{3}$ 
Aaron J. Romanowsky,$^{4,5}$ 
\newauthor
Michael Balogh,$^{1,2}$ 
Pieter van Dokkum,$^{6}$ 
Duncan A. Forbes,$^{7}$ 
Jean Brodie,$^{7}$ 
\newauthor
Christopher Martin,$^{8}$
and Matt Matuszewski$^{8}$
\\
$^{1}$Waterloo Centre for Astrophysics, University of Waterloo, Waterloo, Ontario, N2L3G1, Canada \\
$^{2}$Department of Physics and Astronomy, University of Waterloo, Waterloo, Ontario N2L 3G1, Canada \\
$^{3}$IPAC, Mail Code 314-6, Caltech, 1200 E. California Blvd., Pasadena, CA 91125, USA \\
$^{4}$Department of Physics \& Astronomy, San Jos\'e State University, One Washington Square, San Jose, CA 95192, USA \\
$^{5}$Department of Astronomy and Astrophysics, University of California Santa Cruz, 1156 High Street, Santa Cruz, CA 95064, CA, USA \\
$^{6}$Astronomy Department, Yale University, 52 Hillhouse Avenue, New Haven, CT 06511, USA \\
$^{7}$Centre for Astrophysics \& Supercomputing, Swinburne University, Hawthorn VIC 3112, Australia \\
$^{8}$Cahill Center for Astrophysics, California Institute of Technology, 1216 East California Boulevard, Mail Code 278-17, Pasadena, CA 91125, USA
}

%% file: webb2022_df44_sfhs.bbl
\begin{thebibliography}{}
\makeatletter
\relax
\def\mn@urlcharsother{\let\do\@makeother \do\$\do\&\do\#\do\^\do\_\do\%\do\~}
\def\mn@doi{\begingroup\mn@urlcharsother \@ifnextchar [ {\mn@doi@}
  {\mn@doi@[]}}
\def\mn@doi@[#1]#2{\def\@tempa{#1}\ifx\@tempa\@empty \href
  {http://dx.doi.org/#2} {doi:#2}\else \href {http://dx.doi.org/#2} {#1}\fi
  \endgroup}
\def\mn@eprint#1#2{\mn@eprint@#1:#2::\@nil}
\def\mn@eprint@arXiv#1{\href {http://arxiv.org/abs/#1} {{\tt arXiv:#1}}}
\def\mn@eprint@dblp#1{\href {http://dblp.uni-trier.de/rec/bibtex/#1.xml}
  {dblp:#1}}
\def\mn@eprint@#1:#2:#3:#4\@nil{\def\@tempa {#1}\def\@tempb {#2}\def\@tempc
  {#3}\ifx \@tempc \@empty \let \@tempc \@tempb \let \@tempb \@tempa \fi \ifx
  \@tempb \@empty \def\@tempb {arXiv}\fi \@ifundefined
  {mn@eprint@\@tempb}{\@tempb:\@tempc}{\expandafter \expandafter \csname
  mn@eprint@\@tempb\endcsname \expandafter{\@tempc}}}

\bibitem[\protect\citeauthoryear{Alabi et~al.,}{Alabi et~al.}{2018}]{alabi2018}
Alabi A.,  et~al., 2018, \mn@doi [\mnras] {10/gd6577}, 479, 3308

\bibitem[\protect\citeauthoryear{Alabi, Romanowsky, Forbes, Brodie  \&
  Okabe}{Alabi et~al.}{2020}]{alabi2020}
Alabi A.~B.,  Romanowsky A.~J.,  Forbes D.~A.,  Brodie J.~P.,   Okabe N.,
  2020, \mn@doi [\mnras] {10/gm8wwf}, 496, 3182

\bibitem[\protect\citeauthoryear{Allanson, Hudson, Smith  \& Lucey}{Allanson
  et~al.}{2009}]{allanson2009}
Allanson S.~P.,  Hudson M.~J.,  Smith R.~J.,   Lucey J.~R.,  2009, \mn@doi
  [\apj] {10/b6gt56}, 702, 1275

\bibitem[\protect\citeauthoryear{Amorisco \& Loeb}{Amorisco \&
  Loeb}{2016}]{amorisco2016}
Amorisco N.~C.,  Loeb A.,  2016, \mn@doi [\mnras] {10/gm2h5w}, 459, L51

\bibitem[\protect\citeauthoryear{Asplund, Grevesse, Sauval  \& Scott}{Asplund
  et~al.}{2009}]{asplund2009}
Asplund M.,  Grevesse N.,  Sauval A.~J.,   Scott P.,  2009, \mn@doi [\araa]
  {10.1146/annurev.astro.46.060407.145222}, 47, 481

\bibitem[\protect\citeauthoryear{Barbosa et~al.,}{Barbosa
  et~al.}{2020}]{barbosa2020}
Barbosa C.~E.,  et~al., 2020, \mn@doi [\apjs] {10/gnb26n}, 247, 46

\bibitem[\protect\citeauthoryear{Behroozi, Wechsler, Hearin  \&
  Conroy}{Behroozi et~al.}{2019}]{behroozi2019b}
Behroozi P.,  Wechsler R.~H.,  Hearin A.~P.,   Conroy C.,  2019, \mn@doi
  [\mnras] {10.1093/mnras/stz1182}, 488, 3143

\bibitem[\protect\citeauthoryear{Bell \& {de Jong}}{Bell \& {de
  Jong}}{2001}]{bell2001}
Bell E.~F.,  {de Jong} R.~S.,  2001, \mn@doi [\apj] {10/ds58j5}, 550, 212

\bibitem[\protect\citeauthoryear{Benavides et~al.,}{Benavides
  et~al.}{2021}]{benavides2021}
Benavides J.~A.,  et~al., 2021, \mn@doi [Nature Astronomy] {10/gmp53s}, pp~1--6

\bibitem[\protect\citeauthoryear{Bogd{\'a}n}{Bogd{\'a}n}{2020}]{bogdan2020}
Bogd{\'a}n {\'A}.,  2020, \mn@doi [\apj] {10/gk8xfs}, 901, L30

\bibitem[\protect\citeauthoryear{Boselli \& Gavazzi}{Boselli \&
  Gavazzi}{2006}]{boselli2006}
Boselli A.,  Gavazzi G.,  2006, \mn@doi [Publications of the Astronomical
  Society of the Pacific] {10.1086/500691}, 118, 517

\bibitem[\protect\citeauthoryear{{Cabrera-Ziri} \& Conroy}{{Cabrera-Ziri} \&
  Conroy}{2022}]{cabrera-ziri2022}
{Cabrera-Ziri} I.,  Conroy C.,  2022, \mn@doi [\mnras] {10/gn7jn8}

\bibitem[\protect\citeauthoryear{Calzetti, Armus, Bohlin, Kinney, Koornneef  \&
  Storchi-Bergmann}{Calzetti et~al.}{2000}]{calzetti2000}
Calzetti D.,  Armus L.,  Bohlin R.~C.,  Kinney A.~L.,  Koornneef J.,
  Storchi-Bergmann T.,  2000, \mn@doi [\apj] {10.1086/308692}, 533, 682

\bibitem[\protect\citeauthoryear{Carleton, Errani, Cooper, Kaplinghat,
  Pe{\~n}arrubia  \& Guo}{Carleton et~al.}{2019}]{carleton2019}
Carleton T.,  Errani R.,  Cooper M.,  Kaplinghat M.,  Pe{\~n}arrubia J.,   Guo
  Y.,  2019, \mn@doi [\mnras] {10/gmzdwc}, 485, 382

\bibitem[\protect\citeauthoryear{Carleton, Guo, Munshi, Tremmel  \&
  Wright}{Carleton et~al.}{2021}]{carleton2021}
Carleton T.,  Guo Y.,  Munshi F.,  Tremmel M.,   Wright A.,  2021, \mn@doi
  [\mnras] {10/gmxngh}, 502, 398

\bibitem[\protect\citeauthoryear{Carnall, Leja, Johnson, McLure, Dunlop  \&
  Conroy}{Carnall et~al.}{2019}]{carnall2019a}
Carnall A.~C.,  Leja J.,  Johnson B.~D.,  McLure R.~J.,  Dunlop J.~S.,   Conroy
  C.,  2019, \mn@doi [\apj] {10.3847/1538-4357/ab04a2}, 873, 44

\bibitem[\protect\citeauthoryear{Chabrier}{Chabrier}{2003}]{chabrier2003}
Chabrier G.,  2003, \mn@doi [\pasp] {10.1086/376392}, 115, 763

\bibitem[\protect\citeauthoryear{Chan, Kere{\v s}, Wetzel, Hopkins,
  {Faucher-Gigu{\`e}re}, {El-Badry}, {Garrison-Kimmel}  \&
  {Boylan-Kolchin}}{Chan et~al.}{2018}]{chan2018}
Chan T.~K.,  Kere{\v s} D.,  Wetzel A.,  Hopkins P.~F.,  {Faucher-Gigu{\`e}re}
  C.~A.,  {El-Badry} K.,  {Garrison-Kimmel} S.,   {Boylan-Kolchin} M.,  2018,
  \mn@doi [\mnras] {10/gd2vdd}, 478, 906

\bibitem[\protect\citeauthoryear{Charlot \& Fall}{Charlot \&
  Fall}{2000}]{charlot2000}
Charlot S.,  Fall S.~M.,  2000, \mn@doi [\apj] {10.1086/309250}, 539, 718

\bibitem[\protect\citeauthoryear{Choi, Dotter, Conroy, Cantiello, Paxton  \&
  Johnson}{Choi et~al.}{2016}]{choi2016}
Choi J.,  Dotter A.,  Conroy C.,  Cantiello M.,  Paxton B.,   Johnson B.~D.,
  2016, \mn@doi [\apj] {10.3847/0004-637x/823/2/102}, 823, 102

\bibitem[\protect\citeauthoryear{Conroy}{Conroy}{2013}]{conroy2013a}
Conroy C.,  2013, \mn@doi [\araa] {10.1146/annurev-astro-082812-141017}, 51,
  393

\bibitem[\protect\citeauthoryear{Conroy \& Gunn}{Conroy \& Gunn}{2010}]{fsps}
Conroy C.,  Gunn J.~E.,  2010, p. ascl:1010.043

\bibitem[\protect\citeauthoryear{Conroy, Gunn  \& White}{Conroy
  et~al.}{2009}]{conroy2009}
Conroy C.,  Gunn J.~E.,   White M.,  2009, \mn@doi [\apj]
  {10.1088/0004-637x/699/1/486}, 699, 486

\bibitem[\protect\citeauthoryear{Conroy, Villaume, {van Dokkum}  \&
  Lind}{Conroy et~al.}{2018}]{conroy2018}
Conroy C.,  Villaume A.,  {van Dokkum} P.~G.,   Lind K.,  2018, \mn@doi [\apj]
  {10/ggskpb}, 854, 139

\bibitem[\protect\citeauthoryear{Dalcanton, Spergel  \& Summers}{Dalcanton
  et~al.}{1997}]{dalcanton1997}
Dalcanton J.~J.,  Spergel D.~N.,   Summers F.~J.,  1997, \mn@doi [\apj]
  {10/fj5z9s}, 482, 659

\bibitem[\protect\citeauthoryear{{Danieli} et~al.,}{{Danieli}
  et~al.}{2021}]{danieli2021}
{Danieli} S.,  et~al., 2021, arXiv e-prints, \href
  {https://ui.adsabs.harvard.edu/abs/2021arXiv211114851D} {p. arXiv:2111.14851}

\bibitem[\protect\citeauthoryear{Dav{\'e}, Finlator  \& Oppenheimer}{Dav{\'e}
  et~al.}{2012}]{dave2012}
Dav{\'e} R.,  Finlator K.,   Oppenheimer B.~D.,  2012, \mn@doi [\mnras]
  {10.1111/j.1365-2966.2011.20148.x}, 421, 98

\bibitem[\protect\citeauthoryear{Di~Cintio, Brook, Dutton, Macci{\`o}, Obreja
  \& Dekel}{Di~Cintio et~al.}{2017}]{dicintio2017}
Di~Cintio A.,  Brook C.~B.,  Dutton A.~A.,  Macci{\`o} A.~V.,  Obreja A.,
  Dekel A.,  2017, \mn@doi [\mnras] {10/gbxn8j}, 466, L1

\bibitem[\protect\citeauthoryear{Dickey et~al.,}{Dickey
  et~al.}{2021}]{dickey2021}
Dickey C.~M.,  et~al., 2021, \mn@doi [\apj] {10/gnhrdw}, 915, 53

\bibitem[\protect\citeauthoryear{Digby et~al.,}{Digby et~al.}{2019}]{digby2019}
Digby R.,  et~al., 2019, \mn@doi [\mnras] {10/gnsdxp}, 485, 5423

\bibitem[\protect\citeauthoryear{Dotter}{Dotter}{2016}]{dotter2016}
Dotter A.,  2016, \mn@doi [\apjs] {10.3847/0067-0049/222/1/8}, 222, 8

\bibitem[\protect\citeauthoryear{Draine et~al.,}{Draine
  et~al.}{2007}]{draine2007}
Draine B.~T.,  et~al., 2007, \mn@doi [\apj] {10/cnn9wb}, 663, 866

\bibitem[\protect\citeauthoryear{Eigenthaler et~al.,}{Eigenthaler
  et~al.}{2018}]{eigenthaler2018}
Eigenthaler P.,  et~al., 2018, \mn@doi [\apj] {10.3847/1538-4357/aaab60}, 855,
  142

\bibitem[\protect\citeauthoryear{{El-Badry}, Wetzel, Geha, Hopkins, Kere{\v s},
  Chan  \& {Faucher-Gigu{\`e}re}}{{El-Badry} et~al.}{2016}]{el-badry2016}
{El-Badry} K.,  Wetzel A.,  Geha M.,  Hopkins P.~F.,  Kere{\v s} D.,  Chan
  T.~K.,   {Faucher-Gigu{\`e}re} C.-A.,  2016, \mn@doi [\apj] {10/gkngkt}, 820,
  131

\bibitem[\protect\citeauthoryear{Faber et~al.,}{Faber et~al.}{2007}]{faber2007}
Faber S.~M.,  et~al., 2007, \mn@doi [\apj] {10/dtnsk5}, 665, 265

\bibitem[\protect\citeauthoryear{Fazio et~al.,}{Fazio et~al.}{2004}]{fazio2004}
Fazio G.~G.,  et~al., 2004, \mn@doi [\apjs] {10/bxf7xc}, 154, 10

\bibitem[\protect\citeauthoryear{{Ferr{\'e}-Mateu} et~al.,}{{Ferr{\'e}-Mateu}
  et~al.}{2018}]{ferremateu2018}
{Ferr{\'e}-Mateu} A.,  et~al., 2018, \mn@doi [\mnras] {10/gd5xh5}, 479, 4891

\bibitem[\protect\citeauthoryear{Fillingham, Cooper, {Boylan-Kolchin}, Bullock,
  {Garrison-Kimmel}  \& Wheeler}{Fillingham et~al.}{2018}]{fillingham2018}
Fillingham S.~P.,  Cooper M.~C.,  {Boylan-Kolchin} M.,  Bullock J.~S.,
  {Garrison-Kimmel} S.,   Wheeler C.,  2018, \mn@doi [\mnras]
  {10.1093/mnras/sty958}, 477, 4491

\bibitem[\protect\citeauthoryear{Forbes, Gannon, Romanowsky, Alabi, Brodie,
  Couch  \& {Ferr{\'e}-Mateu}}{Forbes et~al.}{2021}]{forbes2021}
Forbes D.~A.,  Gannon J.~S.,  Romanowsky A.~J.,  Alabi A.,  Brodie J.~P.,
  Couch W.~J.,   {Ferr{\'e}-Mateu} A.,  2021, \mn@doi [\mnras] {10/gmxng5},
  500, 1279

\bibitem[\protect\citeauthoryear{Galliano, Galametz  \& Jones}{Galliano
  et~al.}{2018}]{galliano2018}
Galliano F.,  Galametz M.,   Jones A.~P.,  2018, \mn@doi [Annual Review of
  Astronomy and Astrophysics] {10/gjsg26}, 56, 673

\bibitem[\protect\citeauthoryear{Gannon et~al.,}{Gannon
  et~al.}{2022}]{gannon2022}
Gannon J.~S.,  et~al., 2022, \mn@doi [\mnras] {10/gn3fzq}, 510, 946

\bibitem[\protect\citeauthoryear{{Garrison-Kimmel} et~al.,}{{Garrison-Kimmel}
  et~al.}{2019}]{garrisonkimmel2019}
{Garrison-Kimmel} S.,  et~al., 2019, \mn@doi [\mnras] {10.1093/mnras/stz1317},
  487, 1380

\bibitem[\protect\citeauthoryear{Geha, Blanton, Yan  \& Tinker}{Geha
  et~al.}{2012}]{geha2012}
Geha M.,  Blanton M.~R.,  Yan R.,   Tinker J.~L.,  2012, \mn@doi [\apj]
  {10.1088/0004-637x/757/1/85}, 757, 85

\bibitem[\protect\citeauthoryear{Gordon, Clayton, Misselt, Landolt  \&
  Wolff}{Gordon et~al.}{2003}]{gordon2003}
Gordon K.~D.,  Clayton G.~C.,  Misselt K.~A.,  Landolt A.~U.,   Wolff M.~J.,
  2003, \mn@doi [\apj] {10/cr6d6s}, 594, 279

\bibitem[\protect\citeauthoryear{Greco, Goulding, Greene, Strauss, Huang, Kim
  \& Komiyama}{Greco et~al.}{2018}]{greco2018b}
Greco J.~P.,  Goulding A.~D.,  Greene J.~E.,  Strauss M.~A.,  Huang S.,  Kim
  J.~H.,   Komiyama Y.,  2018, \mn@doi [\apj] {10/gnb3gz}, 866, 112

\bibitem[\protect\citeauthoryear{Grishin, Chilingarian, Afanasiev, Fabricant,
  Katkov, Moran  \& Yagi}{Grishin et~al.}{2021}]{grishin2021}
Grishin K.,  Chilingarian I.,  Afanasiev A.,  Fabricant D.,  Katkov I.,  Moran
  S.,   Yagi M.,  2021, \mn@doi [Nature Astronomy] {10/gnbbxt}

\bibitem[\protect\citeauthoryear{Gu et~al.,}{Gu et~al.}{2018}]{gu2018b}
Gu M.,  et~al., 2018, \mn@doi [\apj] {10/gksv6f}, 859, 37

\bibitem[\protect\citeauthoryear{Han \& Han}{Han \& Han}{2018}]{han2018}
Han Y.,  Han Z.,  2018, \mn@doi [\apjs] {10.3847/1538-4365/aaeffa}, 240, 3

\bibitem[\protect\citeauthoryear{Higson, Handley, Hobson  \& Lasenby}{Higson
  et~al.}{2019}]{higson2019}
Higson E.,  Handley W.,  Hobson M.,   Lasenby A.,  2019, \mn@doi [Statistics
  and Computing, Vol. 29, No. 5, pp. 891-913] {10/ggkdwn}, 29, 891

\bibitem[\protect\citeauthoryear{Hogg, Bovy  \& Lang}{Hogg
  et~al.}{2010}]{hogg2010b}
Hogg D.~W.,  Bovy J.,   Lang D.,  2010, arXiv, p. arXiv:1008.4686

\bibitem[\protect\citeauthoryear{Hunter}{Hunter}{2007}]{matplotlib}
Hunter J.~D.,  2007, \mn@doi [Computing in Science \& Engineering]
  {10.1109/mcse.2007.55}, 9, 90

\bibitem[\protect\citeauthoryear{Jackson et~al.,}{Jackson
  et~al.}{2021}]{jackson2021b}
Jackson R.~A.,  et~al., 2021, \mn@doi [\mnras] {10/gk8xfh}, 502, 4262

\bibitem[\protect\citeauthoryear{{Jhora99}}{{Jhora99}}{2021}]{imclean}
{Jhora99} 2021, {jhora99/imclean: imclean 20210323}, Zenodo,
  \mn@doi{10.5281/zenodo.4850526}

\bibitem[\protect\citeauthoryear{Jiang, Dekel, Freundlich, Romanowsky, Dutton,
  Macci{\`o}  \& Di~Cintio}{Jiang et~al.}{2019a}]{jiang2019a}
Jiang F.,  Dekel A.,  Freundlich J.,  Romanowsky A.~J.,  Dutton A.~A.,
  Macci{\`o} A.~V.,   Di~Cintio A.,  2019a, \mn@doi [\mnras] {10/gk85v7}, 487,
  5272

\bibitem[\protect\citeauthoryear{Jiang et~al.,}{Jiang
  et~al.}{2019b}]{jiang2019b}
Jiang F.,  et~al., 2019b, \mn@doi [\mnras] {10/gn838w}, 488, 4801

\bibitem[\protect\citeauthoryear{Johnson, Leja, Conroy  \& Speagle}{Johnson
  et~al.}{2019}]{johnson2019}
Johnson B.~D.,  Leja J.~L.,  Conroy C.,   Speagle J.~S.,  2019, p.
  ascl:1905.025

\bibitem[\protect\citeauthoryear{Johnson et~al.,}{Johnson
  et~al.}{2021a}]{pythonfsps}
Johnson B.,  et~al., 2021a, Dfm/Python-Fsps: Python-Fsps v0.4.1rc1, Zenodo,
  \mn@doi{10.5281/zenodo.4737461}, \url {https://zenodo.org/record/4737461}

\bibitem[\protect\citeauthoryear{Johnson, Leja, Conroy  \& Speagle}{Johnson
  et~al.}{2021b}]{johnson2021}
Johnson B.~D.,  Leja J.,  Conroy C.,   Speagle J.~S.,  2021b, \mn@doi [The
  Astrophysical Journal Supplement Series] {10/gj3jqr}, 254, 22

\bibitem[\protect\citeauthoryear{Joshi, Pillepich, Nelson, Zinger, Marinacci,
  Springel, Vogelsberger  \& Hernquist}{Joshi et~al.}{2021}]{joshi2021}
Joshi G.~D.,  Pillepich A.,  Nelson D.,  Zinger E.,  Marinacci F.,  Springel
  V.,  Vogelsberger M.,   Hernquist L.,  2021, \mn@doi [\mnras]
  {10.1093/mnras/stab2573}, 508, 1652

\bibitem[\protect\citeauthoryear{Kadowaki, Zaritsky  \& Donnerstein}{Kadowaki
  et~al.}{2017}]{kadowaki2017}
Kadowaki J.,  Zaritsky D.,   Donnerstein R.~L.,  2017, \mn@doi [\apj]
  {10/gkjsj2}, 838, L21

\bibitem[\protect\citeauthoryear{Kass \& Raftery}{Kass \&
  Raftery}{1995}]{kass1995}
Kass R.~E.,  Raftery A.~E.,  1995, \mn@doi [Journal of the American Statistical
  Association] {10/gdnbw3}, 90, 773

\bibitem[\protect\citeauthoryear{Kriek \& Conroy}{Kriek \&
  Conroy}{2013}]{kriek2013}
Kriek M.,  Conroy C.,  2013, \mn@doi [\apjl] {10/gh27j6}, 775, L16

\bibitem[\protect\citeauthoryear{Lee, {Hodges-Kluck}  \& Gallo}{Lee
  et~al.}{2020}]{lee2020}
Lee C.~H.,  {Hodges-Kluck} E.,   Gallo E.,  2020, \mn@doi [\mnras] {10/gj2pk5},
  497, 2759

\bibitem[\protect\citeauthoryear{Leja, Johnson, Conroy, van Dokkum  \&
  Byler}{Leja et~al.}{2017}]{leja2017}
Leja J.,  Johnson B.~D.,  Conroy C.,  van Dokkum P.~G.,   Byler N.,  2017,
  \mn@doi [\apj] {10.3847/1538-4357/aa5ffe}, 837, 170

\bibitem[\protect\citeauthoryear{Leja, Carnall, Johnson, Conroy  \&
  Speagle}{Leja et~al.}{2019}]{leja2019a}
Leja J.,  Carnall A.~C.,  Johnson B.~D.,  Conroy C.,   Speagle J.~S.,  2019,
  \mn@doi [\apj] {10.3847/1538-4357/ab133c}, 876, 3

\bibitem[\protect\citeauthoryear{Liao et~al.,}{Liao et~al.}{2019}]{liao2019}
Liao S.,  et~al., 2019, \mn@doi [\mnras] {10/gmz8bk}, 490, 5182

\bibitem[\protect\citeauthoryear{Lower, Narayanan, Leja, Johnson, Conroy  \&
  Dav{\'e}}{Lower et~al.}{2020}]{lower2020}
Lower S.,  Narayanan D.,  Leja J.,  Johnson B.~D.,  Conroy C.,   Dav{\'e} R.,
  2020, \mn@doi [\apj] {10/gmzdxt}, 904, 33

\bibitem[\protect\citeauthoryear{Lupi, Volonteri  \& Silk}{Lupi
  et~al.}{2017}]{lupi2017}
Lupi A.,  Volonteri M.,   Silk J.,  2017, \mn@doi [\mnras] {10/gbstq4}, 470,
  1673

\bibitem[\protect\citeauthoryear{MacArthur, Gonz{\'a}lez  \&
  Courteau}{MacArthur et~al.}{2009}]{macarthur2009}
MacArthur L.~A.,  Gonz{\'a}lez J.~J.,   Courteau S.,  2009, \mn@doi [\mnras]
  {10.1111/j.1365-2966.2009.14519.x}, 395, 28

\bibitem[\protect\citeauthoryear{Mao, Geha, Wechsler, Weiner, Tollerud, Nadler
  \& Kallivayalil}{Mao et~al.}{2021}]{mao2021}
Mao Y.-Y.,  Geha M.,  Wechsler R.~H.,  Weiner B.,  Tollerud E.~J.,  Nadler
  E.~O.,   Kallivayalil N.,  2021, \mn@doi [\apj] {10.3847/1538-4357/abce58},
  907, 85

\bibitem[\protect\citeauthoryear{Maraston}{Maraston}{2005}]{maraston2005}
Maraston C.,  2005, \mn@doi [\mnras] {10.1111/j.1365-2966.2005.09270.x}, 362,
  799

\bibitem[\protect\citeauthoryear{Maraston \& Thomas}{Maraston \&
  Thomas}{2000}]{maraston2000}
Maraston C.,  Thomas D.,  2000, \mn@doi [\apj] {10/d39vk2}, 541, 126

\bibitem[\protect\citeauthoryear{{Mart{\'i}n-Navarro}
  et~al.,}{{Mart{\'i}n-Navarro} et~al.}{2019}]{martinnavarro2019}
{Mart{\'i}n-Navarro} I.,  et~al., 2019, \mn@doi [\mnras] {10/gk869v}, 484, 3425

\bibitem[\protect\citeauthoryear{Monaco, Bellazzini, Ferraro  \&
  Pancino}{Monaco et~al.}{2003}]{monaco2003}
Monaco L.,  Bellazzini M.,  Ferraro F.~R.,   Pancino E.,  2003, \mn@doi [\apj]
  {10/cqzs95}, 597, L25

\bibitem[\protect\citeauthoryear{Moster, Somerville, Maulbetsch, {van den
  Bosch}, Macci{\`o}, Naab  \& Oser}{Moster et~al.}{2010}]{moster2010}
Moster B.~P.,  Somerville R.~S.,  Maulbetsch C.,  {van den Bosch} F.~C.,
  Macci{\`o} A.~V.,  Naab T.,   Oser L.,  2010, \mn@doi [\apj]
  {2010012604131200}, 710, 903

\bibitem[\protect\citeauthoryear{Moultaka \& Pelat}{Moultaka \&
  Pelat}{2000}]{moultaka2000}
Moultaka J.,  Pelat D.,  2000, \mn@doi [Monthly Notices of the Royal
  Astronomical Society] {10.1046/j.1365-8711.2000.03394.x}, 314, 409

\bibitem[\protect\citeauthoryear{Moultaka, Boisson, Joly  \& Pelat}{Moultaka
  et~al.}{2004}]{moultaka2004}
Moultaka J.,  Boisson C.,  Joly M.,   Pelat D.,  2004, \mn@doi [Astronomy and
  Astrophysics, v.420, p.459-466 (2004)] {10.1051/0004-6361:20034366}, 420, 459

\bibitem[\protect\citeauthoryear{Mowla, {van Dokkum}, Merritt, Abraham, Yagi
  \& Koda}{Mowla et~al.}{2017}]{mowla2017}
Mowla L.,  {van Dokkum} P.,  Merritt A.,  Abraham R.,  Yagi M.,   Koda J.,
  2017, \mn@doi [\apj] {10/gmz8bq}, 851, 27

\bibitem[\protect\citeauthoryear{Noll, Burgarella, Giovannoli, Buat, Marcillac
  \& {Mu{\~n}oz-Mateos}}{Noll et~al.}{2009}]{noll2009}
Noll S.,  Burgarella D.,  Giovannoli E.,  Buat V.,  Marcillac D.,
  {Mu{\~n}oz-Mateos} J.~C.,  2009, \mn@doi [\aap] {10/cpv5zx}, 507, 1793

\bibitem[\protect\citeauthoryear{Ocvirk}{Ocvirk}{2010}]{ocvirk2010}
Ocvirk P.,  2010, \mn@doi [\apj] {10/dsrqgn}, 709, 88

\bibitem[\protect\citeauthoryear{Ocvirk, Pichon, Lan{\c c}on  \&
  Thi{\'e}baut}{Ocvirk et~al.}{2006a}]{ocvirk2006a}
Ocvirk P.,  Pichon C.,  Lan{\c c}on A.,   Thi{\'e}baut E.,  2006a, \mn@doi
  [\mnras] {10/cf985t}, 365, 46

\bibitem[\protect\citeauthoryear{Ocvirk, Pichon, Lan{\c c}on  \&
  Thi{\'e}baut}{Ocvirk et~al.}{2006b}]{ocvirk2006b}
Ocvirk P.,  Pichon C.,  Lan{\c c}on A.,   Thi{\'e}baut E.,  2006b, \mn@doi
  [\mnras] {10/chrfkc}, 365, 74

\bibitem[\protect\citeauthoryear{Pandya et~al.,}{Pandya
  et~al.}{2018}]{pandya2018}
Pandya V.,  et~al., 2018, \mn@doi [\apj] {10/gkk68n}, 858, 29

\bibitem[\protect\citeauthoryear{Papastergis, Adams  \& Romanowsky}{Papastergis
  et~al.}{2017}]{papastergis2017}
Papastergis E.,  Adams E. a.~K.,   Romanowsky A.~J.,  2017, \mn@doi [\aap]
  {10/gknfm6}, 601, L10

\bibitem[\protect\citeauthoryear{Papovich, Dickinson  \& Ferguson}{Papovich
  et~al.}{2001}]{papovich2001}
Papovich C.,  Dickinson M.,   Ferguson H.~C.,  2001, \mn@doi [\apj]
  {10/dr4bt4}, 559, 620

\bibitem[\protect\citeauthoryear{Paxton, Bildsten, Dotter, Herwig, Lesaffre  \&
  Timmes}{Paxton et~al.}{2011}]{paxton2011}
Paxton B.,  Bildsten L.,  Dotter A.,  Herwig F.,  Lesaffre P.,   Timmes F.,
  2011, \mn@doi [\apjs] {10.1088/0067-0049/192/1/3}, 192, 3

\bibitem[\protect\citeauthoryear{Paxton et~al.,}{Paxton
  et~al.}{2013}]{paxton2013}
Paxton B.,  et~al., 2013, \mn@doi [\apjs] {10.1088/0067-0049/208/1/4}, 208, 4

\bibitem[\protect\citeauthoryear{Paxton et~al.,}{Paxton
  et~al.}{2015}]{paxton2015}
Paxton B.,  et~al., 2015, \mn@doi [\apjs] {10.1088/0067-0049/220/1/15}, 220, 15

\bibitem[\protect\citeauthoryear{Paxton et~al.,}{Paxton
  et~al.}{2018}]{paxton2018}
Paxton B.,  et~al., 2018, \mn@doi [\apjs] {10.3847/1538-4365/aaa5a8}, 234, 34

\bibitem[\protect\citeauthoryear{P{\'e}rez \& Granger}{P{\'e}rez \&
  Granger}{2007}]{ipython}
P{\'e}rez F.,  Granger B.~E.,  2007, \mn@doi [Computing in Science \&
  Engineering] {10.1109/mcse.2007.53}, 9, 21

\bibitem[\protect\citeauthoryear{P{\'e}roux \& Howk}{P{\'e}roux \&
  Howk}{2020}]{peroux2020}
P{\'e}roux C.,  Howk J.~C.,  2020, \mn@doi [Annual Review of Astronomy and
  Astrophysics] {10.1146/annurev-astro-021820-120014}, 58, annurev

\bibitem[\protect\citeauthoryear{Polzin, {van Dokkum}, Danieli, Greco  \&
  Romanowsky}{Polzin et~al.}{2021}]{polzin2021}
Polzin A.,  {van Dokkum} P.,  Danieli S.,  Greco J.~P.,   Romanowsky A.~J.,
  2021, \mn@doi [\apj] {10.3847/2041-8213/ac024f}, 914, L23

\bibitem[\protect\citeauthoryear{Renzini}{Renzini}{2006}]{renzini2006}
Renzini A.,  2006, \mn@doi [\araa] {10.1146/annurev.astro.44.051905.092450},
  44, 141

\bibitem[\protect\citeauthoryear{Rinaldi, Caputi, {van Mierlo}, Ashby, Caminha
  \& Iani}{Rinaldi et~al.}{2021}]{rinaldi2021}
Rinaldi P.,  Caputi K.~I.,  {van Mierlo} S.,  Ashby M. L.~N.,  Caminha G.~B.,
  Iani E.,  2021, arXiv:2112.03935 [astro-ph]

\bibitem[\protect\citeauthoryear{Rong, Guo, Gao, Liao, Xie, Puzia, Sun  \&
  Pan}{Rong et~al.}{2017}]{rong2017}
Rong Y.,  Guo Q.,  Gao L.,  Liao S.,  Xie L.,  Puzia T.~H.,  Sun S.,   Pan J.,
  2017, \mn@doi [\mnras] {10/gbvjnw}, 470, 4231

\bibitem[\protect\citeauthoryear{{Ruiz-Lara} et~al.,}{{Ruiz-Lara}
  et~al.}{2015}]{ruiz-lara2015}
{Ruiz-Lara} T.,  et~al., 2015, \mn@doi [\aap] {10.1051/0004-6361/201526752},
  \href {https://ui.adsabs.harvard.edu/abs/2015A&A...583A..60R} {583, A60}

\bibitem[\protect\citeauthoryear{{Ruiz-Lara} et~al.,}{{Ruiz-Lara}
  et~al.}{2018}]{ruizlara2018}
{Ruiz-Lara} T.,  et~al., 2018, \mn@doi [\mnras] {10/gdzwvt}, 478, 2034

\bibitem[\protect\citeauthoryear{Saifollahi, Trujillo, Beasley, Peletier  \&
  Knapen}{Saifollahi et~al.}{2021}]{saifollahi2021}
Saifollahi T.,  Trujillo I.,  Beasley M.~A.,  Peletier R.~F.,   Knapen J.~H.,
  2021, \mn@doi [\mnras] {10/gk8xfd}, 502, 5921

\bibitem[\protect\citeauthoryear{Saifollahi, Zaritsky, Trujillo, Peletier,
  Knapen, Amorisco, Beasley  \& Donnerstein}{Saifollahi
  et~al.}{2022}]{saifollahi2022}
Saifollahi T.,  Zaritsky D.,  Trujillo I.,  Peletier R.~F.,  Knapen J.~H.,
  Amorisco N.,  Beasley M.~A.,   Donnerstein R.,  2022, arXiv:2201.11750
  [astro-ph]

\bibitem[\protect\citeauthoryear{Sales, Navarro, Pe{\~n}afiel, Peng, Lim  \&
  Hernquist}{Sales et~al.}{2020}]{sales2020}
Sales L.~V.,  Navarro J.~F.,  Pe{\~n}afiel L.,  Peng E.~W.,  Lim S.,
  Hernquist L.,  2020, \mn@doi [\mnras] {10/gmzdvp}, 494, 1848

\bibitem[\protect\citeauthoryear{{Salvador-Rusi{\~n}ol}, Vazdekis, La~Barbera,
  Beasley, Ferreras, Negri  \& Dalla~Vecchia}{{Salvador-Rusi{\~n}ol}
  et~al.}{2020}]{salvador-rusinol2020}
{Salvador-Rusi{\~n}ol} N.,  Vazdekis A.,  La~Barbera F.,  Beasley M.~A.,
  Ferreras I.,  Negri A.,   Dalla~Vecchia C.,  2020, \mn@doi [Nature Astronomy]
  {10.1038/s41550-019-0955-0}, 4, 252

\bibitem[\protect\citeauthoryear{{S{\'a}nchez-Bl{\'a}zquez}
  et~al.,}{{S{\'a}nchez-Bl{\'a}zquez} et~al.}{2006}]{sanchez-blazquez2006}
{S{\'a}nchez-Bl{\'a}zquez} P.,  et~al., 2006, \mn@doi [\mnras]
  {10.1111/j.1365-2966.2006.10699.x}, \href
  {https://ui.adsabs.harvard.edu/abs/2006MNRAS.371..703S} {371, 703}

\bibitem[\protect\citeauthoryear{{S{\'a}nchez-Bl{\'a}zquez}, Ocvirk, Gibson,
  P{\'e}rez  \& Peletier}{{S{\'a}nchez-Bl{\'a}zquez}
  et~al.}{2011}]{sanchezblazquez2011}
{S{\'a}nchez-Bl{\'a}zquez} P.,  Ocvirk P.,  Gibson B.~K.,  P{\'e}rez I.,
  Peletier R.~F.,  2011, \mn@doi [\mnras] {10.1111/j.1365-2966.2011.18749.x},
  415, 709

\bibitem[\protect\citeauthoryear{Schaye et~al.,}{Schaye
  et~al.}{2010}]{schaye2010}
Schaye J.,  et~al., 2010, \mn@doi [\mnras] {10.1111/j.1365-2966.2009.16029.x},
  402, 1536

\bibitem[\protect\citeauthoryear{Schiavon}{Schiavon}{2007}]{schiavon2007}
Schiavon R.~P.,  2007, \mn@doi [\apjs] {10/bt8gsx}, 171, 146

\bibitem[\protect\citeauthoryear{Schiavon, Rose, Courteau  \&
  MacArthur}{Schiavon et~al.}{2004}]{schiavon2004}
Schiavon R.~P.,  Rose J.~A.,  Courteau S.,   MacArthur L.~A.,  2004, \mn@doi
  [\apj] {10/d3vtgm}, 608, L33

\bibitem[\protect\citeauthoryear{Schiavon, Rose, Courteau  \&
  MacArthur}{Schiavon et~al.}{2005}]{schiavon2005}
Schiavon R.~P.,  Rose J.~A.,  Courteau S.,   MacArthur L.~A.,  2005, \mn@doi
  [\apjs] {10/dp8wkn}, 160, 163

\bibitem[\protect\citeauthoryear{Schlafly \& Finkbeiner}{Schlafly \&
  Finkbeiner}{2011}]{schlafly2011}
Schlafly E.~F.,  Finkbeiner D.~P.,  2011, \mn@doi [\apj]
  {10.1088/0004-637X/737/2/103}, 737, 103

\bibitem[\protect\citeauthoryear{Serra \& Trager}{Serra \&
  Trager}{2007}]{serra2007}
Serra P.,  Trager S.~C.,  2007, \mn@doi [\mnras] {10/dg48fh}, 374, 769

\bibitem[\protect\citeauthoryear{Skilling}{Skilling}{2004}]{skilling2004}
Skilling J.,  2004, \mn@doi [] {10/d9p42j}, 735, 395

\bibitem[\protect\citeauthoryear{Speagle}{Speagle}{2020}]{speagle2020}
Speagle J.~S.,  2020, \mn@doi [\mnras] {10/ghjwgz}, 493, 3132

\bibitem[\protect\citeauthoryear{Tacchella et~al.,}{Tacchella
  et~al.}{2021}]{tacchella2021}
Tacchella S.,  et~al., 2021, arXiv, 2102, arXiv:2102.12494

\bibitem[\protect\citeauthoryear{{The Astropy Collaboration} et~al.,}{{The
  Astropy Collaboration} et~al.}{2013}]{astropy2}
{The Astropy Collaboration} et~al., 2013, \mn@doi [\aap]
  {10.1051/0004-6361/201322068}, 558, A33

\bibitem[\protect\citeauthoryear{{The Astropy Collaboration} et~al.,}{{The
  Astropy Collaboration} et~al.}{2018}]{astropy1}
{The Astropy Collaboration} et~al., 2018, \mn@doi [\aj]
  {10.3847/1538-3881/aabc4f}, 156, 123

\bibitem[\protect\citeauthoryear{Thomas, Maraston, Bender  \& {de
  Oliveira}}{Thomas et~al.}{2005}]{thomas2005}
Thomas D.,  Maraston C.,  Bender R.,   {de Oliveira} C.~M.,  2005, \mn@doi
  [\apj] {10.1086/426932}, 621, 673

\bibitem[\protect\citeauthoryear{Thomas, Maraston, Schawinski, Sarzi  \&
  Silk}{Thomas et~al.}{2010}]{thomas2010}
Thomas D.,  Maraston C.,  Schawinski K.,  Sarzi M.,   Silk J.,  2010, \mn@doi
  [\mnras] {10.1111/j.1365-2966.2010.16427.x}

\bibitem[\protect\citeauthoryear{Trager, Faber, Worthey  \&
  Gonz{\'a}lez}{Trager et~al.}{2000}]{trager2000b}
Trager S.~C.,  Faber S.~M.,  Worthey G.,   Gonz{\'a}lez J.~J.,  2000, \mn@doi
  [\aj] {10.1086/301442}, 120, 165

\bibitem[\protect\citeauthoryear{Tremmel, Wright, Brooks, Munshi, Nagai  \&
  Quinn}{Tremmel et~al.}{2020}]{tremmel2020}
Tremmel M.,  Wright A.~C.,  Brooks A.~M.,  Munshi F.,  Nagai D.,   Quinn T.~R.,
   2020, \mn@doi [\mnras] {10/gk8xfk}, 497, 2786

\bibitem[\protect\citeauthoryear{{Trujillo-Gomez}, {Reina-Campos}  \&
  Kruijssen}{{Trujillo-Gomez} et~al.}{2019}]{trujillo-gomez2019}
{Trujillo-Gomez} S.,  {Reina-Campos} M.,   Kruijssen J. M.~D.,  2019, \mn@doi
  [\mnras] {10/gpbzhg}, 488, 3972

\bibitem[\protect\citeauthoryear{Van~Nest, Munshi, Wright, Tremmel, Brooks,
  Nagai  \& Quinn}{Van~Nest et~al.}{2022}]{vannest2022}
Van~Nest J.~D.,  Munshi F.,  Wright A.~C.,  Tremmel M.,  Brooks A.~M.,  Nagai
  D.,   Quinn T.,  2022, \mn@doi [\apj] {10.3847/1538-4357/ac43b7}, 926, 92

\bibitem[\protect\citeauthoryear{Vazdekis}{Vazdekis}{1999}]{vazdekis1999}
Vazdekis A.,  1999, \mn@doi [\apj] {10.1086/306843}, 513, 224

\bibitem[\protect\citeauthoryear{Villaume et~al.,}{Villaume
  et~al.}{2022}]{villaume2022}
Villaume A.,  et~al., 2022, \mn@doi [\apj] {10/gn7jmn}, 924, 32

\bibitem[\protect\citeauthoryear{Virtanen et~al.,}{Virtanen
  et~al.}{2020}]{scipy}
Virtanen P.,  et~al., 2020, \mn@doi [Nature Methods]
  {10.1038/s41592-019-0686-2}, 17, 261

\bibitem[\protect\citeauthoryear{Wasserman et~al.,}{Wasserman
  et~al.}{2019}]{wasserman2019}
Wasserman A.,  et~al., 2019, \mn@doi [\apj] {10/gnh3k4}, 885, 155

\bibitem[\protect\citeauthoryear{Wechsler \& Tinker}{Wechsler \&
  Tinker}{2018}]{wechsler2018}
Wechsler R.~H.,  Tinker J.~L.,  2018, \mn@doi [\araa]
  {10.1146/annurev-astro-081817-051756}, 56, 435

\bibitem[\protect\citeauthoryear{Weisz et~al.,}{Weisz
  et~al.}{2011}]{weisz2011a}
Weisz D.~R.,  et~al., 2011, \mn@doi [\apj] {10.1088/0004-637X/739/1/5}, 739, 5

\bibitem[\protect\citeauthoryear{Werner et~al.,}{Werner
  et~al.}{2004}]{werner2004}
Werner M.~W.,  et~al., 2004, \mn@doi [\apjs] {10/dm6jrx}, 154, 1

\bibitem[\protect\citeauthoryear{White \& Frenk}{White \&
  Frenk}{1991}]{white1991}
White S. D.~M.,  Frenk C.~S.,  1991, \mn@doi [\apj] {10/bkstjz}, 379, 52

\bibitem[\protect\citeauthoryear{Worthey}{Worthey}{1994}]{worthey1994}
Worthey G.,  1994, \mn@doi [\apjs] {10.1086/192096}, 95, 107

\bibitem[\protect\citeauthoryear{Wright, Tremmel, Brooks, Munshi, Nagai, Sharma
   \& Quinn}{Wright et~al.}{2021}]{wright2021}
Wright A.~C.,  Tremmel M.,  Brooks A.~M.,  Munshi F.,  Nagai D.,  Sharma R.~S.,
    Quinn T.~R.,  2021, \mn@doi [\mnras] {10/gmz8bm}, 502, 5370

\bibitem[\protect\citeauthoryear{Yagi, Koda, Komiyama  \& Yamanoi}{Yagi
  et~al.}{2016}]{yagi2016}
Yagi M.,  Koda J.,  Komiyama Y.,   Yamanoi H.,  2016, \mn@doi [\apjs]
  {10/gmzdv8}, 225, 11

\bibitem[\protect\citeauthoryear{Yozin \& Bekki}{Yozin \&
  Bekki}{2015}]{yozin2015a}
Yozin C.,  Bekki K.,  2015, \mn@doi [\mnras] {10/f7q6dc}, 452, 937

\bibitem[\protect\citeauthoryear{{da Cunha}, Charlot  \& Elbaz}{{da Cunha}
  et~al.}{2008}]{dacunha2008}
{da Cunha} E.,  Charlot S.,   Elbaz D.,  2008, \mn@doi [\mnras] {10/bj2rqd},
  388, 1595

\bibitem[\protect\citeauthoryear{{van Dokkum}, {Abraham}, {Merritt}, {Zhang},
  {Geha}  \& {Conroy}}{{van Dokkum} et~al.}{2015}]{vandokkum2015_udg}
{van Dokkum} P.~G.,  {Abraham} R.,  {Merritt} A.,  {Zhang} J.,  {Geha} M.,
  {Conroy} C.,  2015, \mn@doi [\apjl] {10.1088/2041-8205/798/2/L45}, \href
  {https://ui.adsabs.harvard.edu/abs/2015ApJ...798L..45V} {798, L45}

\bibitem[\protect\citeauthoryear{{van Dokkum} et~al.,}{{van Dokkum}
  et~al.}{2016}]{vandokkum2016}
{van Dokkum} P.,  et~al., 2016, \mn@doi [\apj] {10/bpzf}, 828, L6

\bibitem[\protect\citeauthoryear{{van Dokkum} et~al.,}{{van Dokkum}
  et~al.}{2017}]{vandokkum2017}
{van Dokkum} P.,  et~al., 2017, \mn@doi [\apj] {10/gkxzb4}, 844, L11

\bibitem[\protect\citeauthoryear{{van Dokkum} et~al.,}{{van Dokkum}
  et~al.}{2019}]{vandokkum2019}
{van Dokkum} P.,  et~al., 2019, \mn@doi [\apj] {10/gj2kbd}, 880, 91

\bibitem[\protect\citeauthoryear{{van de Schoot} et~al.,}{{van de Schoot}
  et~al.}{2021}]{vandeschoot2021}
{van de Schoot} R.,  et~al., 2021, \mn@doi [Nature Reviews Methods Primers]
  {10/ghs29x}, 1, 1

\bibitem[\protect\citeauthoryear{{van der Hulst}, {Terlouw}, {Begeman},
  {Zwitser}  \& {Roelfsema}}{{van der Hulst} et~al.}{1992}]{gipsy}
{van der Hulst} J.~M.,  {Terlouw} J.~P.,  {Begeman} K.~G.,  {Zwitser} W.,
  {Roelfsema} P.~R.,  1992, in {Worrall} D.~M.,  {Biemesderfer} C.,   {Barnes}
  J.,  eds,  Astronomical Society of the Pacific Conference Series Vol. 25,
  Astronomical Data Analysis Software and Systems I. p.~131

\bibitem[\protect\citeauthoryear{van~der Walt, Colbert  \& Varoquaux}{van~der
  Walt et~al.}{2011}]{numpy}
van~der Walt S.,  Colbert S.~C.,   Varoquaux G.,  2011, \mn@doi [Computing in
  Science \& Engineering] {10.1109/mcse.2011.37}, 13, 22

\makeatother
\end{thebibliography}
